\documentclass[twocolumn]{aastex63}

\usepackage{xcolor}

\def\Angstrom{\textup{\AA}}

\newcommand{\LCO}{\affiliation{Las Cumbres Observatory, 6740 Cortona Drive, Suite 102, Goleta, CA 93117-5575, USA}}
\newcommand{\UCSB}{\affiliation{Department of Physics, University of California, Santa Barbara, CA 93106-9530, USA}}

\newcommand{\UCD}{\affiliation{Department of Physics, University of California, 1 Shields Avenue, Davis, CA 95616-5270, USA}}

\newcommand{\QUB}{\affiliation{Astrophysics Research Centre, School of Mathematics and Physics, Queen's University Belfast, Belfast BT7 1NN, UK}}

\newcommand{\Southampton}{\affiliation{Department of Physics and Astronomy, University of Southampton, Southampton SO17 1BJ, UK}}

\newcommand{\CfA}{\affiliation{Center for Astrophysics \textbar{} Harvard \& Smithsonian, 60 Garden Street, Cambridge, MA 02138-1516, USA}}

\newcommand{\TAU}{\affiliation{School of Physics and Astronomy, Tel Aviv University, Tel Aviv 69978, Israel}}
\newcommand{\CIFAR}{\affiliation{CIFAR Azrieli Global Scholars program, CIFAR, Toronto, Canada}}
\newcommand{\NAOJ}{\affiliation{National Astronomical Observatory of Japan, National Institutes of Natural Sciences, 2-21-1 Osawa, Mitaka, Tokyo 181-8588, Japan}}
\newcommand{\Monash}{\affiliation{School of Physics and Astronomy, Faculty of Science, Monash University, Clayton, VIC 3800, Australia}}
\newcommand{\ESO}{\affiliation{European Southern Observatory, Alonso de C\'{o}rdova 3107, Casilla 19, Santiago, Chile}}
\newcommand{\Kyoto}{\affiliation{Department of Astronomy, Kyoto University, Kitashirakawa-Oiwake-cho, Sakyo-ku, Kyoto 606-8502, Japan}}
\newcommand{\IPMU}{\affiliation{Kavli Institute for the Physics and Mathematics of the Universe (WPI), The University of Tokyo Institutes for Advanced Study, The University of Tokyo, 5-1-5 Kashiwanoha, Kashiwa, Chiba 277-8583, Japan}}
\newcommand{\Granada}{\affiliation{Departamento de F\'{i}sica Te\'{o}rica y del Cosmos, Universidad de Granada, 18071 Granada, Spain}}
\newcommand{\Aarhus}{\affiliation{Department of Physics and Astronomy, Aarhus University, Ny Munkegade 120, DK-8000 Aarhus C, Denmark}}
\newcommand{\FSU}{\affiliation{Department of Physics, Florida State University, Tallahassee, FL 32306, USA}}
\newcommand{\LasCampanas}{\affiliation{Las Campanas Observatory, Carnegie Observatories, Casilla 601, La Serena, Chile}}
\newcommand{\Warsaw}{\affiliation{Astronomical Observatory, University of Warsaw, Al. Ujazdowskie 4, 00-478 Warszawa, Poland}}
\newcommand{\TAM}{\affiliation{George P. and Cynthia Woods Mitchell Institute for Fundamental Physics and Astronomy, Department of Physics and Astronomy, Texas A\&M University, College Station, TX 77843, USA}}
\newcommand{\TCD}{\affiliation{School of Physics, Trinity College Dublin, The University of Dublin, Dublin 2, Ireland}}
\newcommand{\FINCA}{\affiliation{Finnish Centre for Astronomy with ESO (FINCA), FI-20014 University of Turku, Finland}}
\newcommand{\Tuorla}{\affiliation{Tuorla Observatory, Department of Physics and Astronomy, FI-20014 University of Turku, Finland}}
\newcommand{\IALP}{\affiliation{Instituto de Astrof\'{i}sica de La Plata (IALP), CONICET, Argentina}}
\newcommand{\LaPlata}{\affiliation{Facultad de Ciencias Astron\'{o}micas y Geof\'{i}sicas, Universidad Nacional de La Plata, Paseo del Bosque, B1900FWA, La Plata, Argentina}}

\received{2020 October 29}
\revised{2021 April 9}
\accepted{2021 April 10}
\published{2021 May 25}

\shorttitle{Luminous Type II Short-Plateau Supernovae 2006Y, 2006ai, and 2016egz}
\shortauthors{Hiramatsu et al.}

\begin{document}

\title{\bf \Large Luminous Type II Short-Plateau Supernovae 2006Y, 2006ai, and 2016egz: A Transitional Class from Stripped Massive Red Supergiants}

\author[0000-0002-1125-9187]{Daichi Hiramatsu}
\LCO\UCSB

\correspondingauthor{Daichi~Hiramatsu}
\email{dhiramatsu@lco.global}

\author[0000-0003-4253-656X]{D.~Andrew~Howell}
\LCO\UCSB

\author[0000-0003-1169-1954]{Takashi~J.~Moriya}
\NAOJ\Monash

\author[0000-0003-1012-3031]{Jared~A.~Goldberg}
\UCSB

\author[0000-0002-0832-2974]{Griffin~Hosseinzadeh}
\CfA

\author[0000-0001-7090-4898]{Iair~Arcavi}
\TAU\CIFAR

\author[0000-0003-0227-3451]{Joseph~P.~Anderson}
\ESO

\author[0000-0003-2375-2064]{Claudia~P.~Guti\'{e}rrez}
\Southampton
\FINCA
\Tuorla

\author[0000-0003-0035-6659]{Jamison~Burke}
\LCO\UCSB

\author[0000-0001-5807-7893]{Curtis~McCully}
\LCO\UCSB

\author[0000-0001-8818-0795]{Stefano~Valenti}
\UCD

\author[0000-0002-1296-6887]{Llu\'{i}s~Galbany}
\Granada

\author[0000-0002-1161-9592]{Qiliang~Fang}
\Kyoto

\author[0000-0003-2611-7269]{Keiichi~Maeda}
\Kyoto\IPMU

\author[0000-0001-5247-1486]{Gast\'{o}n~Folatelli}
\IALP\LaPlata\IPMU

\author[0000-0003-1039-2928]{Eric~Y.~Hsiao}
\FSU

\author[0000-0003-2535-3091]{Nidia~I.~Morrell}
\LasCampanas

\author[0000-0003-2734-0796]{Mark~M.~Phillips}
\LasCampanas

\author[0000-0002-5571-1833]{Maximilian~D.~Stritzinger}
\Aarhus

\author[0000-0002-8102-181X]{Nicholas~B.~Suntzeff}
\TAM

\author[0000-0002-1650-1518]{Mariusz~Gromadzki}
\Warsaw

\author[0000-0002-9770-3508]{Kate~Maguire}
\TCD

\author[0000-0003-3939-7167]{Tom\'{a}s~E.~M\"{u}ller-Bravo}
\Southampton

\author[0000-0002-1229-2499]{David~R.~Young}
\QUB

\begin{abstract}

The diversity of Type II supernovae (SNe II) is thought to be driven mainly by differences in their progenitor's hydrogen-rich (H-rich) envelope mass, with SNe IIP having long plateaus ($\sim100$ days) and the most massive H-rich envelopes. 
However, it is an ongoing mystery why SNe II with short plateaus (tens of days) are rarely seen. 
Here, we present optical/near-infrared photometric and spectroscopic observations of luminous Type II short-plateau SNe 2006Y, 2006ai, and 2016egz. 
Their plateaus of about $50$--$70$ days and luminous optical peaks ($\lesssim-18.4$ mag) indicate significant pre-explosion mass loss resulting in partially stripped H-rich envelopes and early circumstellar material (CSM) interaction. 
We compute a large grid of \texttt{MESA}+\texttt{STELLA} single-star progenitor and light-curve models with various progenitor zero-age main-sequence (ZAMS) masses, mass-loss efficiencies, explosion energies, $^{56}$Ni masses, and CSM densities. Our model grid shows a continuous population of SNe IIP--IIL--IIb-like light-curve morphology in descending order of H-rich envelope mass. With large $^{56}$Ni masses ($\gtrsim0.05\,M_\odot$), short-plateau SNe~II lie in a confined parameter space as a transitional class between SNe IIL and IIb.
For SNe~2006Y, 2006ai, and 2016egz, our findings suggest high-mass red supergiant (RSG) progenitors ($M_{\rm ZAMS} \simeq 18$--$22\,M_{\odot}$) with small H-rich envelope masses ($M_{\rm H_{\rm env}} \simeq 1.7\,M_{\odot}$) that have experienced enhanced mass loss ($\dot{M} \simeq 10^{-2}\,M_{\odot}\,{\rm yr}^{-1}$) for the last few decades before the explosion.
If high-mass RSGs result in rare short-plateau SNe II, then these events might ease some of the apparent underrepresentation of higher-luminosity RSGs in observed SN II progenitor samples.

\end{abstract}

\keywords{
\href{https://vocabs.ands.org.au/repository/api/lda/aas/the-unified-astronomy-thesaurus/current/resource.html?uri=http://astrothesaurus.org/uat/1668}{Supernovae (1668)}; 
\href{https://vocabs.ands.org.au/repository/api/lda/aas/the-unified-astronomy-thesaurus/current/resource.html?uri=http://astrothesaurus.org/uat/304}{Core-collapse supernovae (304)}; 
\href{https://vocabs.ands.org.au/repository/api/lda/aas/the-unified-astronomy-thesaurus/current/resource.html?uri=http://astrothesaurus.org/uat/1731}{Type II supernovae (1731)}; 
\href{https://vocabs.ands.org.au/repository/api/lda/aas/the-unified-astronomy-thesaurus/current/resource.html?uri=http://astrothesaurus.org/uat/732}{Massive stars (732)};
\href{https://vocabs.ands.org.au/repository/api/lda/aas/the-unified-astronomy-thesaurus/current/resource.html?uri=http://astrothesaurus.org/uat/1375}{Red supergiant stars (1375)};
}

\section{Introduction} \label{sec:intro}

The majority of massive stars (zero-age main-sequence masses $M_{\mathrm{ZAMS}} \gtrsim8\,M_{\odot}$) end their lives when their iron cores collapse and explode as hydrogen-rich (H-rich), Type II supernovae (SNe II) \citep{Smartt2009, Smartt2015CCSN}. 
The difference in progenitor's H-rich envelope mass at the moment of core collapse likely results in different SN~II subtypes (e.g., \citealt{Nomoto1995,Heger2003,Dessart2011WR,Dessart2016Ibc,Eldridge2017BPASS,Eldridge2018}): SNe IIP (light-curve plateau of $\sim 100$ days); SNe IIL (linear decline light curve); and SNe IIb (spectrum dominated initially by hydrogen and then by helium), in descending order (see \citealt{Arcavi2017} for a review).
SNe IIn show narrow H emission lines, indicating strong circumstellar material (CSM) interaction (see \citealt{Smith2017} for a review).
Based on the direct progenitor identifications in pre-explosion images, the current consensus is that the progenitors are red supergiants (RSGs) for SNe IIP; yellow supergiants in a binary system for SNe IIb; luminous blue variables and RSGs/super-asymptotic giant branch stars for SNe IIn; and RSGs and/or yellow supergiants for SNe IIL, in descending order of confidence (see \citealt{VanDyk2016} for a review).

The division between SNe IIP and IIL is both arbitrary and controversial because it is solely based on the shape of their photospheric-phase optical light curves \citep{Barbon1979}, while SNe IIb and IIn are spectroscopically distinct. 
There have been claims of distinct light-curve populations of SNe IIP and IIL (e.g., \citealt{Arcavi2012,Faran2014}), but larger light-curve samples have increased the support for a more continuous population (e.g., \citealt{Anderson2014,Sanders2015,Galbany2016II,Valenti2016}). 
While SNe IIP and IIL show a continuous range of spectroscopic properties in optical (e.g., \citealt{Gutierrez2017a,Gutierrez2017b}), \cite{Davis2019} recently find a strong dichotomy of near-infrared (NIR) spectroscopic properties between SNe IIP and IIL, which may point to differences in the immediate environment.

In terms of the photospheric plateau duration, it is puzzling that SNe II with short plateaus (tens of days) are rarely observed (see e.g., \citealt{Nakaoka2019} and \citealt{Bostroem2020} for peculiar SNe IIb and IIL), despite analytical and numerical predictions that the plateau duration scales continuously with progenitor and explosion properties \citep{Popov1993,Kasen2009,Sukhbold2016,Goldberg2019}.
In a broader context of SN~II population, it is also an outstanding question whether SNe IIP/L and IIb form a continuum or not (e.g., \citealt{Arcavi2012,Faran2014,Pessi2019}).

The increasing sample size of SNe IIP/L suggests that CSM interaction, resulting from violent pre-explosion mass loss, plays a key role even when their spectra do not show IIn-like Balmer emission lines. By fitting numerical models to the \cite{Valenti2015,Valenti2016} light-curve sample, \cite{Morozova2017,Morozova2018} show that SNe IIP from RSG progenitors with CSM interaction can reproduce SNe IIL. They also show that CSM interaction is required in even normal SNe IIP to reproduce the rapid UV-optical rise in the models (see also \citealt{Moriya2011,Moriya2017,Moriya2018,Forster2018}).

The observed RSG population (in the Milky Way, Magellanic Clouds, M31, and M33) lies in a luminosity range of $4.5 \lesssim {\rm log}_{10}(L/L_{\odot}) \lesssim 5.5$, implying their ZAMS mass range of $\sim9$--$25\,M_{\odot}$ based on theoretical stellar tracks (e.g., \citealt{Levesque2005,Levesque2006,Massey2009,Drout2012,Gordon2016}). However, \cite{Smartt2009,Smartt2015CCSN} show that the best-fit cumulative Salpeter initial mass function (IMF; \citealt{Salpeter1955}) on 26 pre-explosion detections/limits of SNe IIP/L progenitors truncates below the high-luminosity end of RSGs, translating to a ZAMS mass upper limit of $\lesssim18\,M_{\odot}$. This is referred to as the \textit{red supergiant problem}, since there seems to be a lack of SNe II with identified progenitors in the range $\sim18$--$25\,M_\odot$.
Due to the complicated evolution of terminal massive stars and observational uncertainties in dust extinction and bolometric correction, the statistical significance and robustness of the RSG problem has been a highly debated topic (e.g., \citealt{Walmswell2012, Eldridge2013, Kochanek2014, Meynet2015, Sukhbold2016, Adams2017, Davies2018}).

Here, we report optical/NIR photometry and spectroscopy of Type II SNe~2006Y, 2006ai, and 2016egz. 
In Sections \ref{sec:disc} and \ref{sec:obs}, we summarize their discoveries, follow-up observations, and data reduction.
In Section \ref{sec:ana}, we analyze their host galaxies, light curves, and spectra, in addition to producing a large single-star model grid by varying different progenitor and explosion properties. This reveals their transitional nature between SNe IIL and IIb with small H-rich envelope mass, high progenitor ZAMS mass, and dense CSM estimates.
As such, we discuss their formation channel and implications for the RSG problem in Section \ref{sec:dis}.
Finally, we summarize our findings and draw conclusions in Section \ref{sec:con}.

\section{Discoveries} \label{sec:disc}

\cite{Luckas2006Y} discovered SN~2006Y on 2006 February 3.58 (UT dates are used throughout) at 17.7 mag at $\text{R.A.}=07^{\text{h}}13^{\text{m}}17^{\text{s}}.17$ and $\text{Dec.}=-51^{\circ}41'18".8$ with a subsequent detection on 2006 February 7.60 at 17.3 mag and last non-detection limit on 2006 January 27.59 at 18.5 mag, using the unfiltered $35$ cm Tenagra telescope at Perth, Australia. 
With the same instrumental setup, \cite{Luckas2006ai} discovered SN~2006ai on 2006 February 17.54 at 16.2 mag at $\text{R.A.}=07^{\text{h}}29^{\text{m}}52^{\text{s}}.16$ and $\text{Dec.}=-84^{\circ}02'20".5$ with a subsequent detection on 2006 February 19.52 at 16.0 mag and last non-detection limit on 2005 December 16.79 at 18.5 mag.
\cite{Morrell2006} obtained optical spectra of SNe~2006Y and 2006ai on February 27.14 and March 5.12, respectively, with the Las Campanas $2.5$ m du Pont telescope through the Carnegie Supernova Project-I (CSP-I; \citealt{Hamuy2006}), classifying them as SNe~II. CSP-I also obtained optical spectra of the host galaxies of SNe~2006Y and 2006ai and measured redshifts of $z=0.0336\pm0.0001$ and $0.0158\pm0.0001$, respectively.

The All-Sky Automated Survey for Supernovae (ASAS-SN; \citealt{Shappee2014}) discovered SN 2016egz (ASASSN-16hn) on 2016 July 24.32 at 16.1 mag at $\text{R.A.}=00^{\text{h}}04^{\text{m}}03^{\text{s}}.854$ and $\text{Dec.}=-34^{\circ}48'51".87$ with a last non-detection limit on 2016 July 17.23 at 17.4 mag, using the \textit{V}-band $14$ cm ASAS-SN Cassius telescope at Cerro Tololo, Chile \citep{Brown2016}. 
A prediscovery detection on 2016 July 21.26 at 15.5 mag with the same instrumental setup was retrieved via the ASAS-SN light-curve server\footnote{\url{https://asas-sn.osu.edu/}} \citep{Shappee2014,Kochanek2017}. \cite{PESSTO2016} obtained an optical spectrum of SN~2016egz on 2016 July 26.25 with the European Southern Observatory (ESO) $3.58$ m New Technology Telescope (NTT) through the Public ESO Spectroscopic Survey for Transient Objects (PESSTO; \citealt{PESSTO2015}), classifying it as a young SN~II at $z=0.0232\pm0.0003$ of the host galaxy, GALEXASC J000403.88-344851.6 \citep{Colless2003}.\footnote{Via the NASA/IPAC Extragalactic Database (NED): \url{http://ned.ipac.caltech.edu/}} 

Given the tight last non-detection limits, we estimate the explosion epochs of SNe~2006Y and 2016egz by simply taking the midpoint of the last non-detection and the first detection with the error being the estimated explosion epoch minus the last non-detection. This yields $\text{MJD}_0=53766.1\pm3.4$ and $57588.2\pm2.0$ for SNe~2006Y and 2016egz, respectively.
As there is no constraining last non-detection limit for SN~2006ai, we adopt the explosion epoch estimate $\text{MJD}_0=53781.6\pm5.0$ ($\sim1.9$ days before the discovery) from the spectral matching technique of \cite{Anderson2014} and \cite{Gutierrez2017a}. This is reasonable given the early rising light curves (see $\S$\ref{sec:obs}).
For each SN, we use the explosion epoch as a reference epoch for all phases. 
We assume a standard Lambda cold dark matter cosmology with $H_0=71.0$\, km\,s$^{-1}$\,Mpc$^{-1}$, $\Omega_{\Lambda}=0.7$, and $\Omega_m=0.3$, and convert the redshifts to luminosity distances: $d_L=146$ Mpc ($\mu=35.8$ mag), $67.5$ Mpc ($34.1$ mag), and $100$ Mpc ($35.0$ mag), respectively, for SNe~2006Y, 2006ai, and 2016egz.

\section{Observations and Data Reduction} \label{sec:obs}

\begin{figure*}
 \centering
 \gridline{\fig{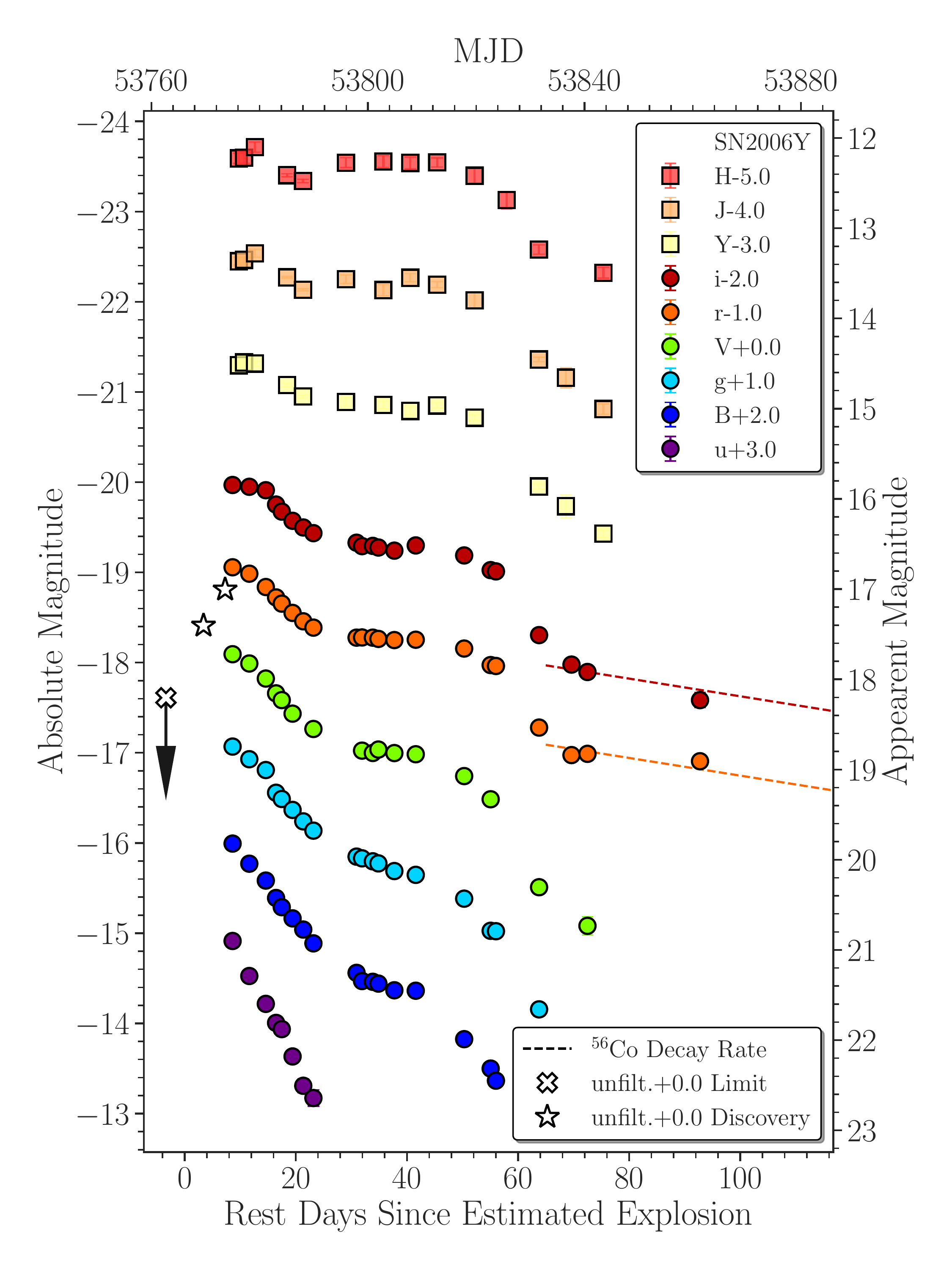}{0.33\textwidth}{\textbf{(a)} SN~2006Y: Swope + du Pont light curves}
            \fig{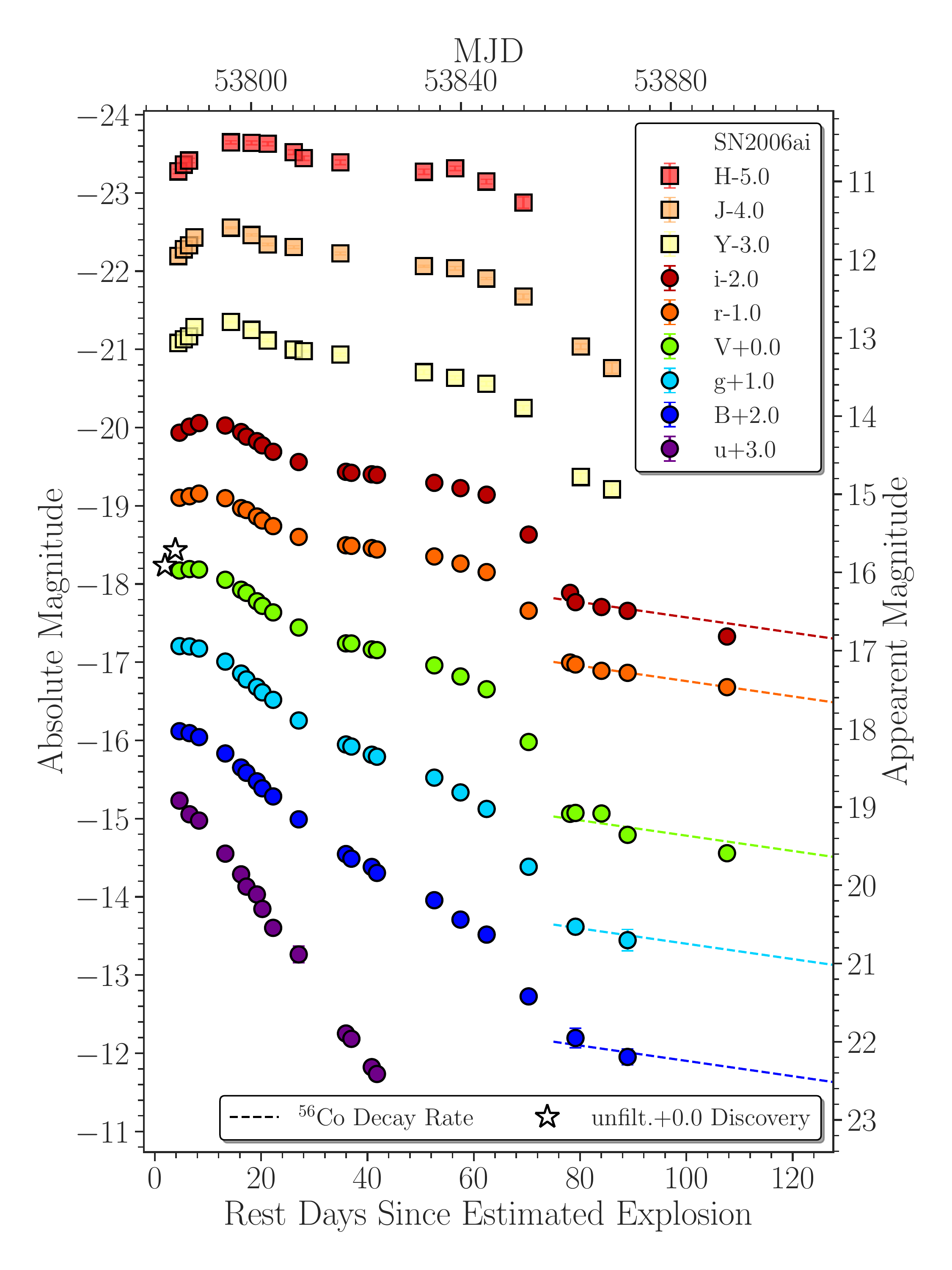}{0.33\textwidth}{\textbf{(b)} SN~2006ai: Swope + du Pont light curves}
            \fig{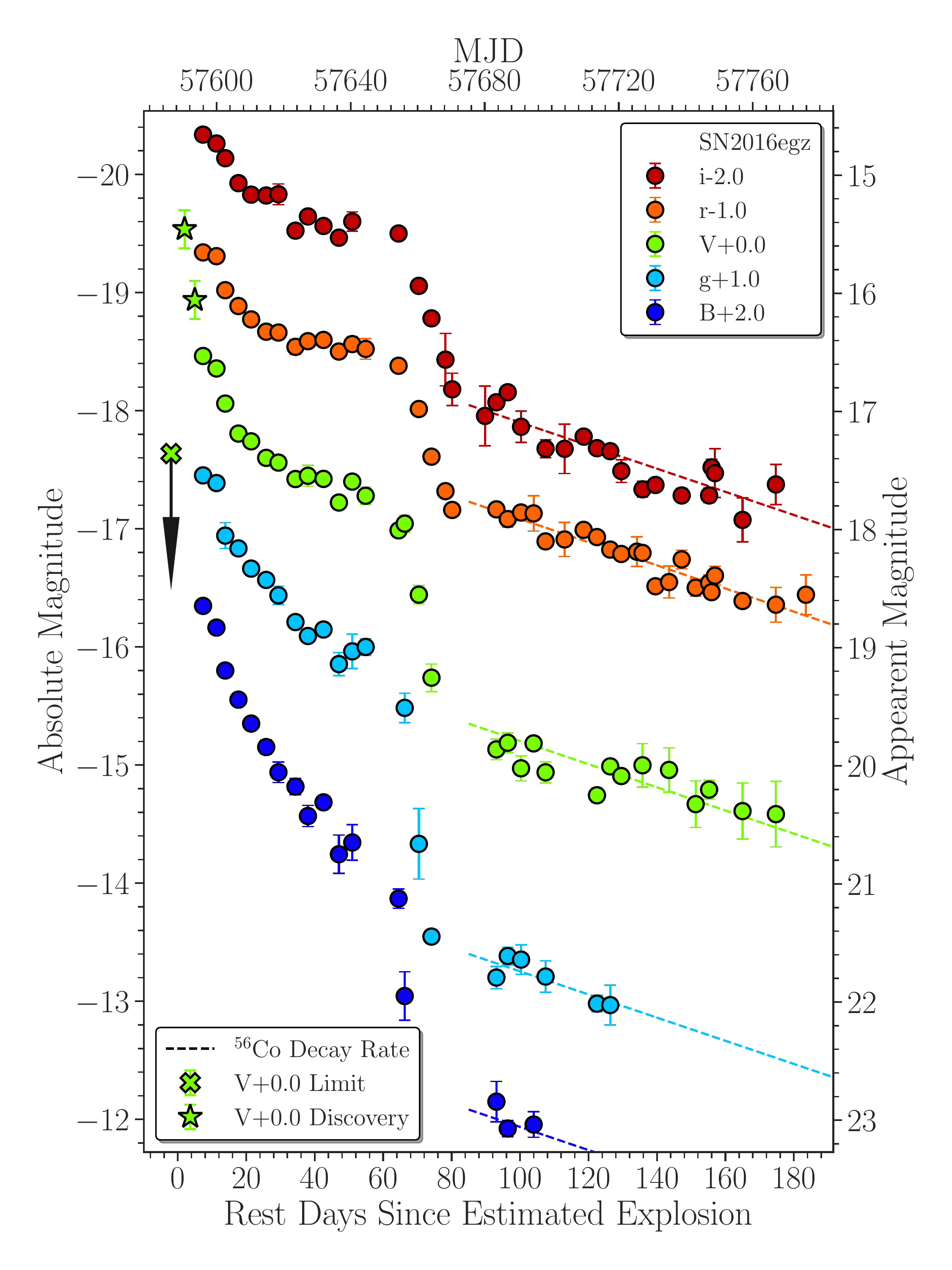}{0.33\textwidth}{\textbf{(c)} SN~2016egz: Las Cumbres light curves}}
  \caption{Host-subtracted and extinction-corrected light curves of SNe~2006Y, 2006ai, and 2016egz. 
  Error bars denote $1\sigma$ uncertainties and are sometimes smaller than the marker size. 
  Note the similar luminous \textit{V}-band peaks ($\lesssim-18.2$ mag) and short \textit{V}-band plateaus ($\sim50$--$70$ days). The tail of SN~2006Y is not well sampled, while those of SNe~2006ai and 2016egz are roughly consistent with $^{56}$Co decay.
  } 
  \label{fig:LC}
\end{figure*}

\begin{figure*}
 \centering
 \gridline{\fig{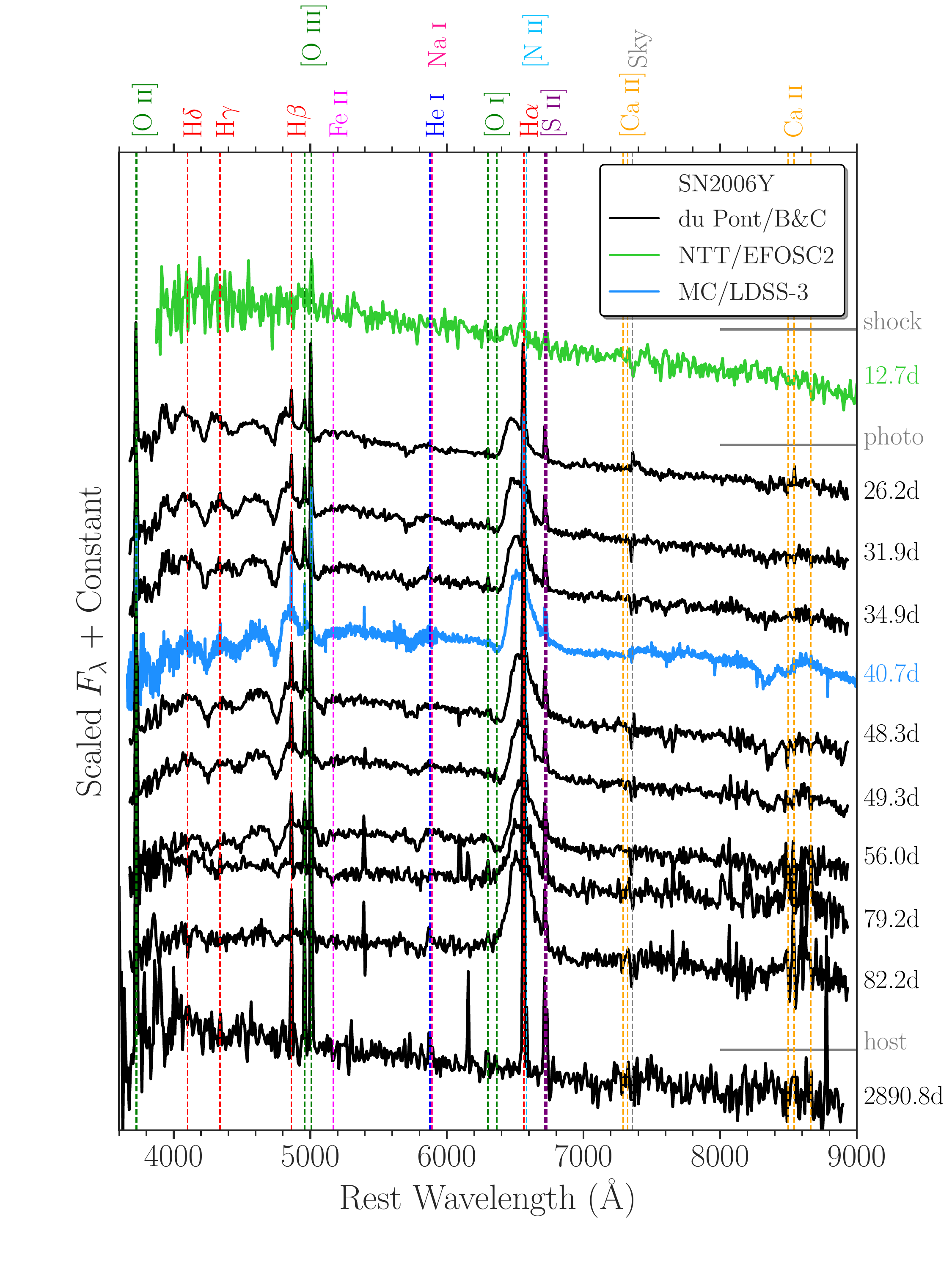}{0.33\textwidth}{\textbf{(a)} SN~2006Y: du Pont + NTT + Magellan Clay (MC) spectral series}
            \fig{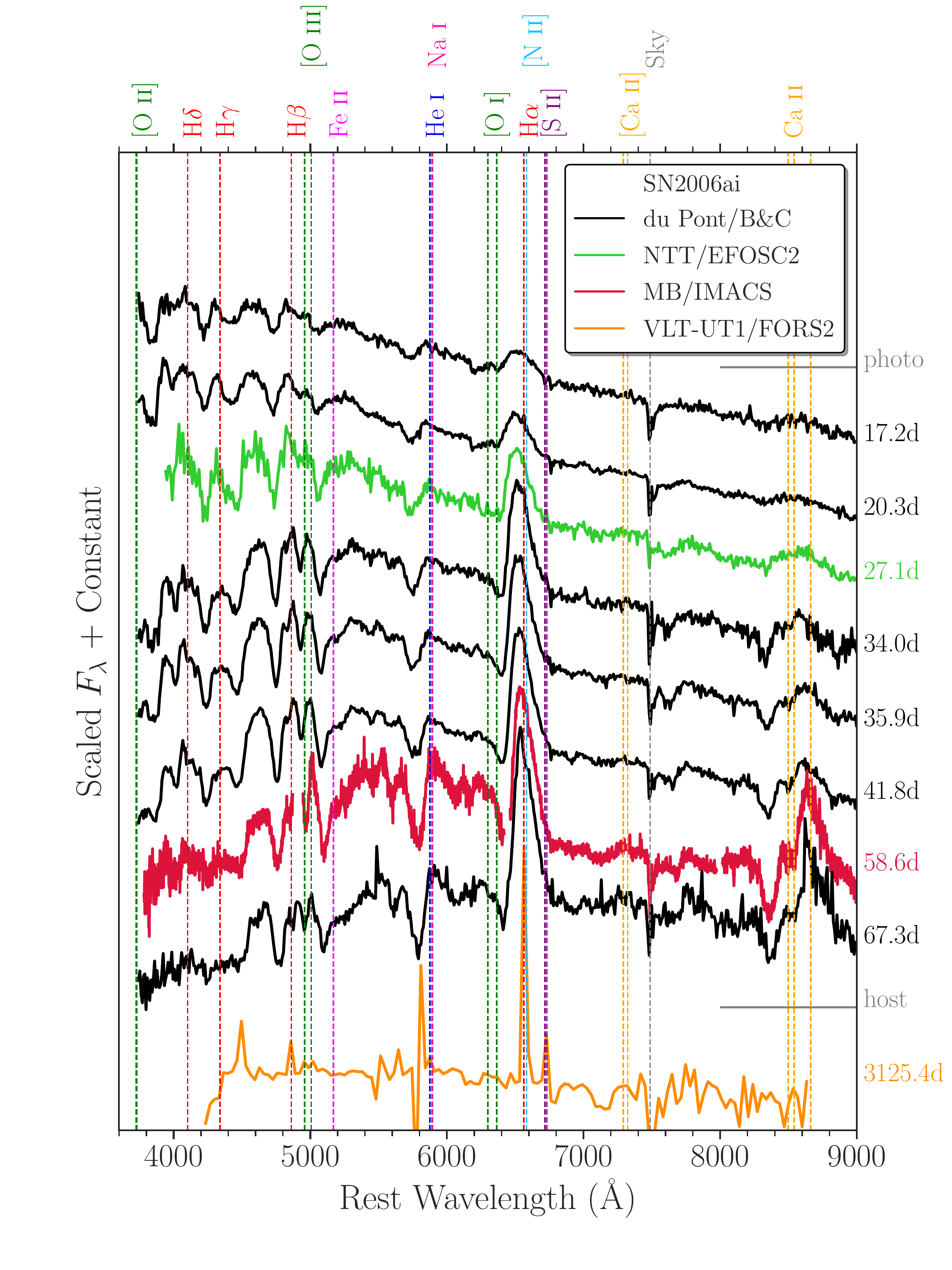}{0.33\textwidth}{\textbf{(b)} SN~2006ai: du Pont + NTT + Magellan Baade (MB) + Very Large Telescope (VLT) spectral series}
            \fig{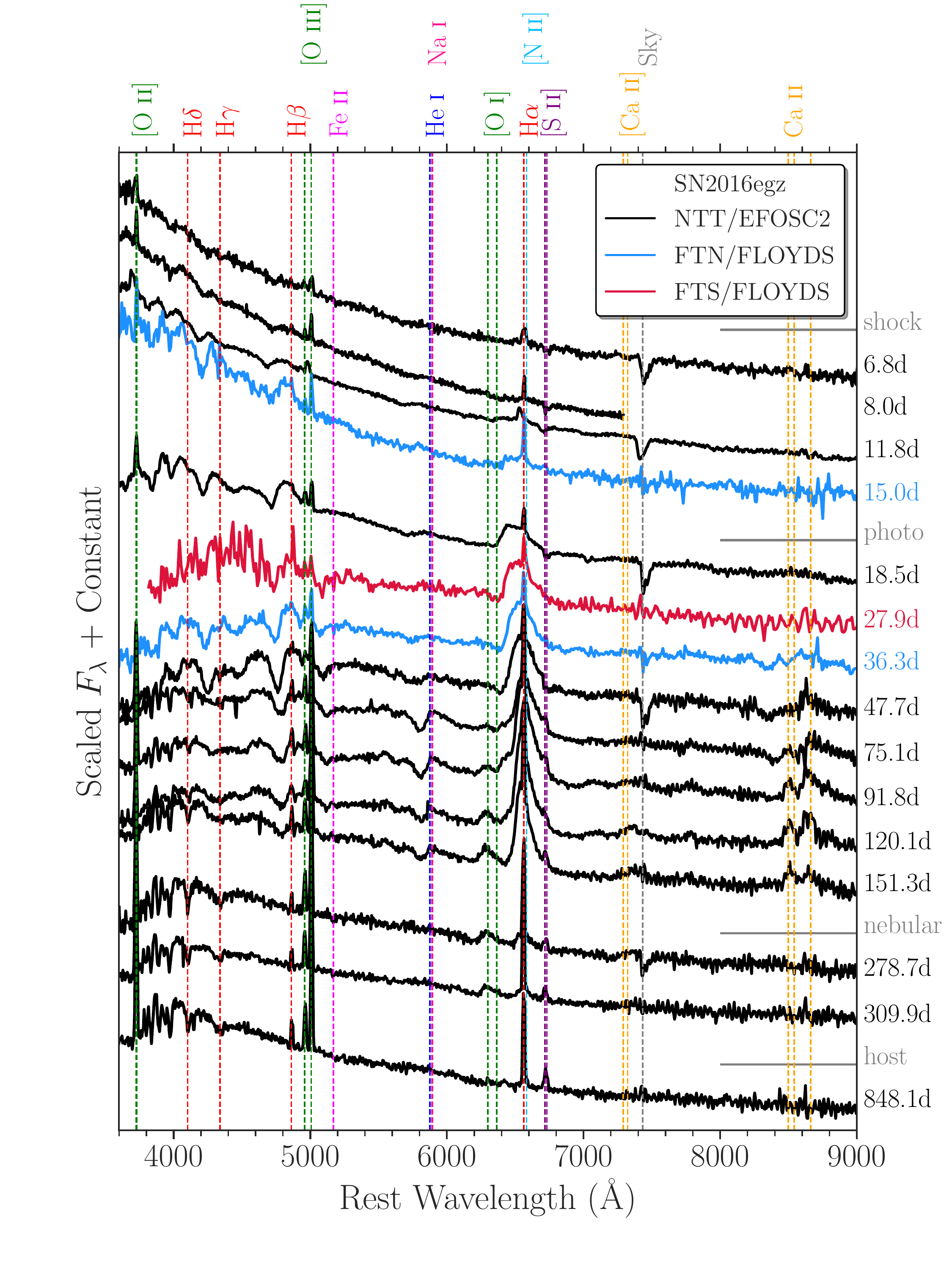}{0.33\textwidth}{\textbf{(c)} SN~2016egz: NTT + FTN/S spectral series}}
  \caption{Extinction-corrected spectral series of SNe~2006Y, 2006ai, and 2016egz. 
  Approximate evolutionary phases are given as: \textit{shock} for shock-cooling phase with a mostly blue featureless continuum; \textit{photo} for optically thick photospheric phase with prominent hydrogen P Cygni features; and \textit{nebular} for optically thin nebular phase with forbidden emission lines (e.g., [O~{\sc i}] $\lambda\lambda6300,6364$).
  Note that the narrow emission lines (e.g., H$\alpha$ and H$\beta$) and the late-time ($\gtrsim 70$\,d) blue continuum with the $4000\,\Angstrom$ break are host contaminants, as seen in the host (i.e., the last) spectra.
  } 
  \label{fig:spec}
\end{figure*}

For SNe~2006Y and 2006ai, \textit{uBgVri} optical and \textit{YJH} NIR photometry were obtained through CSP-I. Standard reduction techniques were applied to all images (e.g., \citealt{Stritzinger2011}). Then deep template observations obtained once the SN had sufficiently faded from detection were used to subtract the underlying host galaxy emission. Photometry of the SN was computed differentially with respect to a local sequence of stars, together with definitive photometry in the standard \textit{ugri} \citep{Smith2002}, \textit{BV} \citep{Landolt1992}, and \textit{YJH} \citep{Persson1998} photometric systems, and calibrated to standard star fields observed on photometric nights (see \citealt{Krisciunas2017} for a detailed description of the above). 
The \textit{V}-band light curves presented here are an updated version to those included in the \cite{Anderson2014} sample.
CSP-I spectroscopy of SNe~2006Y and 2006ai has already been published in \cite{Gutierrez2017a}, and the reader is referred to that publication for more details.

For SN~2016egz, Las Cumbres Observatory \textit{BgVri}-band data were obtained with the SBIG and Sinistro cameras on the network of 1-m telescopes at Sutherland (South Africa), the Cerro Tololo Inter-American Observatory (Chile), and Siding Spring (Australia) \citep{Brown2013}, through the Supernova Key Project and Global Supernova Project. Using \texttt{lcogtsnpipe}\footnote{\url{https://github.com/LCOGT/lcogtsnpipe}} \citep{Valenti2016}, a PyRAF-based photometric reduction pipeline, point-spread function fitting was performed. Reference images were obtained with a Sinistro camera after the SN faded, and image subtraction was performed using \texttt{PyZOGY}\footnote{\url{https://github.com/dguevel/PyZOGY}} \citep{Guevel2017}, an implementation in Python of the subtraction algorithm described in \cite{Zackay2016}. 
{\it BV}- and {\it gri}-band data were calibrated to Vega \citep{Stetson2000} and AB \citep{Albareti2017} magnitudes, respectively, using standard fields observed on the same night by the same telescope as the SN.

Las Cumbres Observatory optical spectra for SN~2016egz were taken with the FLOYDS spectrographs mounted on the 2m Faulkes Telescope North (FTN) and South (FTS) at Haleakala (USA) and Siding Spring (Australia), respectively, through the Supernova Key Project and Global Supernova Project. A $2\arcsec$ slit was placed on the target along the parallactic angle \citep{Filippenko1982}. One-dimensional spectra were extracted, reduced, and calibrated following standard procedures using \texttt{floyds\_pipeline}\footnote{\url{https://github.com/LCOGT/floyds_pipeline}} \citep{Valenti2014}. 
Additional optical spectra of the SN and host galaxy were obtained by PESSTO and extended PESSTO (ePESSTO) with NTT ($+$EFOSC2). EFOSC2 spectra were reduced and calibrated in a standard manner using a custom built pipeline for the PESSTO project \citep{PESSTO2015}.

All photometry and spectroscopy of SNe~2006Y, 2006ai, and 2016egz are presented in Figures~\ref{fig:LC} and \ref{fig:spec}, respectively, and will be available for download via the Open Supernova Catalog \citep{Guillochon2017} and the Weizmann Interactive Supernova Data Repository (WISeREP; \citealt{Yaron2012}).
For SNe~2006Y, 2006ai, and 2016egz, no Na~{\sc i}~D absorption is seen at the host redshift (Figure~\ref{fig:spec}), indicating low host extinction at the SN position. Thus, we correct all photometry and spectroscopy only for the Milky Way extinction \citep{Schlafly2011}\footnote{Via the NASA/IPAC Infrared Science Archive (IRSA): \url{https://irsa.ipac.caltech.edu/applications/DUST/}} of $A_V=0.347$, $0.337$, and $0.042$ mag for SNe~2006Y, 2006ai, and 2016egz, respectively, assuming the \cite{Fitzpatrick1999} reddening law with $R_V=3.1$.

\section{Analysis} \label{sec:ana}

\subsection{Host Galaxies} \label{sec:host}

\begin{deluxetable*}{ccccccccc}
\tablecaption{Host Galaxy Line Fluxes ($10^{-15}$\,erg\,s$^{-1}$\,cm$^{-2}$)\label{tab:hostline}}
\tablehead{
\colhead{SN} & \colhead{[O~{\sc ii}] $\lambda3727$} & \colhead{H$\beta$} & \colhead{[O~{\sc iii}] $\lambda4959$} & \colhead{[O~{\sc iii}] $\lambda5007$} & \colhead{H$\alpha$} & \colhead{[N~{\sc ii}] $\lambda6583$} & \colhead{[S~{\sc ii}] $\lambda6717$} & \colhead{[S~{\sc ii}] $\lambda6731$}
}
\startdata
2006Y & $13.9\pm0.3$ & $2.89\pm0.04$ & $3.08\pm0.07$ & $9.87\pm0.02$ & $10.37\pm0.04$ & $0.86\pm0.01$ & $1.52\pm0.04$ & $1.06\pm0.03$ \\
2006ai & -- & $37\pm2$ & $7.2\pm0.3$ & $18.4\pm0.6$ & $181\pm3$ & $54\pm1$ & $7.8\pm0.6$ & $45\pm3$ \\
2016egz & $5.08\pm0.04$ & $1.03\pm0.04$ & $1.95\pm0.04$ & $4.87\pm0.02$ & $4.62\pm0.03$ & $0.14\pm0.01$ & $0.13\pm0.02$ & $1.13\pm0.02$ \\
\enddata
\end{deluxetable*}

\begin{deluxetable*}{cccccccc}
\tablecaption{Host Galaxy Properties\label{tab:hostclass}}
\tablehead{
\colhead{SN} & \colhead{$M_B$\tablenotemark{a}} & \colhead{Galaxy Class\tablenotemark{b}} & \colhead{${\rm SFR}(L_{\rm H\alpha})$} & \colhead{${\rm SFR}(L_{\rm [O~{\sc II}]})$} & \colhead{${\rm SFR}(L_{\rm GALEX})$} & \colhead{$12+ {\rm log}_{10}({\rm O/H})$\tablenotemark{c}} & \colhead{$Z$} \\[-6pt]
\colhead{} & \colhead{(mag)} & \colhead{}  & \colhead{($M_\odot \, {\rm yr}^{-1}$)} & \colhead{($M_\odot \, {\rm yr}^{-1}$)} & \colhead{($M_\odot \, {\rm yr}^{-1}$)} & \colhead{} & \colhead{($Z_\odot$)}
}
\startdata
2006Y & $-16.32$ & Star forming & $0.2$ & $0.5$ & -- & $8.32$ ($7.72$--$8.72$) & $0.35$\\
2006ai & $-19.22$ & Star forming & $0.8$ & -- & -- & $8.71$ ($8.50$--$8.91$) & $0.87$\\
2016egz & $-16.36$ & Star forming  & $0.04$ & $0.08$ & $0.15$ & $8.20$ ($7.81$--$8.69$) & $0.27$\\
\enddata
\tablenotetext{a}{From \cite{Gutierrez2018}}
\tablenotetext{b}{Based on the BPT diagrams (Figure~\ref{fig:BPT})}
\tablenotetext{c}{In the format of \textit{weighted average (range from various estimates)} using \texttt{PyMCZ} \citep{Bianco2016}}
\end{deluxetable*}

\begin{figure*}
    \centering
    \includegraphics[width=0.99\textwidth]{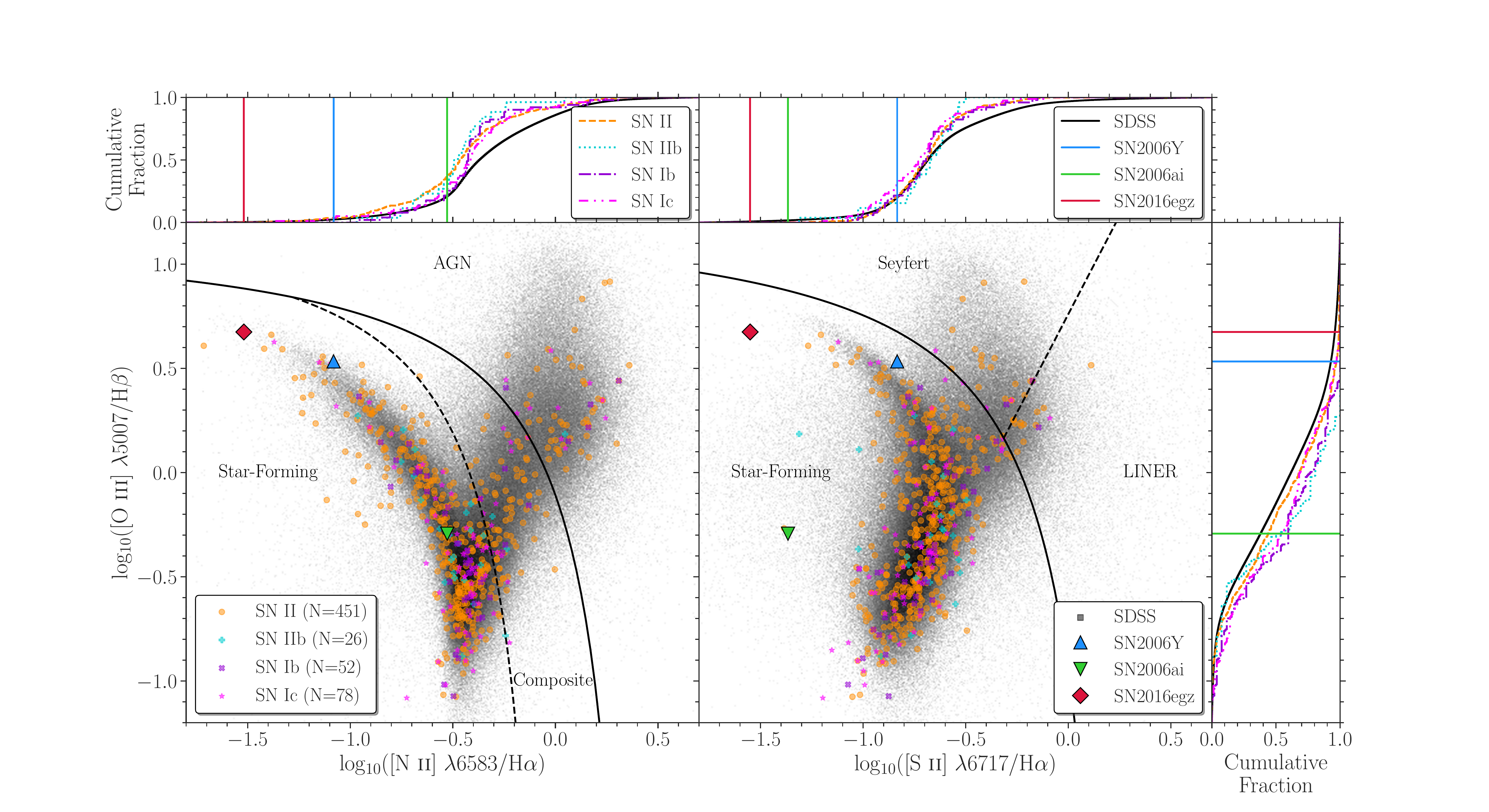}
    \caption{Comparison of the host galaxy BPT diagrams and cumulative fractions with respect to each axis of SNe~2006Y, 2006ai, and 2016egz with the SN sample of \cite{Graham2019} and the Eighth Data Release of the Sloan Digital Sky Survey (SDSS DR8; \citealt{SDSS-III,SDSS-DR8}).
    The galaxy classification scheme is adopted from \cite{Kewley2006}.
    Note that the host galaxies of SNe~2006Y, 2006ai, and 2016egz lie in the star-forming region with relatively low [N~{\sc ii}]/H$\alpha$ ($\lesssim21\%$ and $\lesssim37\%$ of SDSS and SN II host galaxies, respectively), [S~{\sc ii}]/H$\alpha$ ($\lesssim21\%$ of SDSS and SN II host galaxies), and moderate-to-high [O~{\sc iii}]/H$\beta$ ($\lesssim62\%$ and $\lesssim56\%$ of SDSS and SN II host galaxies, respectively).
    }
    \label{fig:BPT}
\end{figure*}

We measure host galaxy line fluxes by fitting a Gaussian profile to each line. 
The measurements are summarized in Table~\ref{tab:hostline}.
In Figure~\ref{fig:BPT}, we place the host galaxies in Baldwin-Phillips-Terlevich (BPT) diagrams \citep{Baldwin1981} based on the line ratios of [O~{\sc iii}] $\lambda5007$/H$\beta$, [N~{\sc ii}] $\lambda6583$/H$\alpha$, and [S~{\sc ii}] $\lambda6717$/H$\alpha$. According to the \cite{Kewley2006} classification scheme, the host galaxies of SNe~2006Y, 2006ai, and 2016egz lie in the star-forming region in the BPT diagrams. 
Thus, we estimate star-formation rates (SFRs) from the H$\alpha$ and [O~{\sc ii}] $\lambda3727$ fluxes using the calibrations summarized in \cite{Kennicutt1998} and from the Galaxy Evolution Explorer (GALEX) photometry (\citealt{Seibert2012}, retrieved via NED) using the \cite{Salim2007} calibration where the measurements are available. These SFR estimates yield a range of $0.04$--$0.8\, M_\odot \, {\rm yr}^{-1}$. 
We also estimate host galaxy metallicities from the measured line ratios and various estimates using \texttt{PyMCZ} \citep{Bianco2016}. The weighted averages of $12+ {\rm log}_{10}(\rm O/H) = 8.20$--$8.71$ roughly correspond to $0.3$--$0.9\,Z_{\odot}$ \citep{Asplund2009}. 
The measured host galaxy properties are summarized in Table~\ref{tab:hostclass}.

The host galaxies of SNe~2006Y, 2006ai, and 2016egz are star-forming galaxies at subsolar metallicities.
They show relatively low [N~{\sc ii}]/H$\alpha$, [S~{\sc ii}]/H$\alpha$, and moderate-to-high [O~{\sc iii}]/H$\beta$ in the BPT comparisons with the SDSS and core-collapse supernova (CCSN) host galaxies (Figure~\ref{fig:BPT}).
Compared to CCSN host galaxy samples, the host SFRs are relatively low ($\lesssim25\%$ of CCSNe; \citealt{Galbany2014}), while the host metallicities span a wide range ($\sim90\%$--$100\%$ of CCSNe; \citealt{,Anderson2016, Galbany2016host}).
SNe~2006Y, 2006ai, and 2016egz are also included in the \cite{Gutierrez2018} SN~II sample in low-luminosity host galaxies, in which they find that low-luminosity galaxies generally host SNe II with slower declining light curves and weaker absorption lines, but did not find strong correlations with plateau lengths or expansion velocities. Thus, short-plateau SNe, like 2006Y, 2006ai, and 2016egz, do not seem to have strong environmental preferences, although this merits future investigations with bigger samples given the rarity of these short-plateau SNe.

\subsection{\textit{V}-band and Bolometric Light Curves} \label{sec:V_bolo}

\begin{figure*}
 \centering
   \gridline{\fig{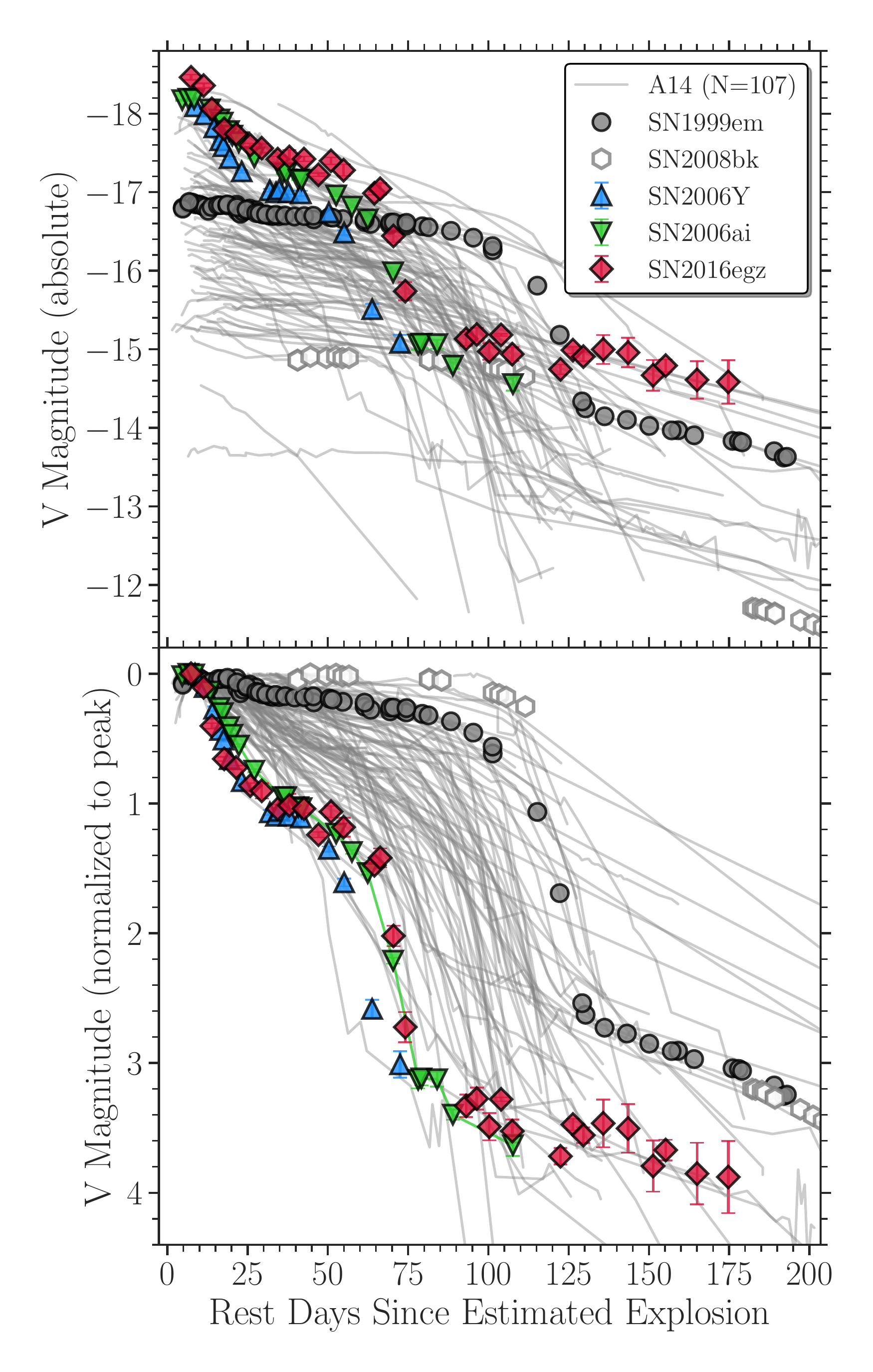}{0.45\textwidth}{\textbf{(a)} \textit{V}-band sample of \citet[A14]{Anderson2014}}
          \fig{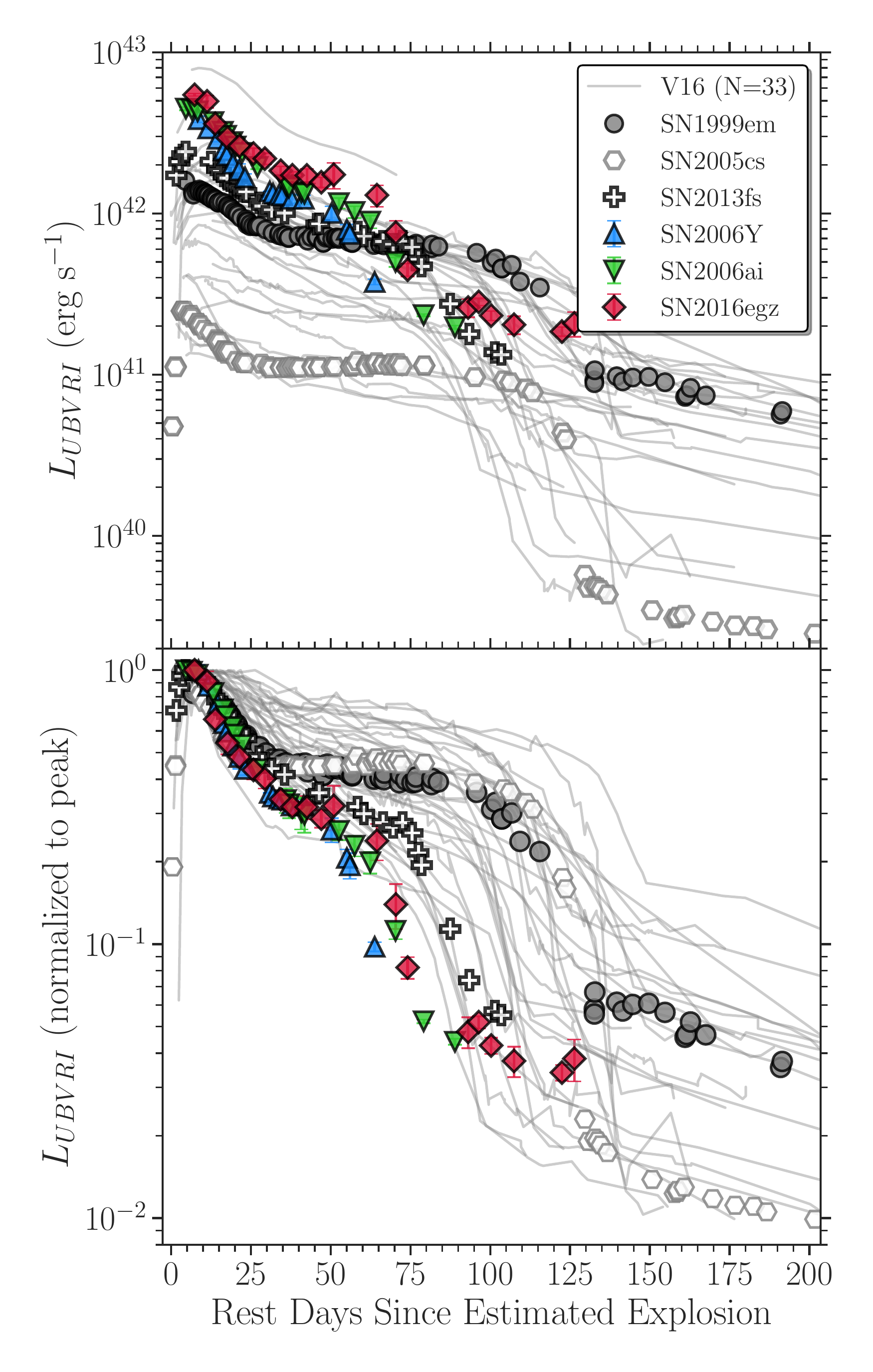}{0.45\textwidth}{\textbf{(b)} Pseudobolometric sample of \citet[V16]{Valenti2016}}
          }
\caption{Comparison of the unnormalized (top) and normalized to peak (bottom) light curves of SNe~2006Y, 2006ai, and 2016egz with SN~II samples (gray transparent lines), including the archetypal SN~IIP~1999em, the low-luminosity SNe~IIP~2005cs and 2008bk, and the early-flash SN~2013fs. Error bars denote $1\sigma$ uncertainties and are sometimes smaller than the marker size. 
Note the similar high luminosity and short plateaus of SNe~2006Y, 2006ai, and 2016egz, which could be more pronounced, as their peaks are lower limits (i.e., not observed).
  } 
  \label{fig:LC_sample}
\end{figure*}

We fit a blackbody spectral energy distribution (SED) to every epoch of photometry containing at least three filters (excluding the {\it r} band owing to strong H$\alpha$ contamination) obtained within $0.3$ days of each other to estimate blackbody temperature and radius.\footnote{The observed SED peaks are bluer than our wavelength coverage during the first $\sim10$\ days for SNe~2006Y, 2006ai, and 2016egz, potentially underestimating the blackbody temperatures.} Then we integrate the fitted blackbody SED over the full (and \textit{UBVRI}: $3250$--$8900\Angstrom$) wavelength range to obtain bolometric (and pseudobolometric) luminosity at each epoch. 
Comparing the luminosity on the $^{56}$Co tail to that of SN~1987A \citep{Hamuy2003}, we estimate $^{56}$Ni masses of $0.062\pm0.002\,M_{\odot}$ and $0.090\pm0.005\,M_{\odot}$ for SNe~2006ai and 2016egz, respectively. Although the tail of SN~2006Y is not well sampled, we put a rough $^{56}$Ni mass constraint of $0.06$--$0.09\,M_{\odot}$ based on the last \textit{V}-band point and \textit{r}- and \textit{i}-band tail luminosity in between those of SNe~2006ai and 2016egz (Figure~\ref{fig:LC}).
These $^{56}$Ni mass estimates are among the highest in the \cite{Anderson2014} and \cite{Valenti2016} samples.

The comparisons of the \textit{V}-band and pseudobolometric light curves of SNe~2006Y, 2006ai, and 2016egz, respectively, with the \cite{Anderson2014} and \cite{Valenti2016} samples are shown in Figure~\ref{fig:LC_sample}. \cite{Anderson2014} include SNe~2006Y and 2006ai in their sample analysis and identify SN~2006Y as an outlier to many observed trends; it shows the fastest decline from the bright maximum followed by the shortest plateau length. The \textit{V}-band light curves of SNe~2006ai and 2016egz show peak and plateau characteristics similar to SN~2006Y. The similarities stand out even more when their light curves are normalized to peak, showing one of the largest peak-to-tail luminosity contrasts and the shortest (optically thick) photospheric durations (Figure~\ref{fig:LC_sample}(a)). The peculiarities of SNe~2006Y and 2016egz could be even more extreme (compared to \textit{normal} SN~II population), given that their light-curve peaks are lower limits (not observed).

The similar characteristics can also be seen in the pseudobolometric light-curve comparison in Figure~\ref{fig:LC_sample}(b). In addition, the peaks and the following decline rates of SNe~2006Y, 2006ai, and 2016egz are brighter and steeper, respectively, than those of SN~2013fs \citep{Yaron2017} whose early time ($\lesssim5$\,d) spectra show \textit{flash features} (high-ionization CSM emission lines excited by the SN shock-breakout radiation; \citealt{Gal-Yam2014,Khazov2016,Bruch2021}). Various flash spectral and light-curve modeling have inferred high mass-loss rates for SN~2013fs, ranging from $\sim0.001$--$0.1\,M_{\odot}\,{\rm yr}^{-1}$ for the last few years to decades before the explosion (e.g., \citealt{Moriya2017, Morozova2017, Yaron2017}, but see also \citealt{Dessart2017,Soker2021} and \citealt{Kochanek2019} for the possible alternatives from an extended envelope and binary interaction, respectively). Thus, the brighter and steeper peaks of SNe~2006Y, 2006ai, and 2016egz likely indicate the presence of similar or even denser CSM. However, we still do not see flash features, probably because the SN ejecta had already overrun the CSM by the time of our first spectra (Figure~\ref{fig:spec}).

As first-order estimates for progenitor and explosion properties, we use the SN~IIP light-curve scaling relations of \cite{Goldberg2019} that give degenerate parameter space for progenitor radius, ejecta mass ($M_{\rm ej}$), and explosion energy ($E_{\rm exp}$) based on the observed luminosity at day 50, plateau duration, and $^{56}$Ni mass. 
We caution that these relations are not calibrated to short-plateau SNe, such as SNe~2006Y, 2006ai, and 2016egz, whose light curves start to drop from the plateaus around day 50, but nonetheless they should provide some crude estimates. 
The extra heating from $^{56}$Ni extends plateau duration, but it is not the lack of $^{56}$Ni that causes the short plateaus of SNe~2006Y, 2006ai, and 2016egz, given their high $^{56}$Ni mass estimates.
If we assume a typical RSG radius of $800\,R_{\odot}$, then H-rich $M_{\rm ej}\sim1$, $2$, and $4\,M_{\odot}$ and $E_{\rm exp}\sim0.3, 0.4, 0.8\times10^{51}$\,erg can be inferred, respectively, for SNe~2006Y, 2006ai, and 2016egz from the light-curve scaling relations. This suggests significant progenitor H-rich envelope stripping.

\subsection{\texttt{MESA}+\texttt{STELLA} Progenitor and Light-curve Modeling} \label{sec:LCmodel}

\begin{figure*}
 \centering
   \gridline{\fig{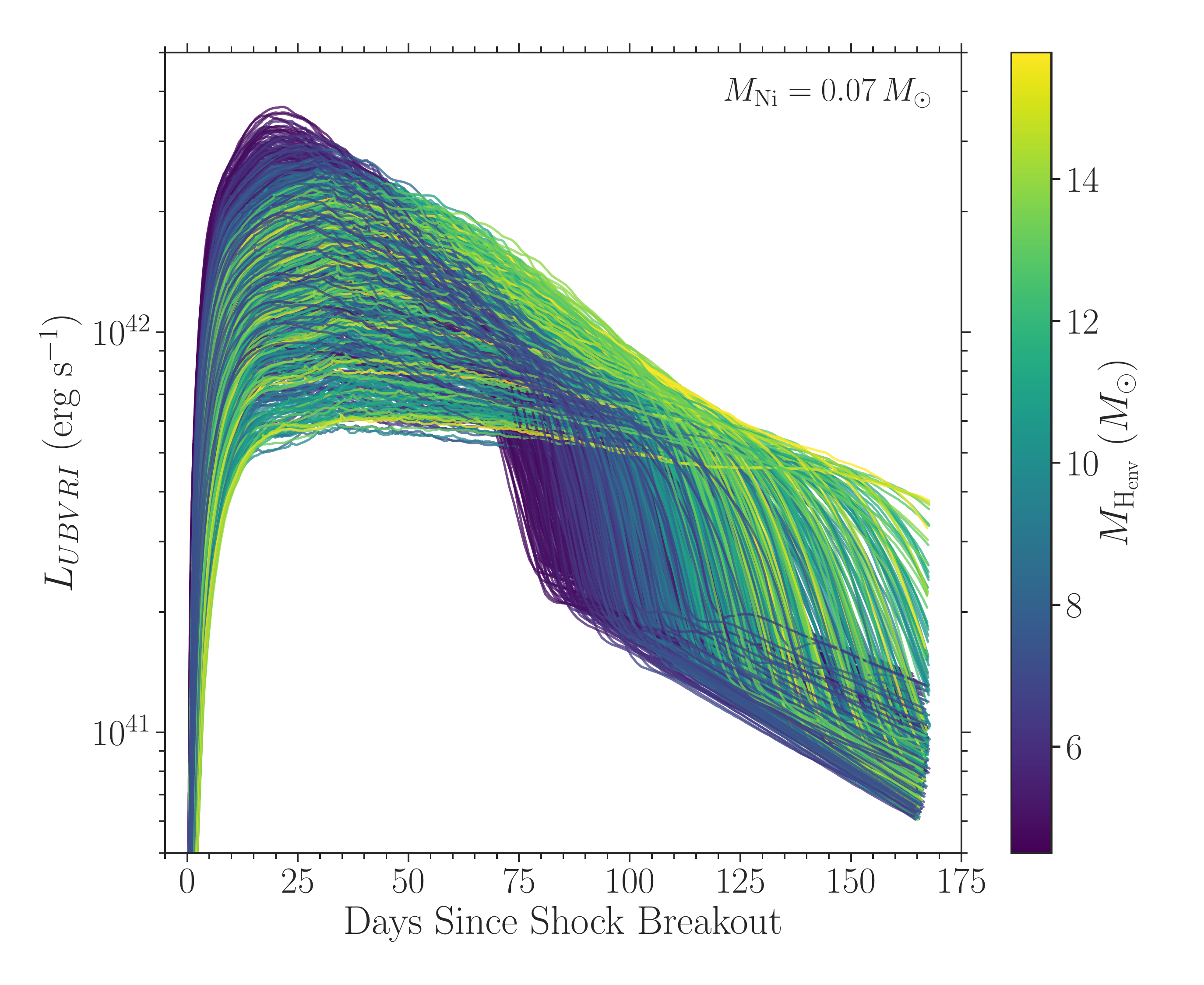}{0.49\textwidth}{\textbf{(a)} IIP ($4.5\,M_{\odot} \leq M_{\rm H_{\rm env}} \leq 15.8\,M_{\odot}$)}
          \fig{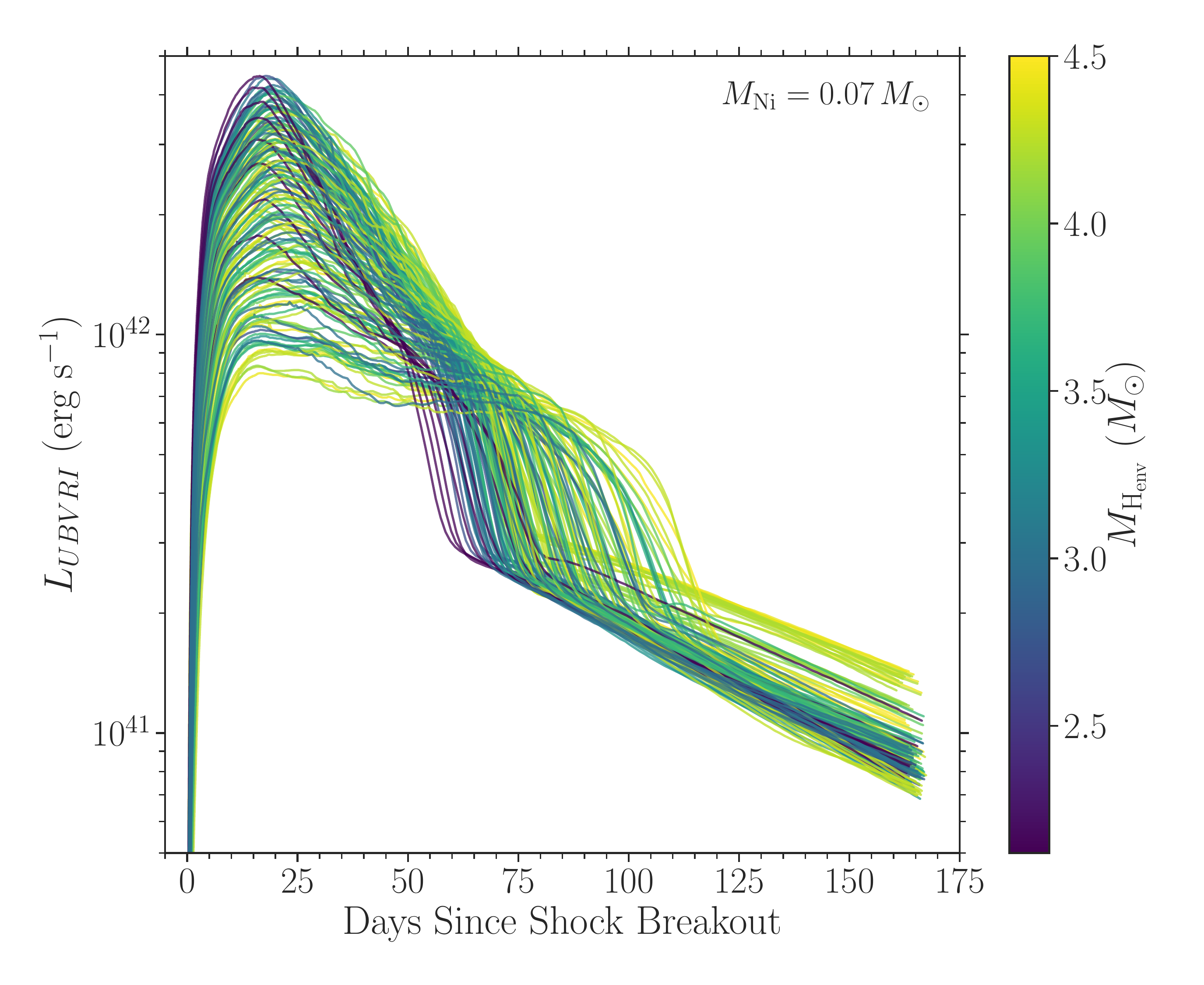}{0.49\textwidth}{\textbf{(b)} IIL ($2.12\,M_{\odot} \leq M_{\rm H_{\rm env}} \leq 4.5\,M_{\odot}$)}}
   \gridline{\fig{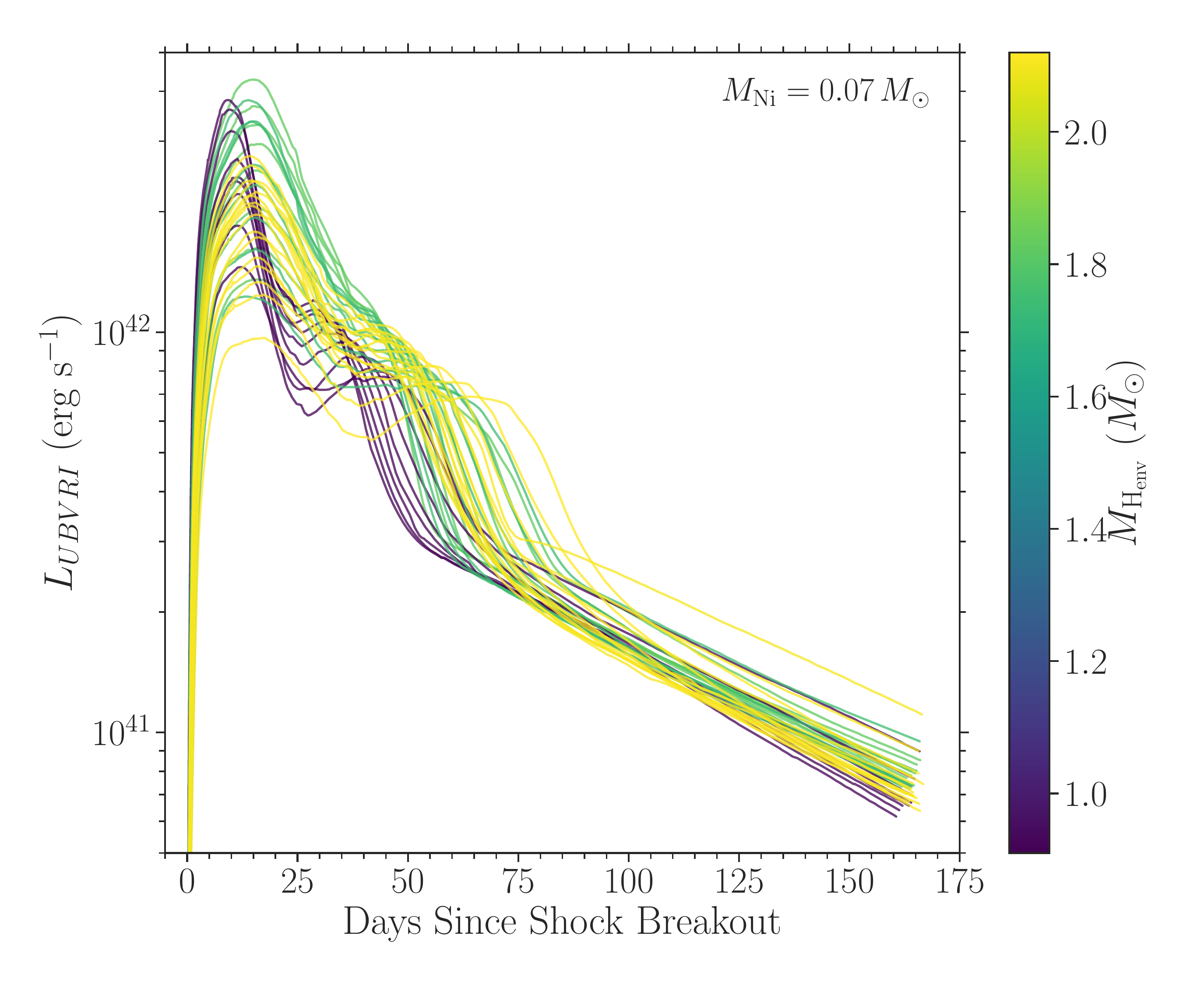}{0.49\textwidth}{\textbf{(c)} Short-Plateau ($0.91\,M_{\odot} \leq M_{\rm H_{\rm env}} \leq 2.12\,M_{\odot}$)}
          \fig{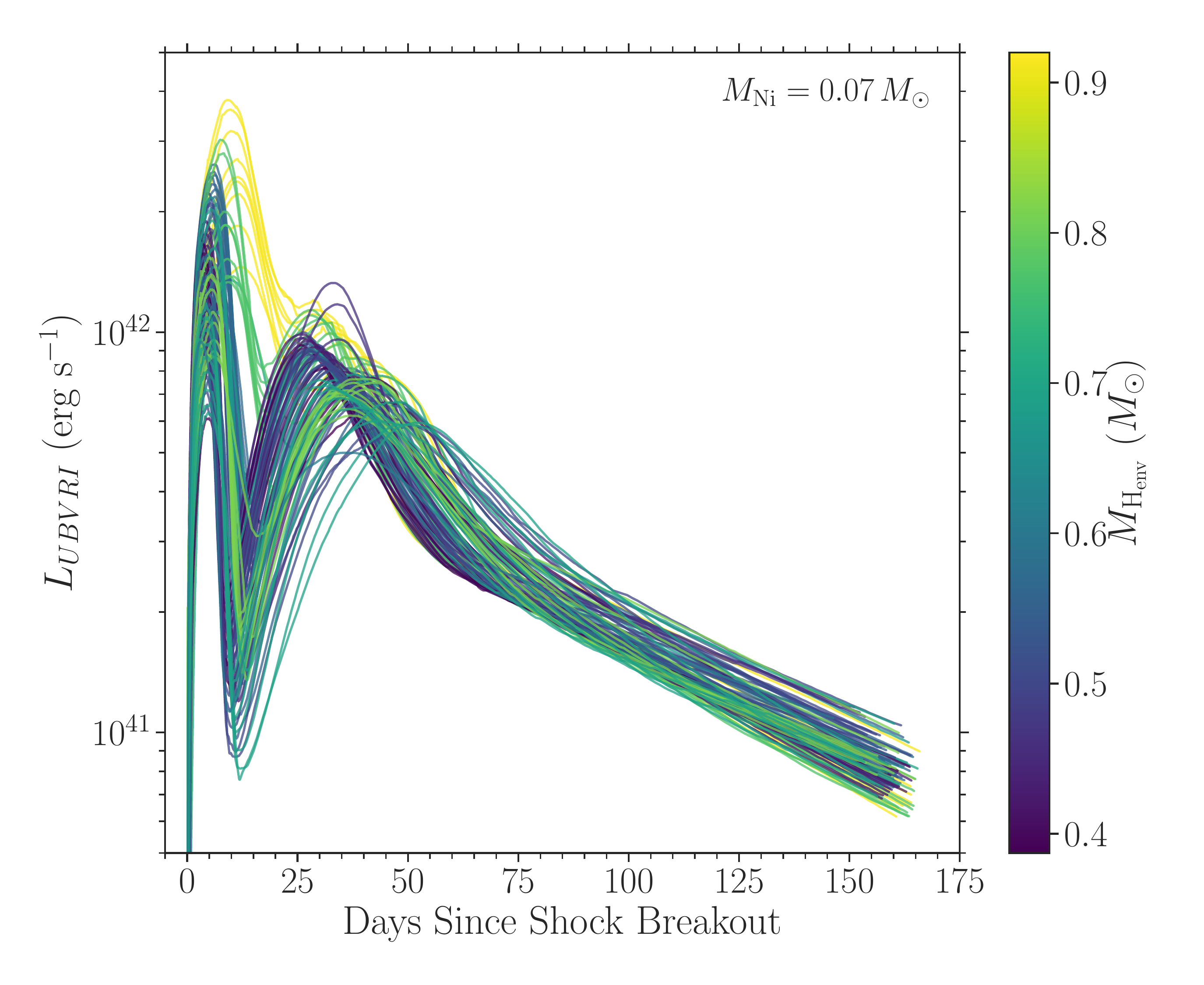}{0.49\textwidth}{\textbf{(d)} IIb ($0.39\,M_{\odot} \leq M_{\rm H_{\rm env}} \leq 0.91\,M_{\odot}$)}}
\caption{\texttt{MESA}+\texttt{STELLA} pseudobolometric light-curve models with a single $^{56}$Ni mass ($M_{\rm Ni}=0.07\,M_{\odot}$), color coded by the H-rich envelope mass ($M_{\rm H_{\rm env}}$) at the core collapse. Arbitrary cuts in $M_{\rm H_{\rm env}}$ are applied to display each SN~II subtype in each panel. Light curves with the same color in each panel come from the same progenitor model exploded with different energies ($E_{\rm exp}$); higher $E_{\rm exp}$ result in brighter, steeper, and shorter photospheric-phase light curves. Note the continuous population of SNe IIP--IIL--short-plateau--IIb in descending order of $M_{\rm H_{\rm env}}$ and the narrow $M_{\rm H_{\rm env}}$ window that results in short-plateau SNe.
  } 
  \label{fig:Model_Lopt}
\end{figure*}

\begin{figure*}
 \centering
   \gridline{\fig{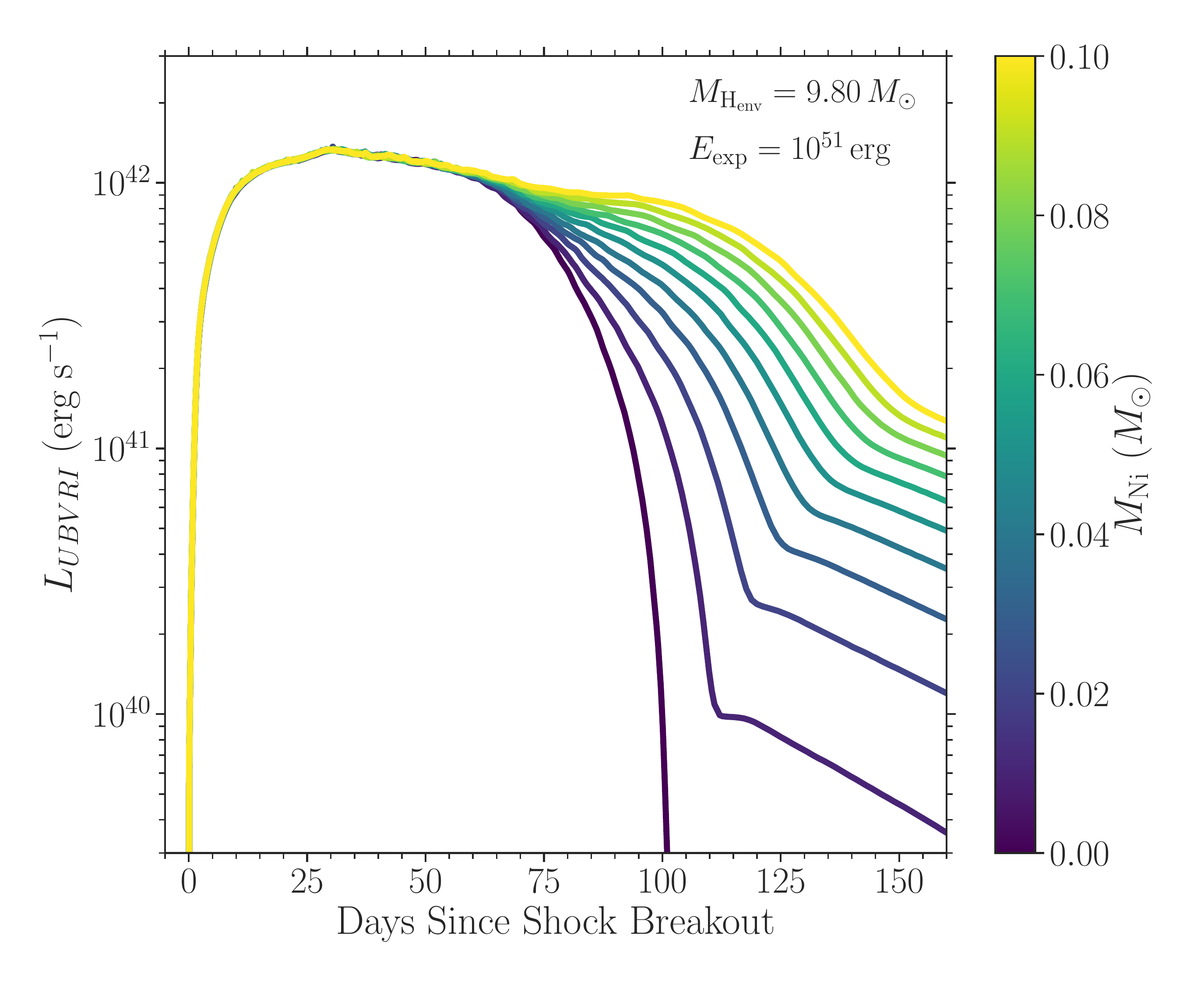}{0.49\textwidth}{\textbf{(a)} IIP ($M_{\rm H_{\rm env}} = 9.80\,M_{\odot}$)}
          \fig{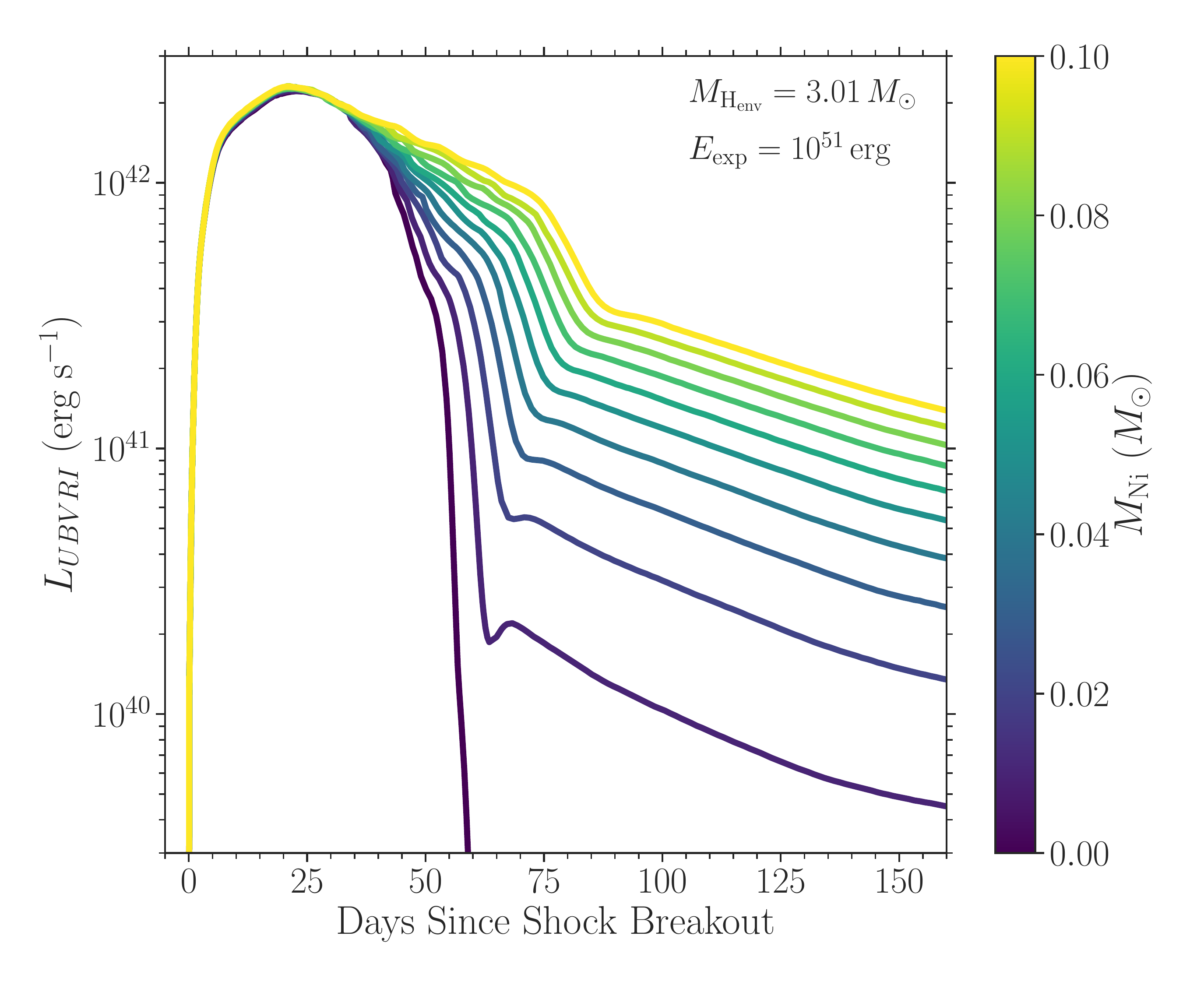}{0.49\textwidth}{\textbf{(b)} IIL ($M_{\rm H_{\rm env}} = 3.01\,M_{\odot}$))}}
   \gridline{\fig{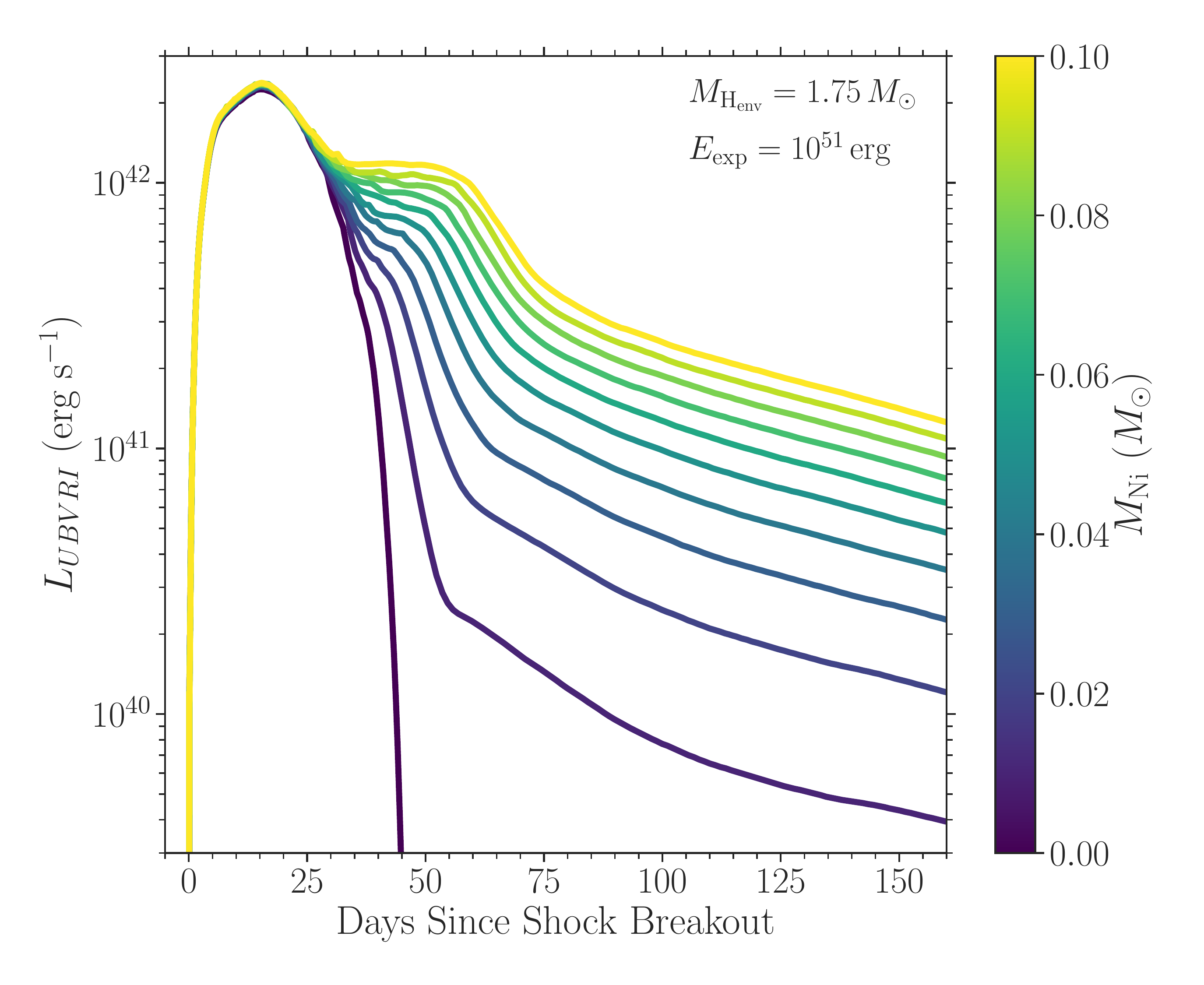}{0.49\textwidth}{\textbf{(c)} Short-Plateau ($M_{\rm H_{\rm env}} = 1.75\,M_{\odot}$)}
          \fig{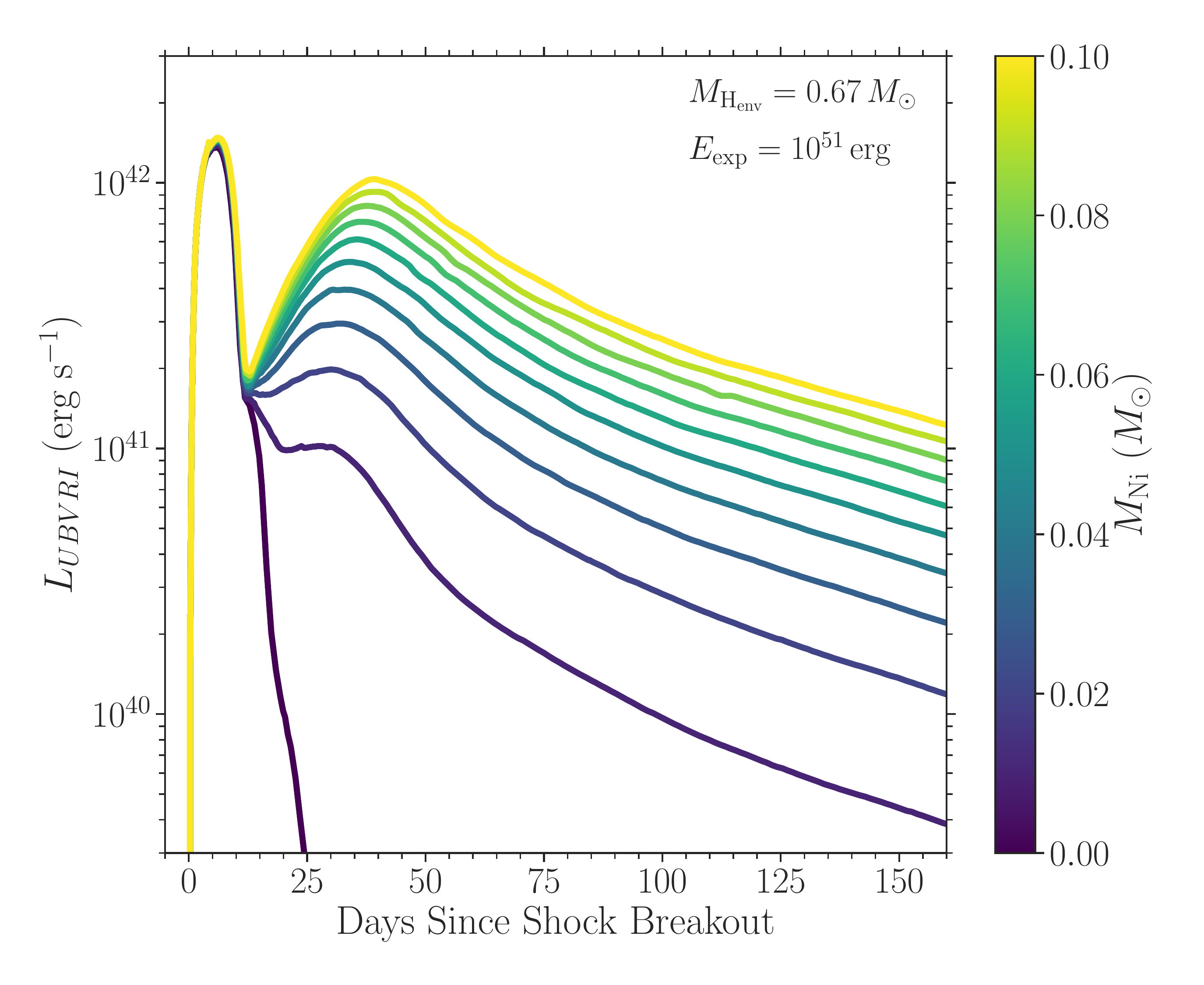}{0.49\textwidth}{\textbf{(d)} IIb ($M_{\rm H_{\rm env}} = 0.67\,M_{\odot}$)}}
\caption{\texttt{MESA}+\texttt{STELLA} pseudobolometric light-curve models with a single explosion energy ($E_{\rm exp}=10^{51}$\,erg) from a representative progenitor model of each SN~II subtype, color coded by the $^{56}$Ni mass ($M_{\rm Ni}$). The initial shock-cooling phase is $M_{\rm Ni}$ independent, while the following photospheric phase is $M_{\rm Ni}$ dependent. Note that in the short-plateau $M_{\rm H_{\rm env}}$ range, the light curves also span IIL morphology with a lower $M_{\rm Ni}$ ($\lesssim0.05\,M_{\odot}$).
} 
  \label{fig:Model_MNI}
\end{figure*}

\begin{figure*}
 \centering
   \gridline{\fig{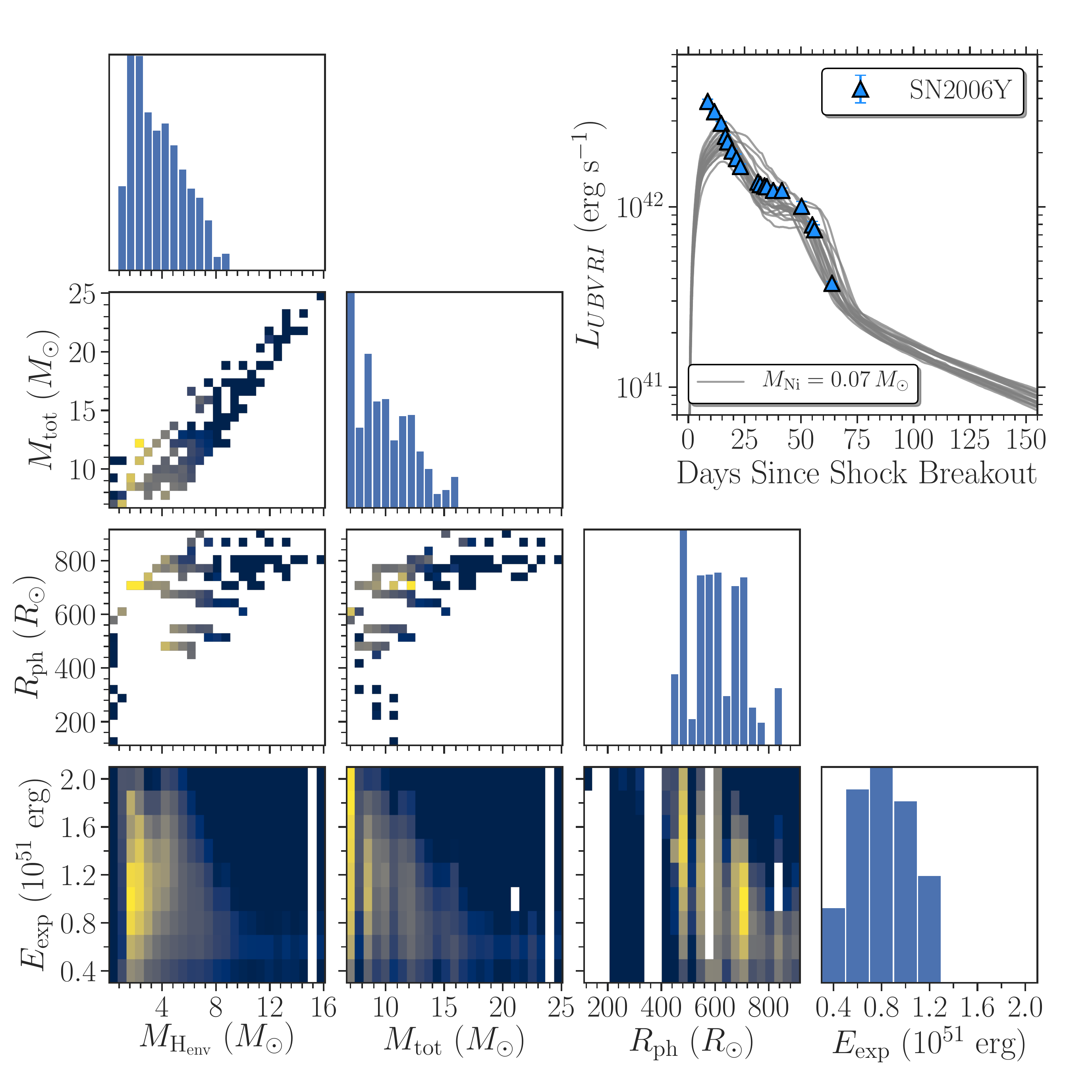}{0.49\textwidth}{\textbf{(a)} SN~2006Y: $M_{\rm H_{\rm env}}\simeq1.7\,M_{\odot}$, $M_{\rm tot}\simeq7.1\,M_{\odot}$, $R_{\rm ph}\simeq480\,R_{\odot}$, $E_{\rm exp}\simeq0.8\times10^{51}$\,erg}
           \fig{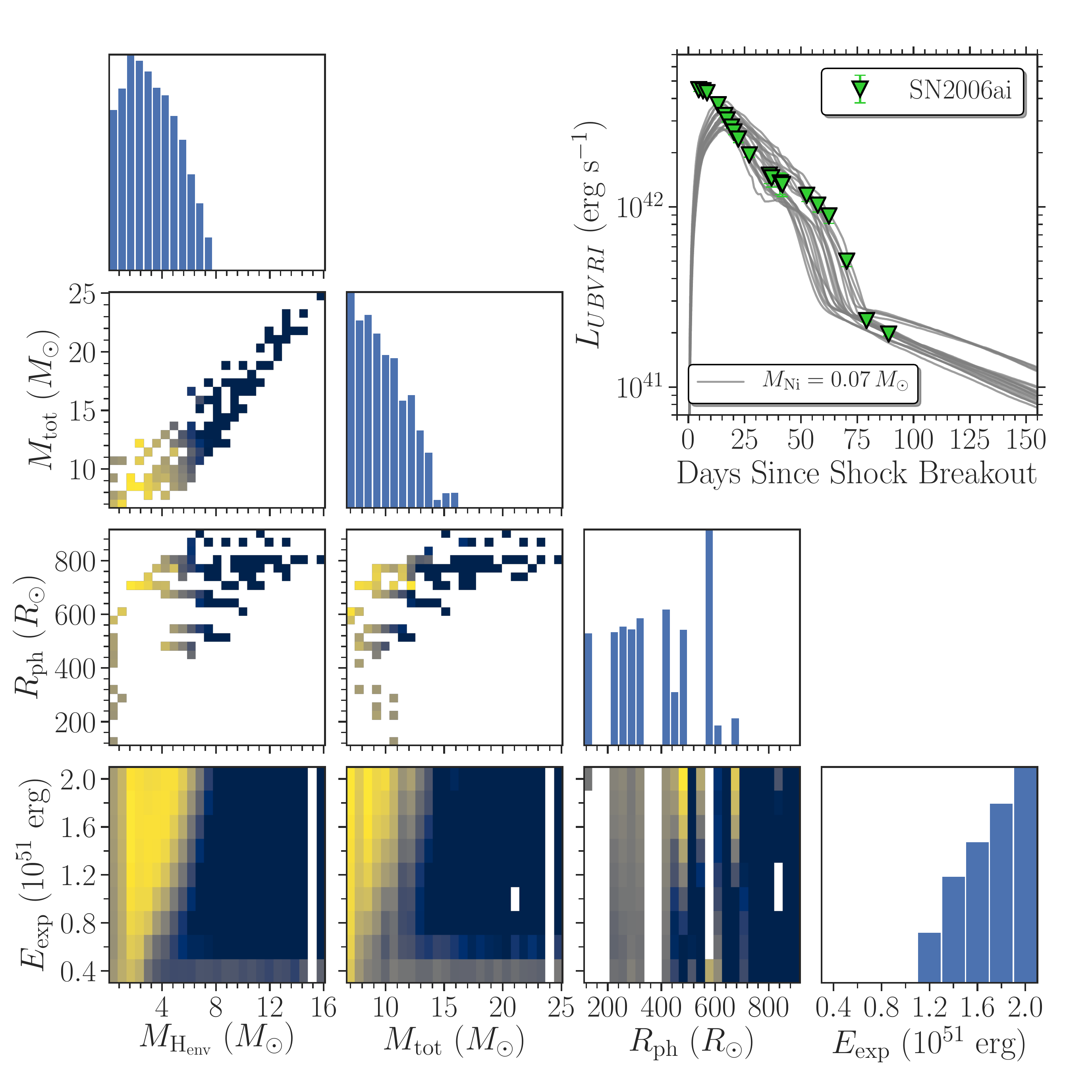}{0.49\textwidth}{\textbf{(b)} SN~2006ai: $M_{\rm H_{\rm env}}\simeq1.7\,M_{\odot}$, $M_{\rm tot}\simeq7.1$--$8.5\,M_{\odot}$, $R_{\rm ph}\simeq580\,R_{\odot}$, $E_{\rm exp}\simeq2.0\times10^{51}$\,erg}}
    \gridline{
          \fig{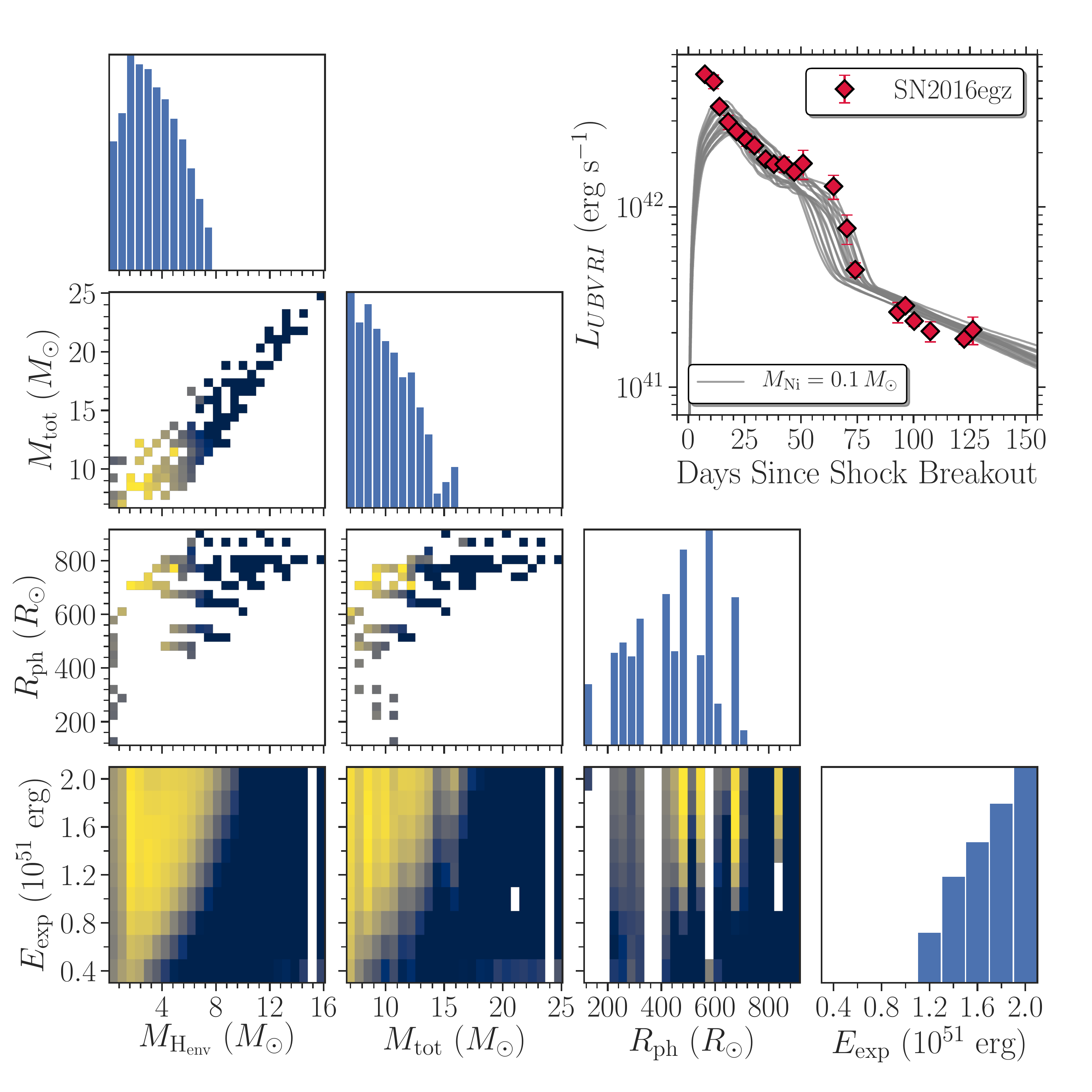}{0.49\textwidth}{\textbf{(c)} SN~2016egz: $M_{\rm H_{\rm env}}\simeq1.7\,M_{\odot}$, $M_{\rm tot}\simeq7.1$--$8.5\,M_{\odot}$, $R_{\rm ph}\simeq580\,R_{\odot}$, $E_{\rm exp}\simeq2.0\times10^{51}$\,erg}}
\caption{Model-grid log likelihood distributions of the explosion energy ($E_{\rm exp}$) and the progenitor properties at the core collapse: H-rich envelope mass ($M_{\rm H_{\rm env}}$), total mass ($M_{\rm tot}$), and photospheric radius ($R_{\rm ph}$), along with the 20 best-fit light-curve models and the maximum likelihood parameters. In the 2D correlation plots, the brighter yellow regions correspond to the higher correlations, while the blank space corresponds to the parameter space that is not covered by the model grid.
The $M_{\rm Ni}=0.07\,M_{\odot}$ grid is used to fit SNe~2006Y and 2006ai, while the $M_{\rm Ni}=0.1\,M_{\odot}$ grid is used to fit SN~2016egz.
The overall light-curve morphology is reasonably well reproduced, except the early ($\lesssim10$ days) excess emission, which we attribute to CSM interaction (Figures~\ref{fig:CSM_Model_Lopt} and \ref{fig:CSM_Model_Lopt_fit}).  
Note the inferred small $M_{\rm H_{\rm env}}$ and high He-core masses ($M_{\rm He_{\rm core}} = M_{\rm tot} - M_{\rm H_{\rm env}}$) of $5.4$--$6.9\,M_{\odot}$ for SNe~2006Y, 2006ai, and 2016egz.
  } 
  \label{fig:Model_Lopt_fit}
\end{figure*}

In order to explore the effect of H-rich envelope stripping in SN~II light curves in more detail and to better extract physical parameters from the short-plateau light curves, we construct a large \texttt{MESA} \citep{Paxton2011,Paxton2013,Paxton2015,Paxton2018,Paxton2019} + \texttt{STELLA} \citep{Blinnikov1998,Blinnikov2000,Blinnikov2006,Blinnikov2004,Baklanov2005} single-star progenitor and light-curve model grid. 
For the \texttt{MESA} progenitor model grid, we vary ZAMS masses ($M_{\rm ZAMS}=10.0$--$25.0\,M_{\odot}$ with $2.5\,M_{\odot}$ increments) and wind scaling factors ($\eta_{\rm wind}=0.0$--$3.0$ with $0.1$ increments), while fixing subsolar ZAMS metallicity ($Z=0.3\,Z_{\odot}$) and no rotation ($\nu/\nu_{\rm crit}=0$). 
For the \texttt{MESA} explosion model grid, we vary explosion energies ($E_{\rm exp}=0.4$--$2.0\times10^{51}$\,erg with $0.2\times10^{51}$\,erg increments) and $^{56}$Ni masses ($M_{\rm Ni}=0.04$, $0.07$, and $0.1\,M_{\odot}$) for each progenitor model.
Then we hand off these explosion models to \texttt{STELLA} to produce synthetic light curves and expansion velocities.
A more detailed description of the model grid is presented in Appendix \ref{sec:extra}.

The full light-curve model grid with $M_{\rm Ni}=0.07\,M_{\odot}$ (totaling 1,303 models)\footnote{The full light-curve model grids with $M_{\rm Ni}=0.04\,M_\odot$ (1,301 models) and $0.1\,M_\odot$ (1,306 models) are shown in Figures~\ref{fig:Model_Lopt_004} and \ref{fig:Model_Lopt_010}, respectively, displaying an SN~II population trend similar to that of the $M_{\rm Ni}=0.07\,M_{\odot}$ grid, albeit with varying plateau duration and tail luminosity.} is shown in Figure~\ref{fig:Model_Lopt}.
This shows SN~II subtypes as part of a continuous population, delineated by their H-rich envelope mass ($M_{\rm H_{\rm env}}$; we use the 20\% mass fraction point $X({\rm H})\geq0.2$).\footnote{The general SN~II population trend is independent of the particular choice of mass fraction point, i.e., $X({\rm H})=0.1$ and $0.5$ show a similar trend, albeit with different $M_{\rm H_{\rm env}}$ cuts.}
Short-plateau SNe represent a transitional class between SNe IIL and IIb in a narrow $M_{\rm H_{\rm env}}$ window ($\Delta M_{\rm H_{\rm env}}\sim1.2\,M_{\odot}$). This is likely why these short-plateau SNe are so rare. 
The SN~II plateau slope correlations with the maximum brightness and plateau duration (e.g., \citealt{Anderson2014,Sanders2015,Galbany2016II,Valenti2016}) are also naturally reproduced with some scatter by varying $M_{\rm H_{\rm env}}$ and $E_{\rm exp}$.
As the population is rather continuous, the applied $M_{\rm H_{\rm env}}$ cuts are somewhat arbitrary and mainly for presentation purposes.

In order to demonstrate the effect of $^{56}$Ni heating on the SN~II light curves, we use a finer $M_{\rm Ni}$ grid spacing ($0.00$--$0.10\,M_\odot$ with $0.01\,M_\odot$ increments) for a representative model of each SN~II subtype and show their light curves in Figure~\ref{fig:Model_MNI}. 
The early phase is $M_{\rm Ni}$ independent, as it is powered by shock cooling. 
For the SN~IIP/L models, the extra heating from $^{56}$Ni extends the photospheric-phase duration \citep{Kasen2009,Goldberg2019,Kozyreva2019}, but does not affect the overall IIP/L light-curve morphology. For the short-plateau SN models, on the other hand, the \textit{plateau} phase is dominantly powered by $^{56}$Ni heating (unlike those of the SN~IIP models, but rather similar to the second peak of the SN~IIb models), and the light curves result in IIL morphology with lower $M_{\rm Ni}$ ($\lesssim0.05\,M_{\odot}$). This suggests the high $M_{\rm Ni}$ preference of short-plateau SNe.

Using the Binary Population and Spectral Synthesis Version 2 (\texttt{BPASS} v2; \citealt{Eldridge2017BPASS}) and the SuperNova
Explosion Code (\texttt{SNEC}; \citealt{Morozova2015}) with a single $E_{\rm exp}=10^{51}$\,erg and $M_{\rm Ni}=0.05\,M_{\odot}$, \cite{Eldridge2018} show a similar SN~II population trend with respect to the total progenitor hydrogen mass ($M_{\rm H}$) and note the small population of short-plateau SNe ($\sim 4.7\%$ of all SNe~II). 
They find a lower $M_{\rm H}$ range for SNe IIL ($0.003$--$0.7\,M_\odot$) than short-plateau ($0.3$--$2.0\,M_\odot$), compared to our model grid of SNe IIL ($1.3$--$3.1\,M_\odot$) and short-plateau ($0.45$--$1.3\,M_\odot$). But we do not consider this as a serious conflict as the subtype classifications are again ambiguous and also sensitive to other physical parameters (e.g., $E_{\rm exp}$ and $M_{\rm Ni}$ as shown in Figures.~\ref{fig:Model_Lopt} and \ref{fig:Model_MNI}).
A more detailed analyses of our \texttt{MESA}+\texttt{STELLA} model grid and its comparisons to observed SN~II samples will be presented in a future publication (Hiramatsu et al., in prep.). In this work, we focus on its application to short-plateau SNe.

The short-plateau SN models come from massive progenitors ($M_{\rm ZAMS} \geq 17.5\,M_{\odot}$) with strong wind mass loss ($\eta_{\rm wind}\geq1.2$) stripping significant amounts of the H-rich envelope ($\gtrsim9\,M_{\odot}$).
We caution that this could partially be due to a modeling bias as the wind mass loss is more sensitive to the choice of $\eta_{\rm wind}$ for more massive progenitors. But we also note that $15.0\,M_{\odot}$ progenitors within the short-plateau $M_{\rm H_{\rm env}}$ range (modeled with $\eta_{\rm wind}=3.4$; not included in the grid) do not result in short-plateau SNe, but IIL, even with $M_{\rm Ni}=0.1\,M_\odot$.
With binary interactions, \cite{Eldridge2018} find a wider $M_{\rm ZAMS}$ range ($7$--$25\,M_\odot$) for short-plateau SNe. 
Thus, it is useful to have independent means of estimating $M_{\rm ZAMS}$ for cross checking (e.g., direct progenitor identification and nebular spectra; see $\S$\ref{sec:neb_spec} for nebular spectral analysis).

For SNe~2006Y, 2006ai, and 2016egz, we perform $\chi^2$ fitting on the observed pseudobolometric light curves with our model grid.
In Figure~\ref{fig:Model_Lopt_fit}, we show the resultant model-grid log likelihood distributions of $E_{\rm exp}$ and the progenitor properties at the core collapse: $M_{\rm H_{\rm env}}$, total mass ($M_{\rm tot}$), and photospheric radius ($R_{\rm ph}$), along with the best-fit light-curve models and the maximum likelihood parameters. 
The parameter choices are based on SNe~IIP light-curve scaling relations \citep{Popov1993,Kasen2009,Sukhbold2016,Goldberg2019}.
But we split the mass parameter into two components: $M_{\rm H_{\rm env}}$ and $M_{\rm tot}$ to estimate He-core mass ($M_{\rm He_{\rm core}} = M_{\rm tot} - M_{\rm H_{\rm env}}$) and then to translate $M_{\rm He_{\rm core}}$ to $M_{\rm ZAMS}$. As we control H-rich envelope stripping by arbitrarily varying $\eta_{\rm wind}$, there is no one-to-one relationship between $M_{\rm tot}$ and $M_{\rm ZAMS}$. Thus, $M_{\rm He_{\rm core}}$--$M_{\rm ZAMS}$ relation is more reliable as it is less sensitive to H-rich envelope stripping and metallicity for $M_{\rm ZAMS}\lesssim30\,M_{\odot}$ \citep{Woosley1995,Woosley2002}, although binary interaction may alter the relation (e.g., \citealt{Zapartas2019,Zapartas2021}). 

Overall, the observed short-plateau light curves of SNe~2006Y, 2006ai, and 2016egz are reasonably well reproduced by the models with $M_{\rm H_{\rm env}}\simeq1.7\,M_{\odot}$, $M_{\rm tot}\simeq7.1$--$8.5\,M_{\odot}$, $R_{\rm ph}\simeq480$--$580\,R_{\odot}$, and $E_{\rm exp}\simeq0.8$--$2.0\times10^{51}$\,erg. 
We also note that there exists some parameter degeneracy \citep{Dessart2019,Goldberg2019,Goldberg2020}. 
Using the $M_{\rm He_{\rm core}}$--$M_{\rm ZAMS}$ relation from the \cite{Sukhbold2016} model grid, we translate $M_{\rm He_{\rm core}}\simeq5.4$--$6.9\,M_{\odot}$ to $M_{\rm ZAMS}\simeq18$--$22\,M_{\odot}$. This suggests partially stripped massive progenitors.
Finally, we note that the observed early ($\lesssim10$ days) emission of SNe~2006Y, 2006ai, and 2016egz are underestimated by the models, indicating the presence of an additional power source to pure shock-cooling emission from the bare stellar atmosphere.

\subsection{\texttt{MESA}+\texttt{STELLA} CSM Light-curve Modeling} \label{sec:CSM_LCmodel}

\begin{figure}
    \centering
    \includegraphics[width=0.45\textwidth]{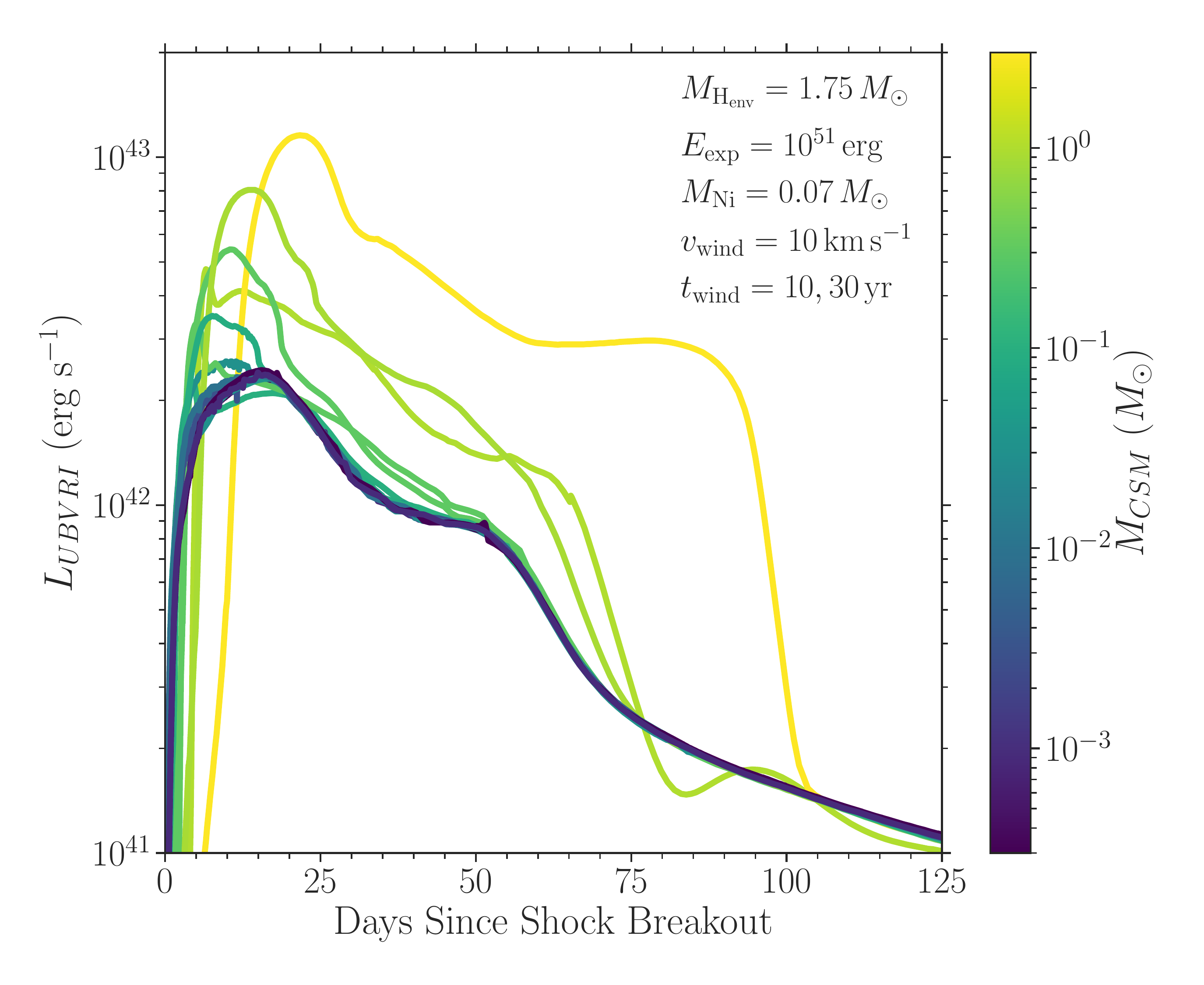}
    \caption{Subset of \texttt{MESA}+\texttt{STELLA} pseudobolometric CSM light-curve models with the single wind velocity ($v_{\rm wind}$) and two different mass-loss duration ($t_{\rm wind}$), color coded by the CSM mass ($M_{\rm CSM} = \dot{M}_{\rm wind} t_{\rm wind}$). Light curves with the same color come from different combinations of $\dot{M}_{\rm wind}$ and $t_{\rm wind}$ that result in the same $M_{\rm CSM}$; higher $\dot{M}_{\rm wind}$ and lower $t_{\rm wind}$ result in sharper light-curve peaks. Note the effect of CSM interaction on both the luminosity and shape.}
    \label{fig:CSM_Model_Lopt}
\end{figure}

\begin{figure*}
 \centering
   \gridline{\fig{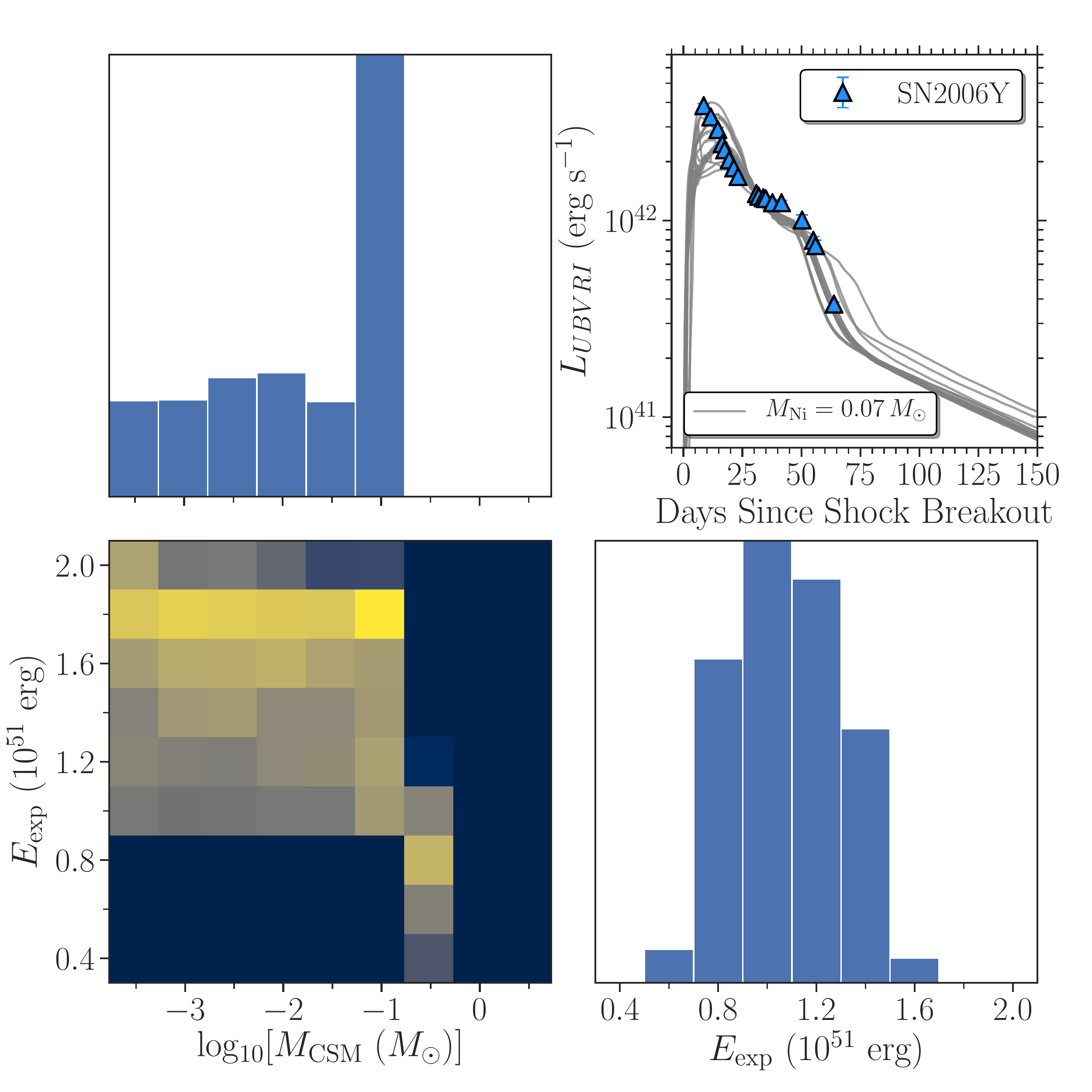}{0.49\textwidth}{\textbf{(a)} SN~2006Y: $M_{\rm CSM} \simeq 0.1\,M_{\odot}$, $E_{\rm exp} \simeq 1.0\times10^{51}$\,erg}
          \fig{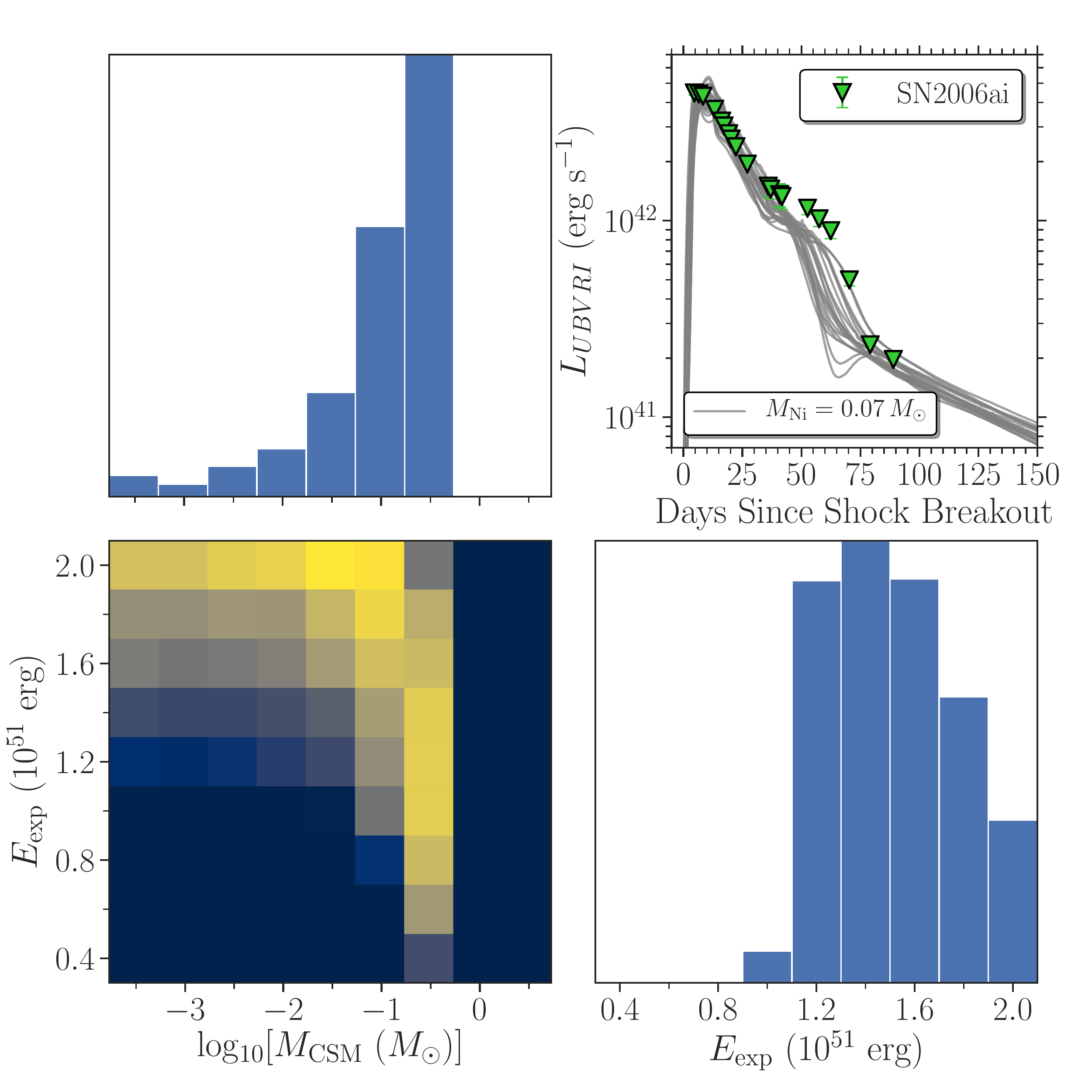}{0.49\textwidth}{\textbf{(b)} SN~2006ai: $M_{\rm CSM} \simeq 0.3\,M_{\odot}$, $E_{\rm exp} \simeq1.4\times10^{51}$\,erg}}
    \gridline{
          \fig{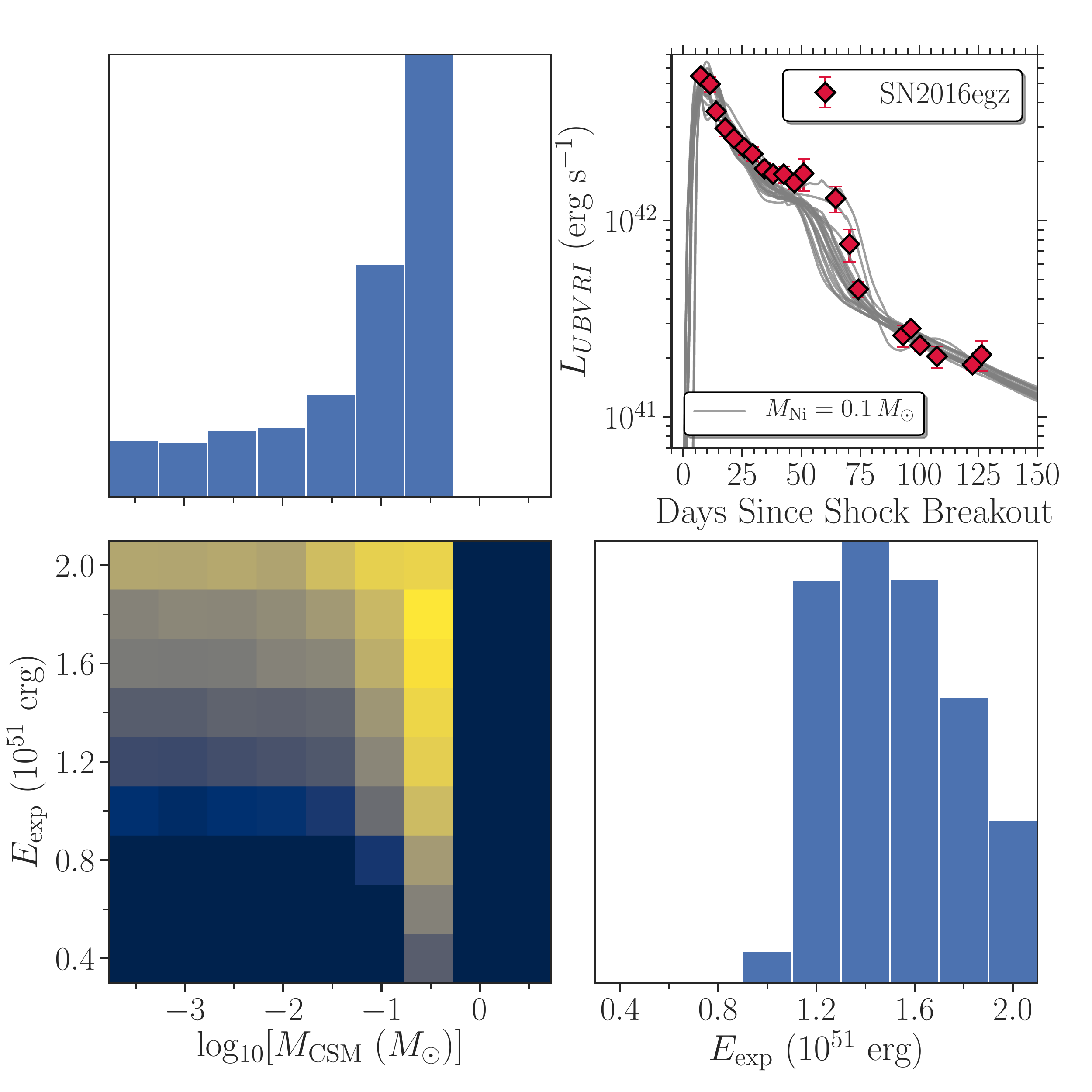}{0.49\textwidth}{\textbf{(c)} SN~2016egz: $M_{\rm CSM} \simeq 0.3\,M_{\odot}$, $E_{\rm exp} \simeq1.4\times10^{51}$\,erg }}
\caption{Similar to Figure~\ref{fig:Model_Lopt_fit}, but with the CSM models. The early ($\lesssim10$\ days) excess emission are reasonably well reproduced. Note the inferred smaller $E_{\rm exp}$ compared to the CSM-free fits in general, and the large $M_{\rm CSM}$ that suggest violent mass loss in the last few decades before the explosion (actual $M_{\rm CSM}$ could be even higher as the light-curve peaks are lower limits). 
  } 
  \label{fig:CSM_Model_Lopt_fit}
\end{figure*}

In order to account for the early excess emission, we propose CSM interaction as a possible power source, as suggested in $\S$\ref{sec:V_bolo}. 
At handoff to \texttt{STELLA}, we affix a wind density profile with $\rho_\mathrm{wind}(r) = \dot{M}_\mathrm{wind}/4\pi r^2 v_\mathrm{wind}$,
where $\dot{M}_\mathrm{wind}$ is a constant wind mass-loss rate, and $v_\mathrm{wind}$ is the wind velocity for time $t_\mathrm{wind}$ (i.e., the CSM mass, $M_\mathrm{CSM} = \dot{M}_\mathrm{wind} t_\mathrm{wind}$), onto the subset of \texttt{MESA} explosion models that result in short-plateau SNe (40 models each with $M_{\rm Ni}=0.04$, $0.07$, and $0.1\,M_{\odot}$; Figures~\ref{fig:Model_Lopt}(c), \ref{fig:Model_Lopt_004}(c), and \ref{fig:Model_Lopt_010}(c)). In addition to the 400 spatial zones for the original SN ejecta, we use 200 spatial zones for the CSM model in \texttt{STELLA}. We construct a grid of CSM models by varying $\dot{M}_\mathrm{wind}$ ($10^{-5}$--$10^{-1}\,M_{\odot}$\,yr$^{-1}$ with $0.5$ dex increments) and $t_\mathrm{wind}$ ($10$ and $30$\,yr) for each \texttt{MESA} short-plateau SN model, assuming a typical RSG $v_\mathrm{wind}=10$\,km\,s$^{-1}$ (\citealt{Moriya2011}). 
The subset of the CSM light-curve model grid is shown in Figure~\ref{fig:CSM_Model_Lopt} where the effect of increasing CSM mass on the light-curve luminosity and shape can be seen. 

In Figure~\ref{fig:CSM_Model_Lopt_fit}, we show the CSM model-grid log likelihood distributions of $M_\mathrm{CSM}$ and $E_{\rm exp}$ from $\chi^2$ fitting on SNe~2006Y, 2006ai, and 2016egz. The observed early excess emission as well as the overall light curves of SNe~2006Y, 2006ai, and 2016egz are reasonably well reproduced by the CSM models with $M_{\rm CSM}\simeq0.1$--$0.3\,M_{\odot}$ ($\dot{M}_\mathrm{wind} \simeq 10^{-2}\,M_{\odot}$\,yr$^{-1}$ with $t_{\rm wind}=10$\,yr for SN~2006Y and $30$\,yr for SNe~2006ai and 2016egz) and $E_{\rm exp}\simeq1.0$--$1.4\times10^{51}$\,erg. This suggests enhanced mass loss (a few orders of magnitude greater than the standard continuous $\dot{M}_{\rm wind}\sim10^{-5}\,M_\odot\,{\rm yr}^{-1}$) in the last few decades before the explosion. The inferred $E_{\rm exp}$ are generally lower than those from the CSM-free fits (Figure~\ref{fig:Model_Lopt_fit}) since the CSM interaction provides additional luminosity, especially around the peak.
The actual CSM could be even denser and more confined for SNe~2006Y, 2006ai, and 2016egz because their light-curve peaks are lower limits.

\subsection{Photospheric Spectra} \label{sec:phot_spec}

\begin{figure}
    \centering
    \includegraphics[width=0.37\textwidth]{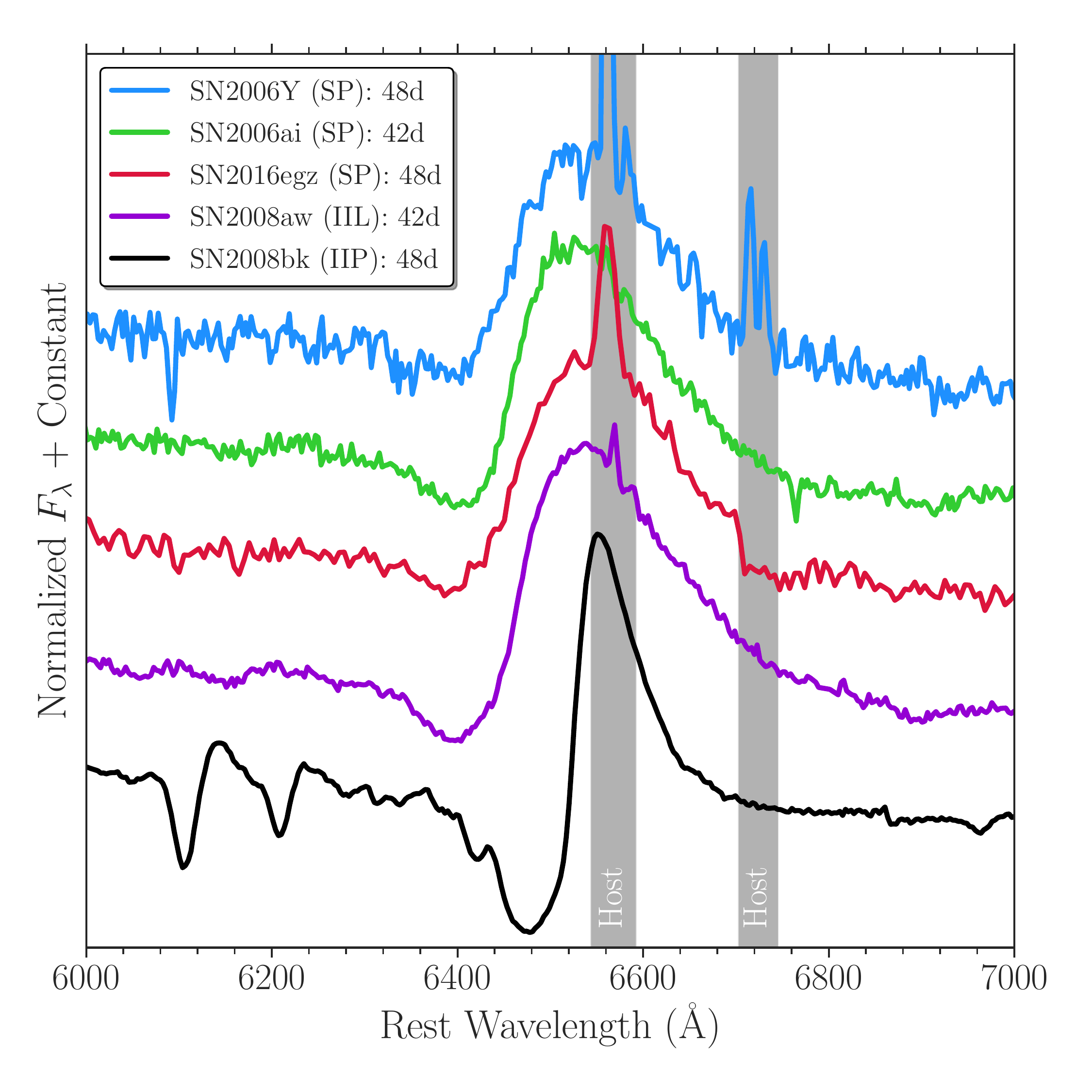}
    \caption{Comparison of the H$\alpha$ P Cygni profiles of short-plateau (SP) SNe~2006Y, 2006ai and 2016egz with the SN~IIL 2008aw and the SN~IIP 2008bk around $50$ days after explosion \citep{Gutierrez2014, Gutierrez2017a}, retrieved via the Open Supernova Catalog. The gray shaded regions represent host galaxy contamination. Each spectrum is normalized by its broad SN H$\alpha$ peak. Note that the H$\alpha$ P Cygni absorption to emission ratios of SNe~2006Y, 2006ai and 2016egz are smaller than those of the SN~IIP 2008bk and even the SN~IIL 2008aw.}
    \label{fig:Ha}
\end{figure}

\begin{figure}
    \centering
    \includegraphics[width=0.45\textwidth]{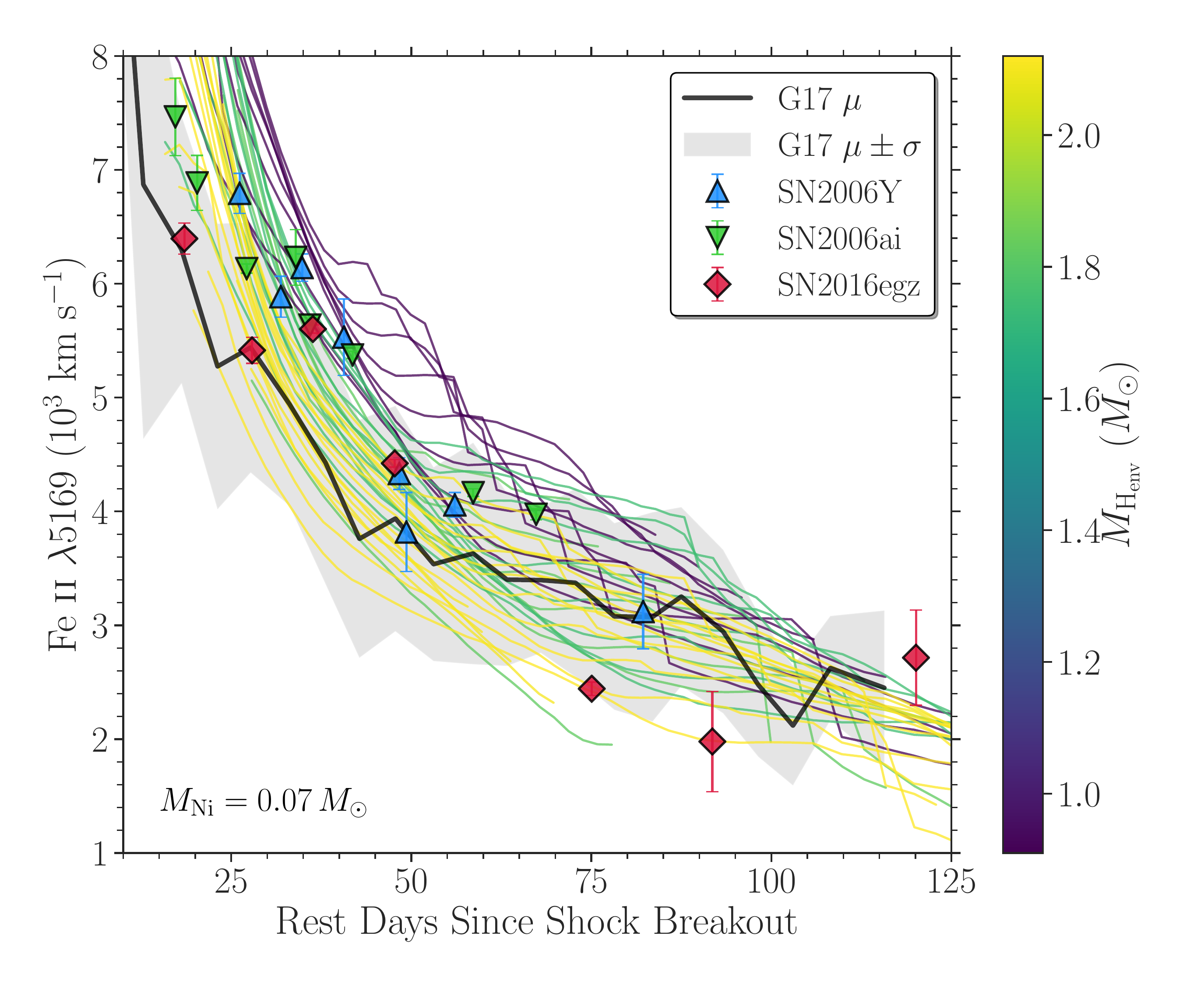}
    \caption{Comparison of Fe~{\sc ii} $\lambda$5169 velocities of SNe~2006Y, 2006ai, and 2016egz with the \citet[G17]{Gutierrez2017a} mean velocity evolution (black line and gray shaded region) and \texttt{MESA}+\texttt{STELLA} short-plateau SN models with $M_{\rm Ni}=0.07\,M_{\odot}$ (defined by the Sobolev optical depth $\tau_{\rm Sob}=1$), color coded by $M_{\rm H_{\rm env}}$ as in Figure~\ref{fig:Model_Lopt}(c). Error bars denote $1\sigma$ uncertainties and are sometimes smaller than the marker size. Velocity models with the same color come from the same progenitor model exploded with different energies; higher explosion energies result in faster velocities. Note the better agreements with the higher $M_{\rm H_{\rm env}}$ models for SNe~2006Y, 2006ai, and 2016egz.}
    \label{fig:VFe}
\end{figure}

As in the \cite{Anderson2014} sample light-curve analysis, \cite{Gutierrez2014,Gutierrez2017a,Gutierrez2017b} also include SNe~2006Y and 2006ai in their sample spectral analysis and identify their spectral peculiarity, namely, the smallest H$\alpha$ P Cygni absorption to emission ratios ($a/e$). They also find a correlation between $a/e$ and light-curve plateau slope (i.e., SNe IIL have shallower absorption components than SNe IIP). Several possible explanations for the smaller $a/e$ for SNe IIL have been proposed, e.g., lower $M_{\rm H_{\rm env}}$ \citep{Schlegel1996,Gutierrez2014,Gutierrez2017b} and steeper envelope density gradients \citep{Dessart2005,Dessart2013,Gutierrez2014}.
In Figure~\ref{fig:Ha}, we show the H$\alpha$ P Cygni profile comparison of SNe~2006Y, 2006ai, and 2016egz with SNe~IIL and IIP. SN~2016egz displays a H$\alpha$ P Cygni profile similar to that of SNe~2006Y and 2006ai with a shallower absorption feature than SNe~IIL and IIP. 
This supports lower $M_{\rm H_{\rm env}}$ as a possible cause of smaller $a/e$ since short-plateau SNe with lower $M_{\rm H_{\rm env}}$ result in smaller $a/e$ than SNe IIP and IIL, as seen in our \texttt{MESA}+\texttt{STELLA} light-curve model grid (Figure~\ref{fig:Model_Lopt}).

For SNe~2006Y, 2006ai, and 2016egz, we measure expansion velocities of Fe~{\sc ii} $\lambda$5169 from the absorption minimum by fitting a P Cygni profile. 
In Figure~\ref{fig:VFe}, we show the velocity comparison of SNe~2006Y, 2006ai, and 2016egz with the \texttt{MESA}+\texttt{STELLA} short-plateau SN models with $M_{\rm Ni}=0.07\,M_\odot$ (Figure~\ref{fig:Model_Lopt}(c)),\footnote{The short-plateau SN models with $M_{\rm Ni}=0.04$ and $0.1\,M_\odot$ are shown in Figure~\ref{fig:VFe_ext}, displaying a velocity evolution similar to that of the $M_{\rm Ni}=0.07\,M_{\odot}$ models, albeit with varying late-time evolution.} as well as the mean velocity evolution from the \cite{Gutierrez2017a} SN~II sample.
The overall velocity evolution is better reproduced with the higher $M_{\rm H_{\rm env}}$ ($\gtrsim1.7\,M_{\odot}$) models. This is in agreement with the values inferred from the light-curve fitting (Figure~\ref{fig:Model_Lopt_fit}).
The Fe~{\sc ii} $\lambda$5169 velocities of SNe~2006Y, 2006ai, and 2016egz are generally higher than the \cite{Gutierrez2017a} mean velocity evolution, but still within $1\sigma$ deviation. Thus, unlike their light curves, their velocities do not appear to be outliers.

However, it is worth noting that the velocity of SN~2016egz is lower than those of SNe~2006Y and 2006ai, despite it exhibiting a more luminous light curve (Figure~\ref{fig:LC_sample}). This is in disagreement with the SN~IIP luminosity--velocity correlation ($L_P \propto v_P^2$) from the homologously expanding photosphere set by H-recombination \citep{Hamuy2002,Kasen2009,Goldberg2019}. By comparing the light curves (Figure~\ref{fig:Model_Lopt}(c)) and velocities (Figure~\ref{fig:VFe}) of short-plateau SN models, we do not see an obvious $L_P$--$v_P$ correlation (e.g., some of the highest and lowest $M_{\rm H_{\rm env}}$ models have similar $L_P$, but the highest $M_{\rm H_{\rm env}}$ models have systematically lower $v_P$). This suggests that the short-plateau light curves are not purely powered by shock energy released at the H-recombination front; $^{56}$Ni heating is also important in shaping their light curves (more so than in typical SN~IIP light curves). This is in agreement with the light-curve analysis with varying $M_{\rm Ni}$ (Figure~\ref{fig:Model_MNI}(c)). 

Finally, we note that both line shape and velocity could also be affected by the extra emission from persistent CSM interaction (e.g, \citealt{Moriya2011,Moriya2018,Dessart2016CSM,Hillier2019}), even though high-velocity H$\alpha$ absorption features are absent \citep{Chugai2007}. Thus, the qualitative analyses, partially based on the light-curve modeling, in this section merit future spectral modeling of short-plateau SNe with and without CSM interaction.

\subsection{Nebular Spectra} \label{sec:neb_spec}

While there are no late-time ($>100$\,d) spectra available for SNe~2006Y and 2006ai, we obtained two nebular-phase ($>200$\,d) spectra for SN~2016egz (Figure~\ref{fig:spec}).
We simultaneously fit a Gaussian profile to the broad SN H$\alpha$ (excluding the narrow host H$\alpha$ region) and a double-Gaussian profile to [O~{\sc i}] $\lambda\lambda6300,6364$ assuming a doublet flux ratio of 3:1 (as not resolved) and single full width at half-maximum (FWHM) velocity.
Due to the low signal-to-noise ratio (S/N) and fringing, we are unable to measure [Ca~{\sc ii}] $\lambda\lambda7291,7323$ fluxes. Instead, we inject a double-Gaussian profile at the [Ca~{\sc ii}] rest wavelengths by assuming a doublet flux ratio of 1:1 and the same single FWHM velocity as the [O~{\sc i}] doublet at each epoch to place a $3\sigma$ flux upper limit. 
The measured nebular line fluxes and FWHM velocities are summarized in Tables~\ref{tab:nebularline} and \ref{tab:nebularvel}.

\begin{deluxetable}{ccccccccc}
\tablecaption{SN~2016egz Nebular Line Fluxes ($10^{-15}$\,erg\,s$^{-1}$\,cm$^{-2}$)\label{tab:nebularline}}
\tablehead{
\colhead{Days} & \colhead{[O~{\sc i}] $\lambda\lambda6300,6364$} & \colhead{H$\alpha$} & \colhead{[Ca~{\sc ii}] $\lambda\lambda7291,7323$}}
\startdata
$278.7$ & $2.1\pm0.4$ & $2.6\pm0.3$ & $\leq2.9$ \\
$309.9$ & $2.0\pm0.3$ & $1.8\pm0.2$ & $\leq2.4$ \\
\enddata
\end{deluxetable}

\begin{deluxetable}{ccccccccc}
\tablecaption{SN~2016egz Nebular FWHM Velocities ($10^3$\,km\,s$^{-1}$)\label{tab:nebularvel}}
\tablehead{
\colhead{Days} & \colhead{[O~{\sc i}] $\lambda\lambda6300,6364$} & \colhead{H$\alpha$} & \colhead{[Ca~{\sc ii}] $\lambda\lambda7291,7323$}}
\startdata
$278.7$ & $3.7\pm0.6$ & $4.0\pm0.5$ & -- \\
$309.9$ & $3.4\pm0.4$ & $3.0\pm0.4$ & -- \\
\enddata
\end{deluxetable}

It is known that [O~{\sc i}] and its ratio to [Ca~{\sc ii}] are insensitive respectively to the SN explosive nucleosynthesis and the SN ejecta density and temperature, so can be used as a proxy for progenitor O-core mass, and so ZAMS mass (e.g., \citealt{Fransson1989,Woosley1995,Woosley2002,Elmhamdi2004,Maeda2007,Dessart2011neb,Dessart2020,Jerkstrand2012, Jerkstrand2014,Jerkstrand2015,Fang2018,Fang2019}). 
Also, since H$\alpha$ and [N~{\sc ii}] are dominantly emitted respectively from the H- and He-rich envelopes, (H$\alpha$, [N~{\sc ii}])/[O~{\sc i}] can be used as a proxy for H/He-rich envelope stripping \citep{Jerkstrand2015,Fang2019,Dessart2020}. 

However, there are a few caveats to note especially for stripped-envelope SNe (SESNe: IIb, Ib, and Ic in descending order of H/He-rich envelope mass).
[N~{\sc ii}]/[O~{\sc i}] could be affected by the progenitor He burning due to more He/N-layer burning in more massive stars \citep{Jerkstrand2015,Fang2019}.
Also, [Ca~{\sc ii}]/[O~{\sc i}] could be affected by the explosion energy due to more emission from the synthesized calcium than primordial calcium in the H-rich envelope \citep{Li1993,Maguire2012,Jerkstrand2015,Jerkstrand2017}.
In principle, the progenitor convective mixing and SN explosive $^{56}$Ni mixing could also alter these line ratios for CCSNe in general \citep{Jerkstrand2017,Dessart2020}.
Because of these possible systematic effects, the direct and quantitative comparison of the nature of the progenitors between different subclasses (e.g., SNe~II vs. SESNe) will require detailed spectral modeling.
Nevertheless, these line ratios should provide rough CCSN progenitor estimates.

\begin{figure}
    \centering
    \includegraphics[width=0.45\textwidth]{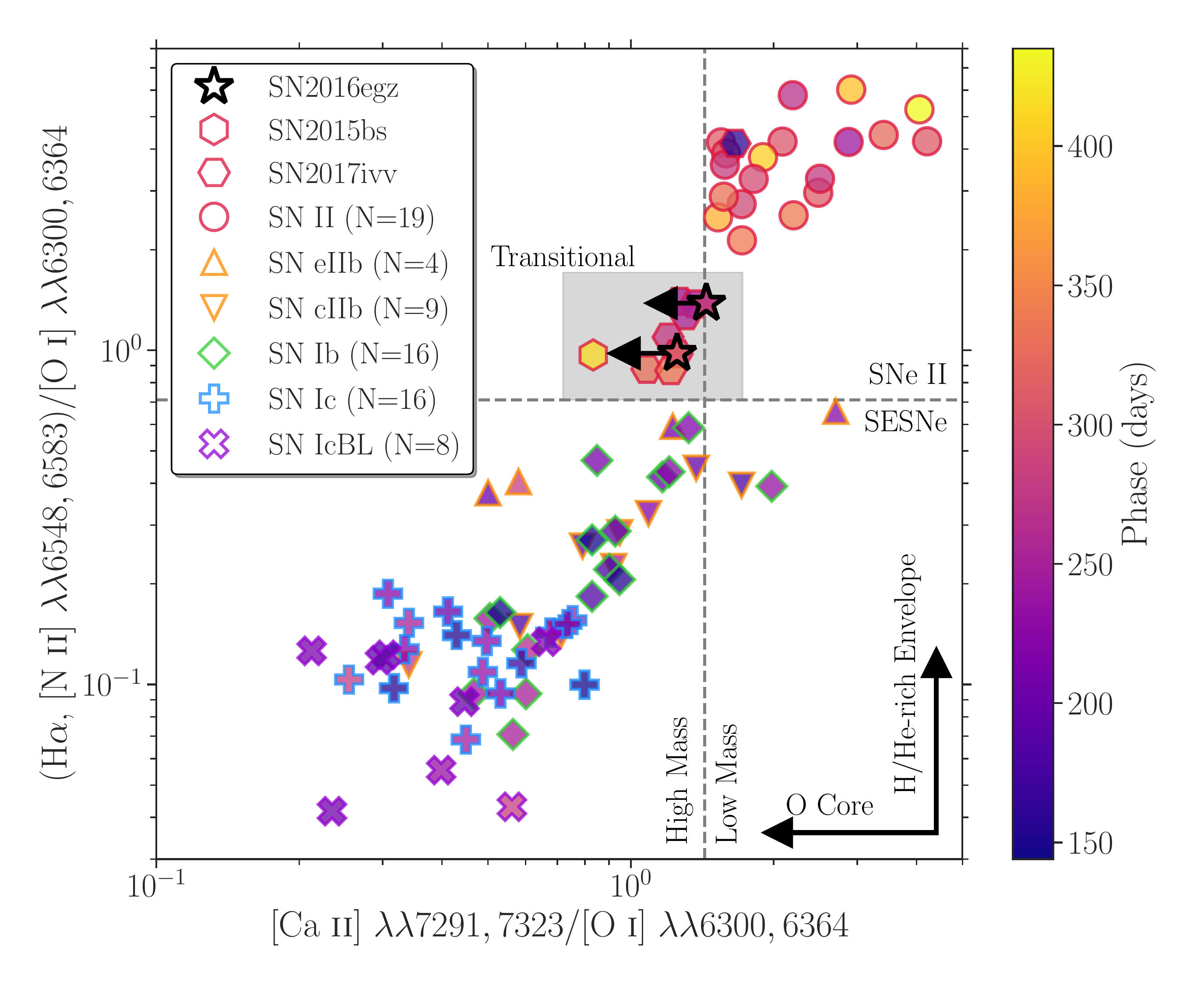}
    \caption{Comparison of the nebular line ratios of SN~2016egz (Table~\ref{tab:nebularline}) with a CCSN sample, color coded by the phase. 
    [Ca~{\sc ii}]/[O~{\sc i}] and (H$\alpha$, [N~{\sc ii}])/[O~{\sc i}] are proxies for progenitor O-core mass and H/He-rich envelope stripping, respectively. The vertical gray dashed line ($=1.43$) is a rough division between SNe II and SESNe adopted in \cite{Gutierrez2020}, and the horizontal gray dashed line ($=0.71$) is the highest value ($+1\sigma$ error) in the SESN sample of \cite{Fang2019}.
    Note the similar transitional nature of SN~2016egz to SNe~II~2015bs and 2017ivv in between SNe~II and SESNe (gray shaded region).
    Figure adapted from \cite{Kuncarayakti2015}, \cite{Fang2019}, and \cite{Gutierrez2020}, and expanded to include additional measurements from this work on the \cite{Kuncarayakti2015} SN~II sample and SNe II 1990E, 1993K, 2012A \citep{Silverman2017}, 1990K \citep{Cappellaro1995}, 1994N \citep{Pastorello2004}, 2002hh \citep{Faran2014}, 2003B, 2008bk \citep{Gutierrez2017a}, 2005cs \citep{Pastorello2009}, 2012aw \citep{Jerkstrand2014}, 2013ej \citep{Silverman2012}, 2015bs \citep{Anderson2018}, and 2016bkv \citep{Hosseinzadeh2018}, retrieved via the Open Supernova Catalog and WISeREP.
    }
    \label{fig:neb_CaOI}
\end{figure}

\begin{figure*}
 \centering
 \gridline{\fig{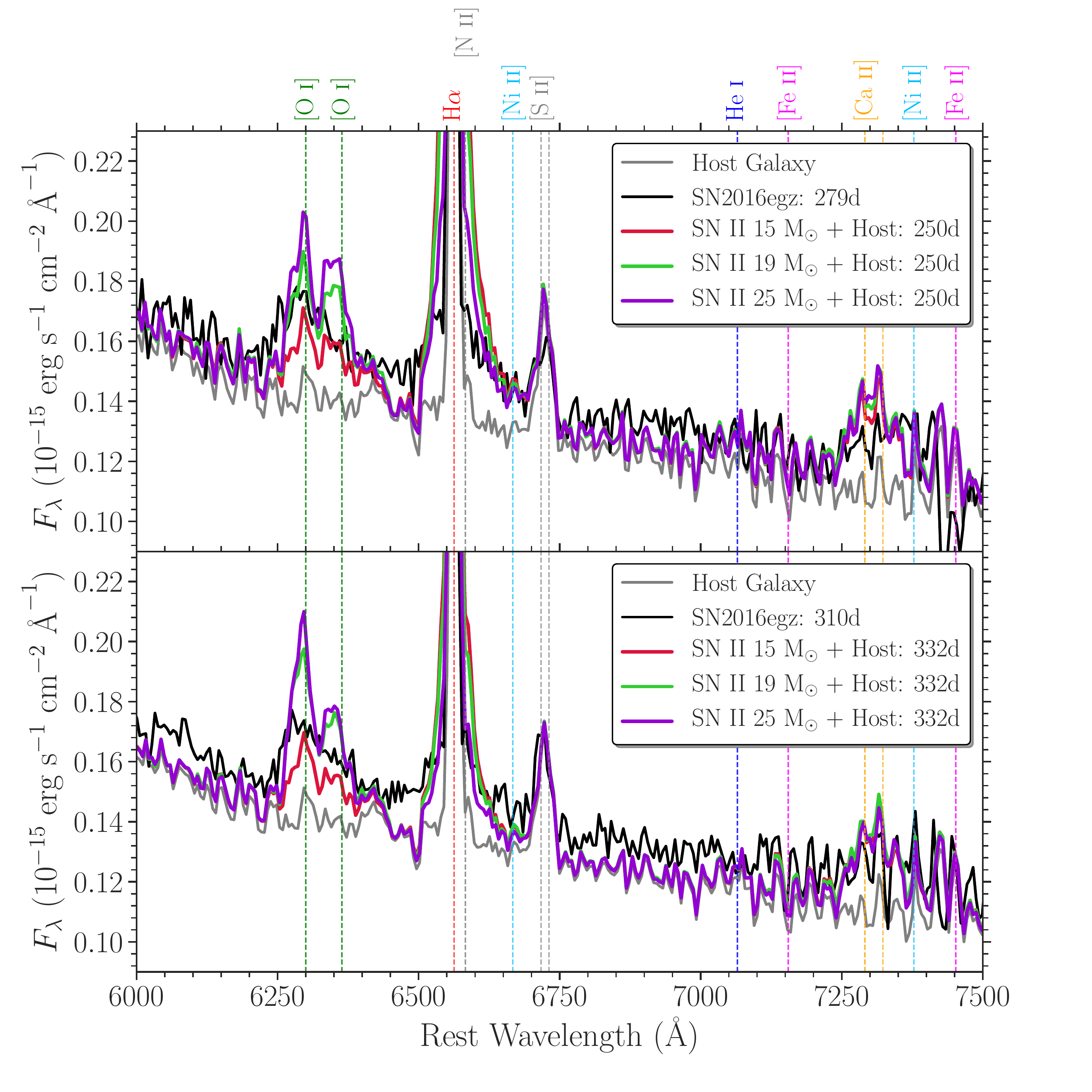}{0.45\textwidth}{\textbf{(a)} SN~II models of \cite{Jerkstrand2014}}
          \fig{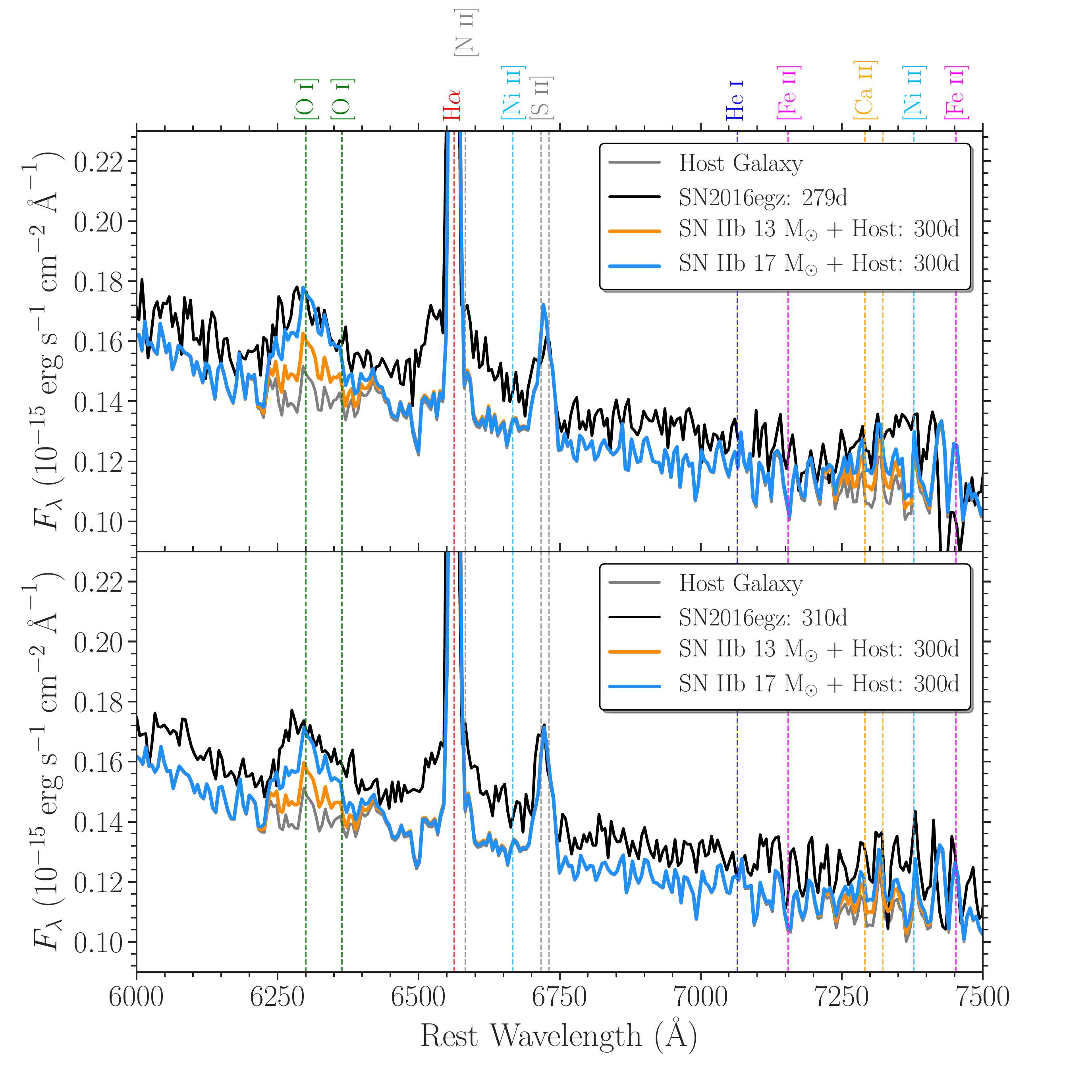}{0.45\textwidth}{\textbf{(b)} SN~IIb models of \cite{Jerkstrand2015}}}
  \caption{Comparison of the nebular spectra of SN~2016egz with the SN~II and IIb models from different progenitor $M_{\rm ZAMS}$. 
  The host spectrum is added to the model spectra that are scaled to match the distance, $^{56}$Ni mass, and phase of the observed SN spectra.
  Due to the significant host contamination and low S/N, only broad [O~{\sc i}] and H$\alpha$ features can be identified.
  Note the observed broad H$\alpha$ and [O~{\sc i}] strengths in between those of SNe II--IIb and $17$--$25\,M_\odot$ models, respectively.
  } 
  \label{fig:neb_spec}
\end{figure*}

In Figure~\ref{fig:neb_CaOI}, we show the comparison of the nebular line ratios of SN~2016egz with a combined CCSN sample from \cite{Kuncarayakti2015}, \cite{Fang2019}, and \cite{Gutierrez2020} and additional SN~II measurements from this work.\footnote{By following the same procedure as the SN~2016egz measurements, except for [Ca~{\sc ii}] where we fit a double-Gaussian profile with a single FWHM velocity.} The general increasing and decreasing trends, respectively, in the progenitor O-core mass and H/He-rich envelope stripping can be seen in the sequence of SNe II--IIb--Ib--Ic--IcBL.
For a given CCSN, [Ca~{\sc ii}]/[O~{\sc i}] is relatively constant over $\sim150$--$400$\,d after the explosion \citep{Elmhamdi2004,Maguire2012,Kuncarayakti2015,Fang2019}, while (H$\alpha$, [N~{\sc ii}])/[O~{\sc i}] decreases with time \citep{Maguire2012,Fang2019}. Since the mean phases of the SN~II and SESN samples are $330$ days and $215$ days, respectively, we expect to see more separation in the (H$\alpha$, [N~{\sc ii}])/[O~{\sc i}] space if they had been all taken at the same epoch.

\cite{Fang2019} divide the SN~IIb sample into extended (eIIb) and compact (cIIb) H-rich envelope classes based on their light-curve morphology.
\citealt{Chevalier2010} suggest $M_{\rm H_{\rm env}}\gtrsim0.1\,M_{\odot}$ for SNe eIIb, and otherwise for SNe cIIb. Given the $M_{\rm H_{\rm env}}$ range, SNe eIIb typically result in a double-peak light curve  (Figure~\ref{fig:Model_Lopt}(d)) where the first and second peaks are powered by shock-cooling envelope and radioactive decay, respectively. Compared to the other SESN types, \cite{Fang2018} and \cite{Fang2019} identify excess emission around [N~{\sc ii}] for SNe eIIb and attribute it to H$\alpha$ from the residual envelope. 

We note that SN~2016egz as well as SNe 2015bs and 2017ivv lie in a somewhat transitional region between SNe II and eIIb--Ib (Figure~\ref{fig:neb_CaOI}).
For SN~II 2015bs, \cite{Anderson2018} infer a high progenitor $M_{\rm ZAMS}$ of $17$--$25\,M_\odot$ based on the nebular line fluxes and velocities. \cite{Dessart2020} also suggest a high $M_{\rm He_{\rm core}}$ and low $M_{\rm H_{\rm env}}$ to better match the broad nebular line profiles ($v_{{\rm H}\alpha,\,{\rm FWHM}}\simeq4200$\,km\,s$^{-1}$; see also \citealt{Dessart2010}). 
For SN~2017ivv, \cite{Gutierrez2020} estimate a $M_{\rm ZAMS}$ of $15$--$17\,M_\odot$ and note the transitional characteristics from SN~II to IIb by analyzing the temporal evolution of the nebular line ratios.
Following this line of reasoning, the nebular line ratios (and upper limits) and velocities of SN~2016egz likely suggest a similar partially stripped massive progenitor. This agrees with the inferred parameters from the light-curve analysis. 

In order to be more quantitative, we compare the nebular spectra of SN~2016egz with the SNe II and IIb models of \cite{Jerkstrand2014,Jerkstrand2015}, respectively, in Figure~\ref{fig:neb_spec} where [O~{\sc i}] fluxes can be used to estimate the progenitor $M_{\rm ZAMS}$. 
We scale the model spectra by the observed distance, $^{56}$Ni mass, and phase of the SN~2016egz nebular spectra according to Equation (2) of \cite{Bostroem2019}. Then, we add the observed host galaxy spectrum to the model spectra to account for the host contamination. 
The broad SN H$\alpha$ (excluding the narrow host component) fluxes are in between those of the SNe II and IIb models, indicating a small $M_{\rm H_{\rm env}}$.
The model [O~{\sc i}] lines start to saturate above $19\,M_{\odot}$ around $300$ days \citep{Jerkstrand2014}, making the $19\,M_{\odot}$ and $25\,M_{\odot}$ models almost indistinguishable. Thus, we put a conservative $M_{\rm ZAMS}$ constraint of $17$--$25\,M_\odot$ based on the observed [O~{\sc i}] fluxes. This is consistent with the expected $M_{\rm ZAMS} \gtrsim 17.5\,M_\odot$ from the light-curve analysis.
One caveat to note is that there is no measurable [O~{\sc i}] $\lambda5577$ in the observed spectra due to the significant host contamination, preventing us from constraining the temperature and directly estimate the O-core mass \citep{Jerkstrand2014,Jerkstrand2015,Jerkstrand2017}.

\section{Discussion} \label{sec:dis}

\subsection{Formation Channel} \label{sec:formation}

The short-plateau SNe~2006Y, 2006ai, and 2016egz most likely come from partially stripped massive progenitors,\footnote{The lack of nebular spectra for SNe~2006Y and 2006ai remains a caveat.} but a remaining question is their exact formation channel. 
If it is single-star evolution as assumed in this work, the main theoretical uncertainties are RSG wind mass-loss rates and stellar rotation (e.g., \citealt{Hirschi2004,Georgy2012,Chieffi2013,Meynet2015,Renzo2017}). We assume no rotation and tweak the wind efficiency by hand, but it is debatable whether such high mass-loss rates are physically plausible. 
Observationally, there is indeed a wide range of measured RSG wind mass-loss rates (e.g., \citealt{deJager1988,vanLoon2005,Mauron2011,Goldman2017,Beasor2020}). In addition, recent observational and theoretical studies on RSGs and SNe II indicate that RSG wind mass-loss rates may be independent from metallicity \citep{Goldman2017,Chun2018,Gutierrez2018}. Thus, it could be possible that the short-plateau SNe~2006Y, 2006ai, and 2016egz originate from single-star evolution.
However, it is unlikely the case if RSG mass-loss rates are metallicity dependent (as in the main-sequence O/B stars; e.g., \citealt{Vink2000,Vink2001,Mokiem2007}), given the estimated subsolar host metallicities (Table~\ref{tab:hostclass}). In such a case, interacting binary evolution is more plausible, as \cite{Eldridge2017BPASS,Eldridge2018} indeed show some interacting binary products also result in short-plateau SNe. 
It is also important to note that any mass-loss models need to reproduce the observed populations of not only SNe~II but also RSGs. For example, \cite{Neugent2020} recently show that the luminosity function of RSGs can be used to constrain their mass-loss rates. Future statistical studies with both RSG and SN~II populations at various metallicities are required to distinguish the formation channels of short-plateau SNe.

Aside from the continuous mass loss, the origin of the confined dense CSM is not well understood. 
In the single-star scenario, a number of possible mechanisms have been proposed, including pulsation-driven superwinds \citep{Heger1997,Yoon2010}, extended stellar envelopes \citep{Dessart2017,Soker2021}, and internal gravity waves \citep{Quataert2012,Shiode2014,Quataert2016,Fuller2017,Leung2020,Morozova2020,Wu2021}. 
Although the expected $M_{\rm ZAMS}$ range ($19$--$25\,M_\odot$), where pulsation-driven superwinds could significantly alter mass loss, matches with our inferred $M_{\rm ZAMS}$ range for SNe~2006Y, 2006ai, and 2016egz, the expected maximum mass-loss rate of $\lesssim10^{-3}\,M_\odot\,{\rm yr}^{-1}$ around $10^4$--$10^6$\,yr before the core collapse may not result in confined CSM. Also, as the expected timescale of dominant CSM emission from an extended envelope (up to about $10$--$100$ stellar radii) is only over a few days after explosion, whether it can reproduce the $\sim10$\,d excess emission seen in SNe~2006Y, 2006ai, and 2016egz is questionable.   
In the binary scenario, the confined dense CSM may be expected from a wind collision interface formed between each binary component if the separation is wide enough \citep{Kochanek2019}. With such a wide separation, however, short-plateau SNe do not result solely from binary interaction \citep{Eldridge2017BPASS,Eldridge2018}.
As such, the enhanced mass loss from internal gravity waves remains more probable, although the light-curve model predictions can be quite flexible by varying the amount and time of nuclear energy injection \citep{Morozova2020}.

\subsection{Implications for the RSG Problem} \label{sec:RSG}

\begin{figure*}
 \centering
 \gridline{\fig{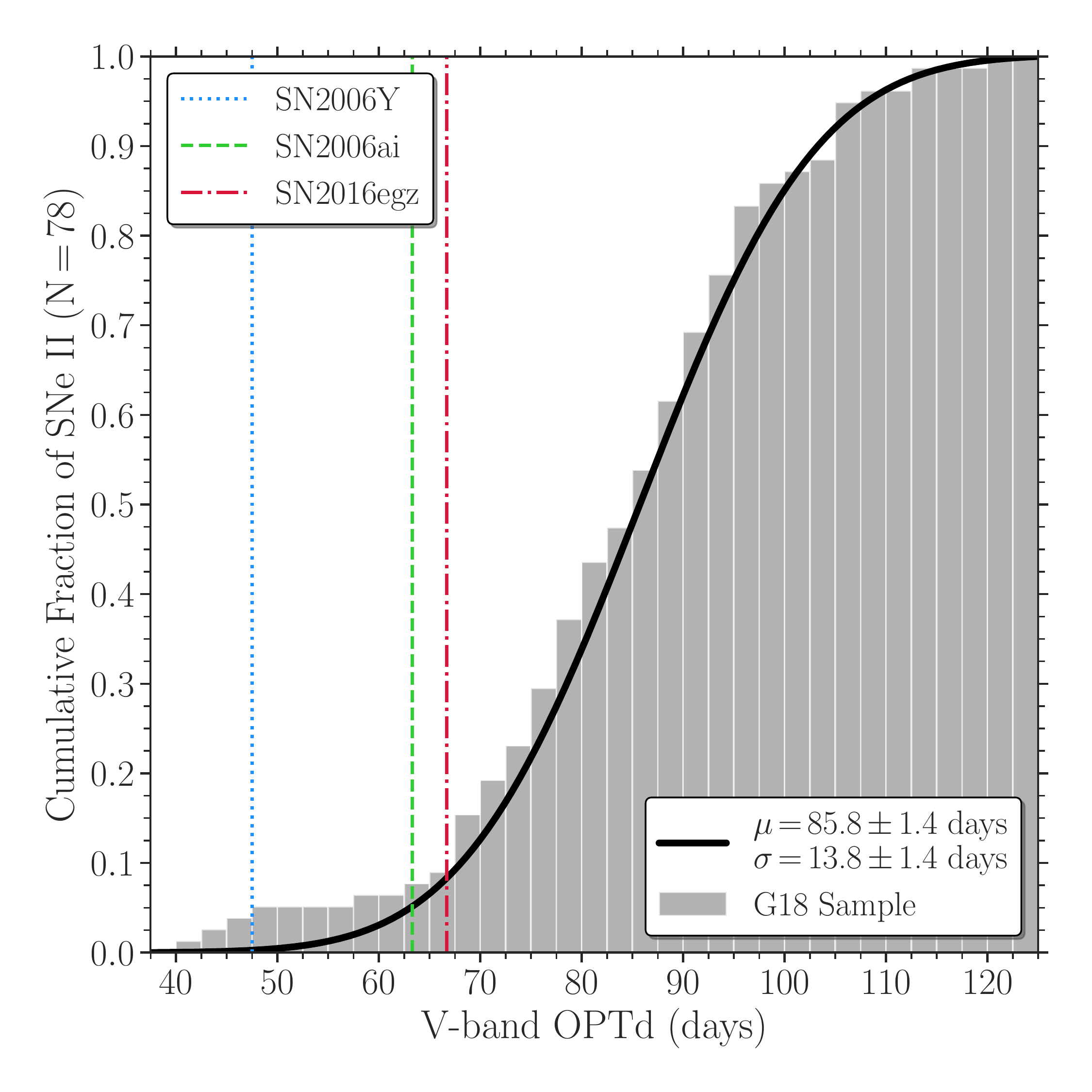}{0.45\textwidth}{\textbf{(a)} Cumulative distribution of \textit{V}-band optically-thick phase duration (OPTd) of the \citet[G18]{Gutierrez2018} sample}
          \fig{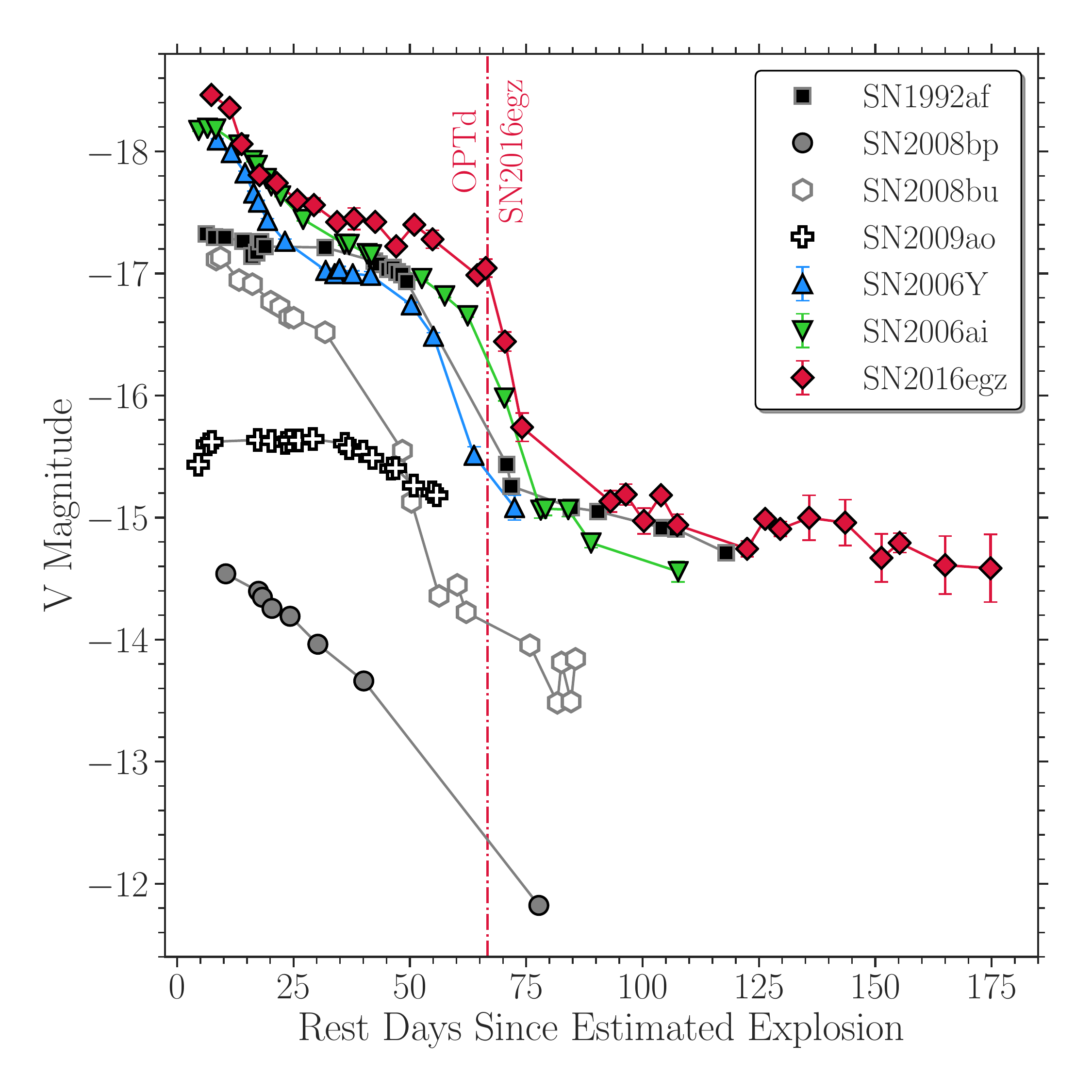}{0.45\textwidth}{\textbf{(b)} SN~II \textit{V}-band light curves with shorter OPTd than SN~2016egz, as in (a), from \cite{Anderson2014}}}
  \caption{Distribution of the SN~II OPTd with a mean $\mu\simeq86$\,days and standard deviation $\sigma\simeq14$\,days. SNe~2006Y, 2006ai, and 2016egz show similar luminous short-plateau light curves, while SNe~2008bp and 2008bu show IIL-like light curves (the extinction and explosion epoch of SN~1992af are not well constrained, and the fall from the plateau of SN~2009ao is not well sampled). Note the small fraction of short-plateau SNe ($3/78\simeq4\%$).
  } 
  \label{fig:SPSN_V_OPTd_LC}
\end{figure*}

Regardless of the exact progenitor formation scenario, our conclusion based on the SN photometric and spectroscopic analyses should still hold, and it is worth discussing the implications of the short-plateau SNe for the RSG problem. 
Some proposed solutions even point toward the nonexistence of the RSG problem, as the high-mass progenitors ($\gtrsim20\,M_\odot$) may not end their lives as RSGs, but as yellow supergiants (or even blue supergiants or Wolf–Rayet stars) resulting in SNe IIb/Ib (or Ib/c) due to binary interaction \citep{Eldridge2008,Eldridge2013,Eldridge2017BPASS,Eldridge2018}, stellar rotation \citep{Hirschi2004,Chieffi2013}, and/or wind mass loss \citep{Georgy2012,Meynet2015,Renzo2017}. 
Others propose the high-mass RSGs do explode as SNe II, but the mistreatment of dust extinction \citep{Walmswell2012,Beasor2016} and/or bolometric correction \citep{Davies2018} in the direct progenitor identifications systematically underestimates the progenitor masses. 
In addition, the statistical significance of the RSG problem has been questioned (see \citealt{Davies2020,Kochanek2020} for recent discussions).
If the RSG problem is indeed real, then the direct collapse of high-mass RSGs to black holes (i.e., failed explosion) is a plausible solution \citep{O'Connor2011, Ugliano2012, Kochanek2014, Gerke2015, Pejcha2015, Sukhbold2016, Adams2017, Basinger2020, Pejcha2020, Kresse2021}. 
In this context, the transitional nature of SNe~2006Y, 2006ai, and 2016egz is important, highlighting the possibility that these partially stripped massive progenitors die as RSGs and explode as short-plateau SNe II, rather than becoming SESNe or directly collapsing to black holes.

As a first-order rate estimate for short-plateau SNe, we use the optically thick phase duration (OPTd; the time between the explosion and the plateau drop) distribution of SNe II (including SNe~2006Y, 2006ai, and 2016egz) from \cite{Gutierrez2018} in Figure~\ref{fig:SPSN_V_OPTd_LC}. The lack of SNe II with ${\rm OPTd}\lesssim40$\,days indicates that there exists an H-rich envelope boundary between SNe II and IIb, as also seen in the light-curve model grid (Figure~\ref{fig:Model_Lopt}). 
On the tail of the smooth Gaussian-like OPTd distribution, there is a small population with shorter OPTd than SN~2016egz (Figure~\ref{fig:SPSN_V_OPTd_LC}). 
SNe~2008bp\footnote{The extinction estimate for SN~2008bp may be significantly underestimated \citep{Anderson2014}, and so the luminosity could be much higher.} and 2008bu show IIL-like light curves with fainter $^{56}$Co decay tails than those of SNe~2006Y, 2006ai, and 2016egz. Their light curves might be reproduced with the high-mass ($\geq17.5\,M_\odot$) progenitor models within the short-plateau $M_{\rm H_{\rm env}}$ range with a lower $M_{\rm Ni}$ and varying $E_{\rm exp}$ (Figures~\ref{fig:Model_Lopt}(c) and \ref{fig:Model_MNI}(c)). 
We do not consider the other SNe as probable short-plateau candidates here because of the uncertainties in extinction and explosion epoch for SN~1992af \citep{Anderson2014,Galbany2016II,Gutierrez2017a} and the poor light-curve sampling at the plateau fall for SN~2009ao.
If we take the short-plateau SN fraction of $3/78$ (or $5/78$ if we naively include SNe~2008bp and 2008bu) at face value, a short-plateau SN rate of $\sim4\%$ (or $\sim6\%$) of all SNe~II can be inferred. This is roughly consistent with the rate estimate from \cite{Eldridge2018} ($\sim 4.7\%$; see $\S$\ref{sec:LCmodel}). 
Recently, \cite{Gofman2020} apply enhanced mass-loss rates to RSGs and found that massive RSGs ($M_{\rm ZAMS}\simeq20$--$21\,M_\odot$) with a similar $M_{\rm H_{\rm env}}$ range ($\sim0.8$--$1.8\,M_\odot$) to that of our short-plateau SN models end up as dust-unobscured SNe~II. They roughly estimate a rate of this class to be $2$--$4\%$ of all SNe~II, which is similar to our estimate of short-plateau SNe. 

Assuming the Salpeter IMF with lower and upper RSG mass limits of $8$ and $25\,M_\odot$, respectively, the inferred ZAMS mass range of the short-plateau SNe~2006Y, 2006ai, and 2016egz ($18$--$22\,M_\odot$) corresponds to $10\%$ of all SNe~II.
It may be possible that the missing fraction ($10\%-4\%=6\%$) does not end up as RSGs, but there also seems to be an absence of promising high-mass SESN candidates \citep{Lyman2016,Taddia2018,Barbarino2020}.
Thus, together with SN~2015bs \citep{Anderson2018} and the possible failed explosion of a $\sim25\,M_\odot$ RSG \citep{Gerke2015,Adams2017,Basinger2020}, this apparent rate mismatch may support the proposed \textit{islands of explodability in a sea of black hole formation} \citep{Sukhbold2016} where there is no single mass cut between SN~II explosions and black hole formations (see also \citealt{O'Connor2011,Ugliano2012,Pejcha2015,Patton2020,Pejcha2020,Sukhbold2020,Kresse2021}).

In order to further test the hypothesis, more detailed progenitor mass and rate calculations of short-plateau SNe are necessary.
The bare photospheres of the short-plateau SN progenitor models in this work lie in the effective temperature and photospheric luminosity ranges of $3.65 < {\rm log}_{10}(T_{\rm eff}/{\rm K}) < 3.72$ and $5.23 < {\rm log}_{10}(L_{\rm ph}/L_{\odot}) < 5.53$, respectively, which are within those observed for luminous RSGs (e.g., \citealt{Levesque2005,Levesque2006,Massey2009,Drout2012,Gordon2016}). Given the high constant mass loss as well as the enhanced mass loss in the last few decades before the explosion, however, significant dust extinction is likely expected for the short-plateau SN progenitors (e.g., \citealt{Nozawa2013,Nozawa2014}), making their direct progenitor identifications unlikely (if they all come from the single-star channel; see \citealt{Eldridge2017BPASS,Eldridge2018} for the interacting binary channel). The circumstellar dust is likely destroyed as the SN shock progresses through.
Therefore, SN photometric and spectroscopic analyses of large samples will be required to better determine their progenitor mass and rate distributions.

\section{Conclusions} \label{sec:con}

We have presented the optical/NIR photometric and spectroscopic observations of luminous Type II short-plateau SNe~2006Y, 2006ai, and 2016egz. 
Based on the \textit{V}-band and pseudobolometric light-curve sample comparisons, the peculiar light curves with the short plateaus and luminous peaks suggest partially stripped H-rich envelopes and CSM interaction at early time.
We have constructed a large \texttt{MESA}+\texttt{STELLA} single-star progenitor and light-curve model grid (with and without CSM interaction) that shows a continuous population of SNe IIP--IIL--IIb-like light-curve morphology in descending order of H-rich envelope mass, with short-plateau SNe lying in a confined parameter space between SNe IIL and IIb with large $^{56}$Ni masses ($\gtrsim0.05\,M_\odot$).
For SNe~2006Y, 2006ai, and 2016egz, our model grid suggests high-mass RSG progenitors ($M_{\rm ZAMS} \simeq 18$--$22\,M_{\odot}$) with small H-rich envelope masses ($M_{\rm H_{\rm env}} \simeq 1.7\,M_{\odot}$) to reproduce the short-plateau light curves, and enhanced mass loss ($\dot{M} \simeq 10^{-2}\,M_{\odot}\,{\rm yr}^{-1}$) for the last few decades before the explosion to capture the early luminous peaks.
In addition, the P Cygni profiles and photospheric velocities likely point toward low H-rich envelope masses, and the nebular line ratios and spectral model comparisons prefer high-mass ($M_{\rm ZAMS} \simeq 17$--$25\,M_{\odot}$) progenitors. 

Although the exact progenitor formation channel remains an open question, the transitional nature of SNe~2006Y, 2006ai, and 2016egz has important implications for the RSG problem, in that these partially stripped massive progenitors end their lives as RSGs and explode as short-plateau SNe.  
We roughly estimate a short-plateau SN rate of $\sim4\%$ of all SNe~II, which is smaller than the IMF expectation of $\sim10\%$. This rate mismatch may support the proposed \textit{islands of explodability} scenario. 
Further verification of this scenario requires the determination of the short-plateau SN progenitor mass and rate distributions through large SN photometric and spectroscopic sample analyses (as done in this work for three objects), as their direct progenitor identifications are expected to be quite challenging.
Current and future large transient surveys are promising to provide the necessary SN samples.

\acknowledgments

We are grateful to Viktoriya Morozova, Anthony L. Piro, Lars Bildsten, Bill Paxton, J. J. Eldridge, Eva Laplace, Sergei I. Blinnikov, Ken'ichi Nomoto, Masaomi Tanaka, and Nozomu Tominaga for their comments and discussions.

D.H., D.A.H., G.H., C.M., and J.B. were supported by NSF grants AST-1313484 and AST-1911225, as well as by NASA grant 80NSSC19kf1639.
D.H. is thankful for support and hospitality by the National Astronomical Observatory of Japan where many discussions of this work took place.
J.A.G. is supported by the NSF GRFP under grant number 1650114.
I.A. is a CIFAR Azrieli Global Scholar in the Gravity and the Extreme Universe Program and acknowledges support from that program, from the European Research Council (ERC) under the European Union's Horizon 2020 research and innovation program (grant agreement number 852097), from the Israel Science Foundation (grant numbers 2108/18 and 2752/19), from the United States - Israel Binational Science Foundation (BSF), and from the Israeli Council for Higher Education Alon Fellowship.
C.P.G acknowledges support from EU/FP7-ERC grant number 615929.
Research by S.V. is supported by NSF grant AST–1813176.
Q.F. acknowledges support from the JSPS Kakenhi grant (20J23342). 
K. Maeda acknowledges support by JSPS KAKENHI grants JP20H00174, JP20H04737, JP18H04585, JP18H05223, and JP17H02864.
G.F. acknowledges the support from CONICET through grant PIP-2015-2017-11220150100746CO and from ANPCyT through grant PICT-2017-3133.
M.D.S. acknowledges support from a project grant (8021-00170B) from the Independent Research Fund Denmark and a generous grant (28021) from the VILLUM FONDEN.
M.G. is supported by the Polish NCN MAESTRO grant 2014/14/A/ST9/00121.
K. Maguire is funded by the EU H2020 ERC grant number 758638.
T.M.B. was funded by the CONICYT PFCHA/DOCTORADOBECAS CHILE/2017-72180113.

This paper made use of data from the Las Cumbres Observatory global network of telescopes through the Supernova Key Project and Global Supernova Project.
This paper includes data gathered with the 6.5 meter Magellan Telescopes located at Las Campanas Observatory, Chile.
The CSP-I was supported by the NSF under grants AST-0306969, AST-0607438 and AST-1008343.
This research has made use of the NASA/IPAC Extragalactic Database (NED), which is funded by NASA and operated by the California Institute of Technology, as well as IRAF, which is distributed by NOAO (operated by AURA, Inc.), under cooperative agreement with NSF.

The authors wish to recognize and acknowledge the very significant cultural role and reverence that the summit of Haleakal$\bar{\text{a}}$ has always had within the indigenous Hawaiian community. We are most fortunate to have the opportunity to conduct observations from the mountain.

Funding for SDSS-III has been provided by the Alfred P. Sloan Foundation, the Participating Institutions, the National Science Foundation, and the U.S. Department of Energy Office of Science. The SDSS-III website is \url{http://www.sdss3.org/}.

SDSS-III is managed by the Astrophysical Research Consortium for the Participating Institutions of the SDSS-III Collaboration, including the University of Arizona, the Brazilian Participation Group, Brookhaven National Laboratory, Carnegie Mellon University, University of Florida, the French Participation Group, the German Participation Group, Harvard University, the Instituto de Astrofisica de Canarias, the Michigan State/Notre Dame/JINA Participation Group, Johns Hopkins University, Lawrence Berkeley National Laboratory, Max Planck Institute for Astrophysics, Max Planck Institute for Extraterrestrial Physics, New Mexico State University, New York University, Ohio State University, Pennsylvania State University, University of Portsmouth, Princeton University, the Spanish Participation Group, University of Tokyo, University of Utah, Vanderbilt University, University of Virginia, University of Washington, and Yale University.

\vspace{5mm}
\facilities{ADS, du Pont (B\&C, WFCCD, WIRC), IRSA, Las Cumbres (FLOYDS, SBIG, Sinistro), Magellan:Baade (IMACS), Magellan:Clay (LDSS2), NED, NTT (EFOSC2), Swope (SITe2K CCD, RetroCam), VLT-UT1 (FORS2)}.

\defcitealias{AstropyCollaboration2018}{Astropy Collaboration 2018}
\software{Astropy \citepalias{AstropyCollaboration2018}, \texttt{floyds\_pipeline} \citep{Valenti2014},  \texttt{lcogtsnpipe} \citep{Valenti2016}, Matplotlib \citep{Hunter2007}, \texttt{MESA} \citep{Paxton2011,Paxton2013,Paxton2015,Paxton2018,Paxton2019}, NumPy \citep{Oliphant2006}, \texttt{PyMCZ} \citep{Bianco2016}, PyRAF \citep{PyRAF2012}, \texttt{PyZOGY} \citep{Guevel2017}, SciPy \citep{SciPy2020}, seaborn \citep{Waskom2020}, \texttt{SExtractor} \citep{Bertin1996}, \texttt{STELLA} \citep{Blinnikov1998,Blinnikov2000,Blinnikov2006,Blinnikov2004,Baklanov2005}}.

\appendix
\restartappendixnumbering
\section{Extra \texttt{MESA}+\texttt{STELLA} Description and Models}\label{sec:extra}

We start with \texttt{MESA} revision 10398 test suites, \texttt{example\_make\_pre\_ccsn} and \texttt{example\_ccsn\_IIp}, respectively, for progenitor evolution from the pre-main sequence to Fe-core infall and SN explosions from the core collapse to near shock breakout (by excising the core and injecting a thermal bomb). Then we use \texttt{STELLA} for SN light-curve and expansion-velocity calculations from the shock breakout to radioactive decay tail.
The reader is referred to \cite{Paxton2018} for more details on the workflow and relevant physical parameters.

For the \texttt{MESA} progenitor model grid, we vary ZAMS masses ($M_{\rm ZAMS}=10.0$--$25.0\,M_{\odot}$ with $2.5\,M_{\odot}$ increments) and wind scaling factors ($\eta_{\rm wind}=0.0$--$3.0$ with $0.1$ increments), while fixing subsolar ZAMS metallicity ($Z=0.3\,Z_{\odot}$), no rotation ($\nu/\nu_{\rm crit}=0$), mixing length in the H-rich envelope ($\alpha_{\rm env}=3.0$), and convective overshooting parameters ($f_{\rm ov}=0.02$ and $0.01$ for $10.0\,M_{\odot}$ and the other ZAMS masses, respectively). 
As the main objective of this work is to observe how SN~II light curves are affected by H-rich envelope stripping, we control the progenitor mass loss by using the \textit{Dutch} prescription (\citealt{Glebbeek2009}, and references therein) and arbitrarily varying $\eta_{\rm wind}$ in single-star evolution (see, e.g., \citealt{Mauron2011,Goldman2017,Beasor2020} for a wide range of observed RSG mass-loss rates), rather than exploring the details of mass-loss mechanisms (e.g., interacting binary evolution; see \citealt{Eldridge2017HSN} for a review). 
With our parameter setup, $\sim20\%$ of the \texttt{MESA} progenitor models do not advance to Fe-core formation, mostly due to some low-mass ($\leq12.5\,M_{\odot}$) models developing highly degenerate cores and fail during off-center burning stages such as neon ignition.

For the \texttt{MESA} explosion model grid, we vary explosion energies ($E_{\rm exp}=0.4$--$2.0\times10^{51}$\,erg with $0.2\times10^{51}$\,erg increments) for each progenitor model. The SN shock propagation is modeled with the \cite{Duffell2016} Rayleigh-Taylor instability mixing prescription until near shock breakout, and the resultant $^{56}$Ni distribution is scaled to match a fixed $^{56}$Ni mass ($M_{\rm Ni}=0.04$, $0.07$, and $0.1\,M_{\odot}$).
Then, we hand off these explosion models to \texttt{STELLA} to produce synthetic light curves and expansion velocities using 400 spatial zones and 40 frequency bins.
Any fallback material is frozen onto the central remnant and excised via a $500$\,km\,s$^{-1}$ velocity cut at the inner boundary at the handoff between \texttt{MESA} and \texttt{STELLA} (i.e., the extra heating from fallback accretion onto the central remnant is not taken into account, but can be relevant for high-mass progenitors with low $E_{\rm exp}$; e.g., \citealt{Lisakov2018,Chan2018,Moriya2019}).
With our parameter setup, $\sim5\%$ each of the \texttt{MESA} explosion models and \texttt{STELLA} light-curve models do not converge numerically, mostly with high-mass ($\geq22.5\,M_{\odot}$) progenitors.
We simply discard the failed models (in Fe-core formation or explosion) and focus on the bulk properties of the model grid in this work.

\begin{figure*}
 \centering
   \gridline{\fig{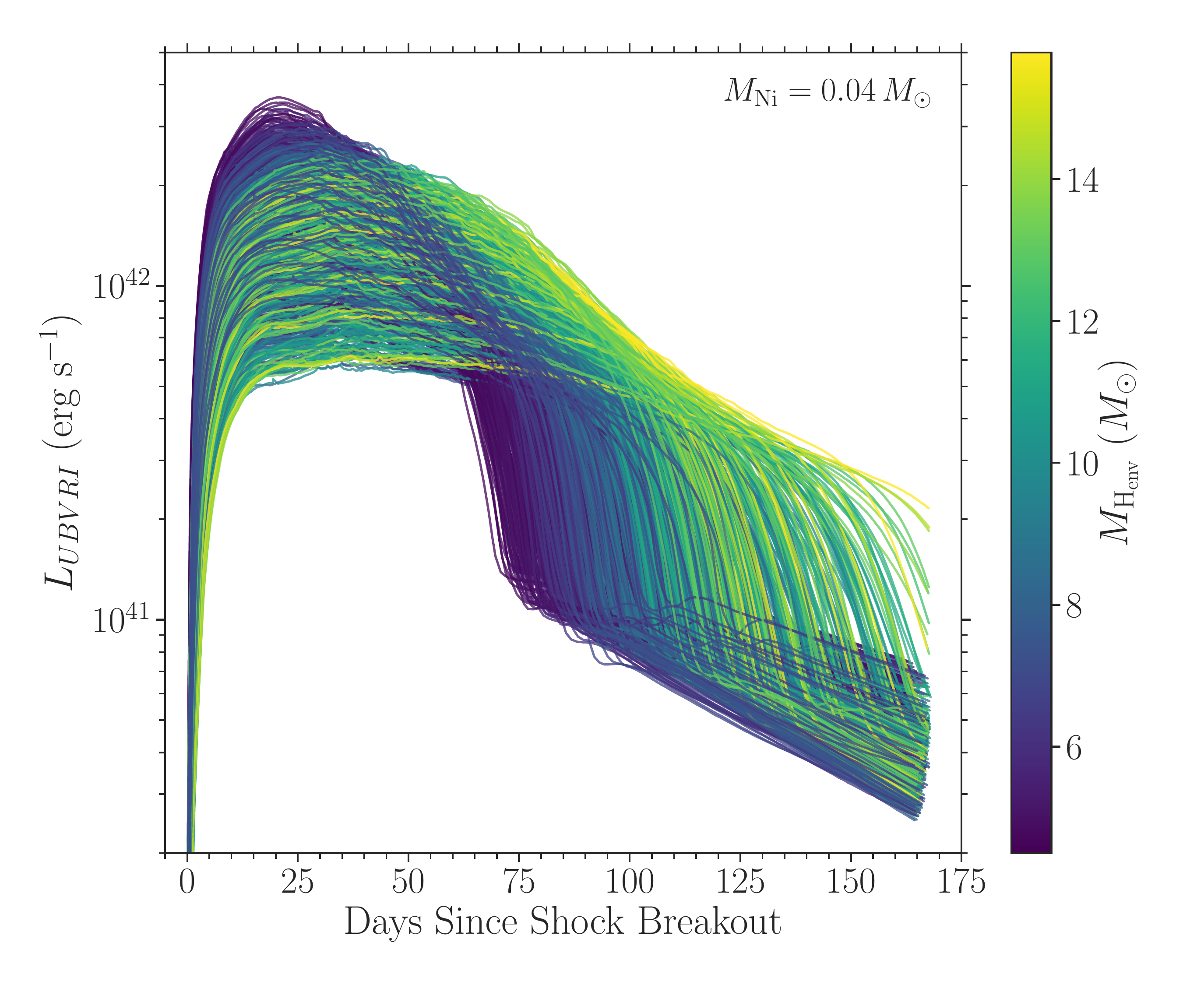}{0.49\textwidth}{\textbf{(a)} IIP ($4.5\,M_{\odot} \leq M_{\rm H_{\rm env}} \leq 15.8\,M_{\odot}$)}
          \fig{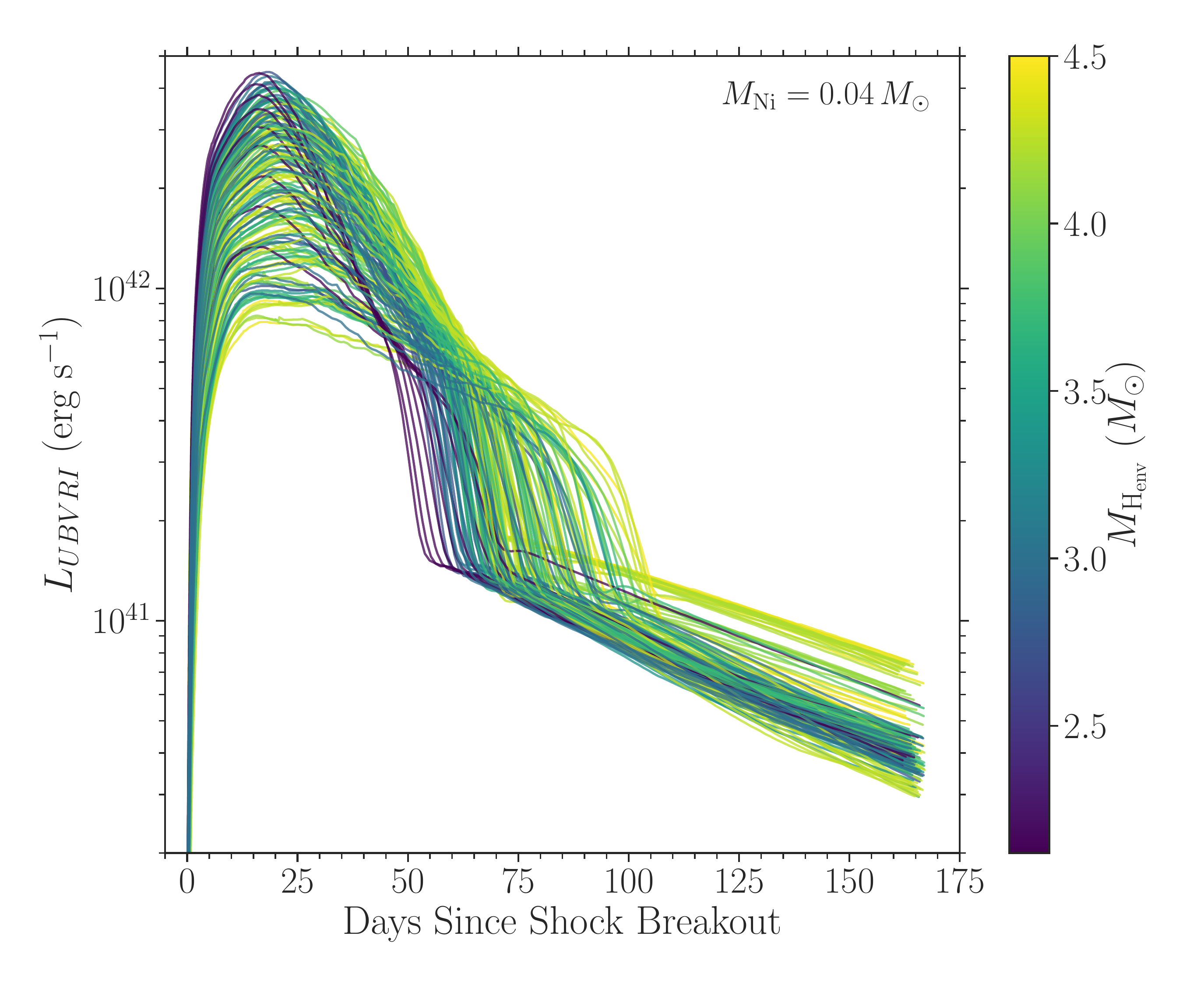}{0.49\textwidth}{\textbf{(b)} IIL ($2.12\,M_{\odot} \leq M_{\rm H_{\rm env}} \leq 4.5\,M_{\odot}$)}}
   \gridline{\fig{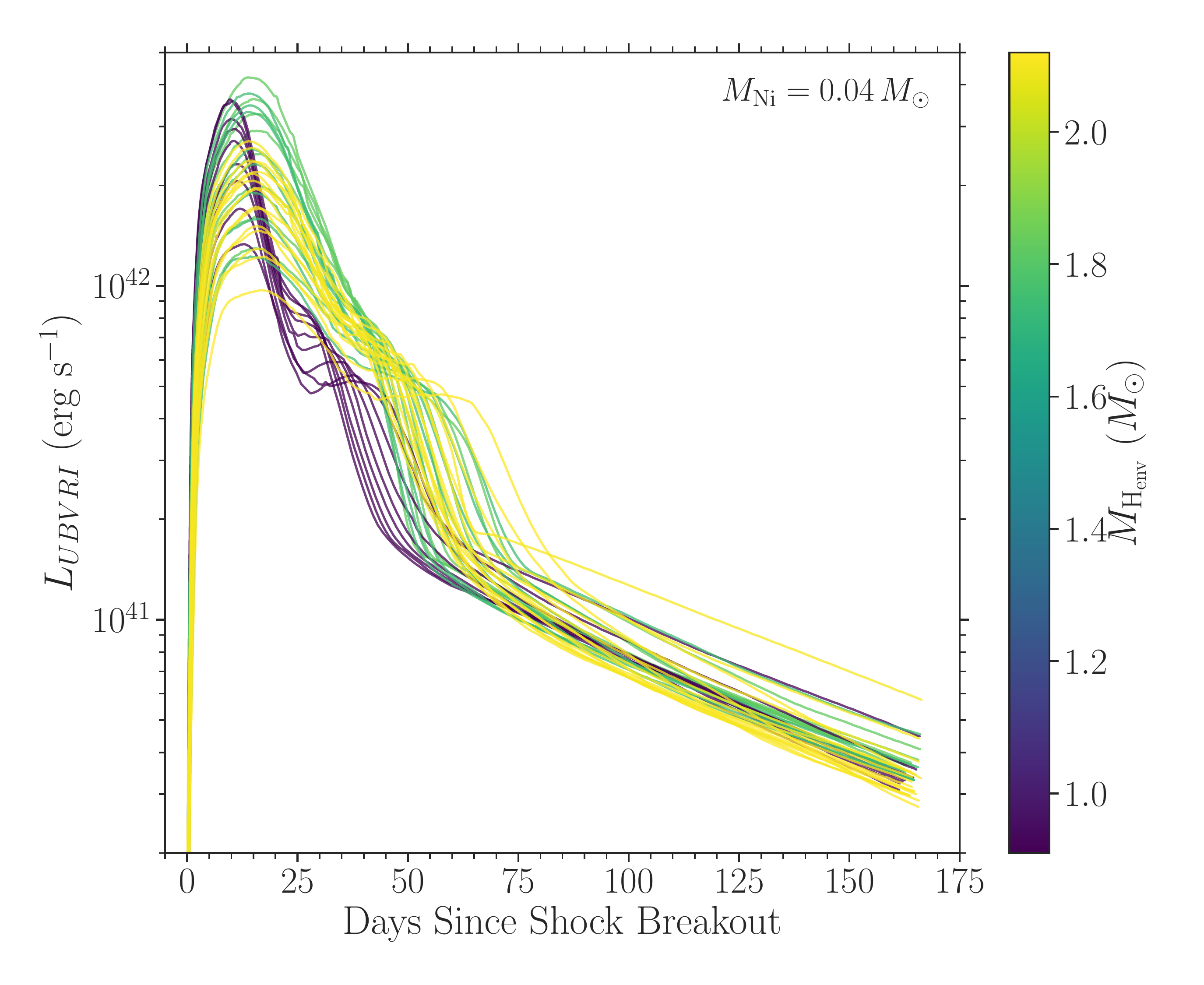}{0.49\textwidth}{\textbf{(c)} Short-Plateau ($0.91\,M_{\odot} \leq M_{\rm H_{\rm env}} \leq 2.12\,M_{\odot}$)}
          \fig{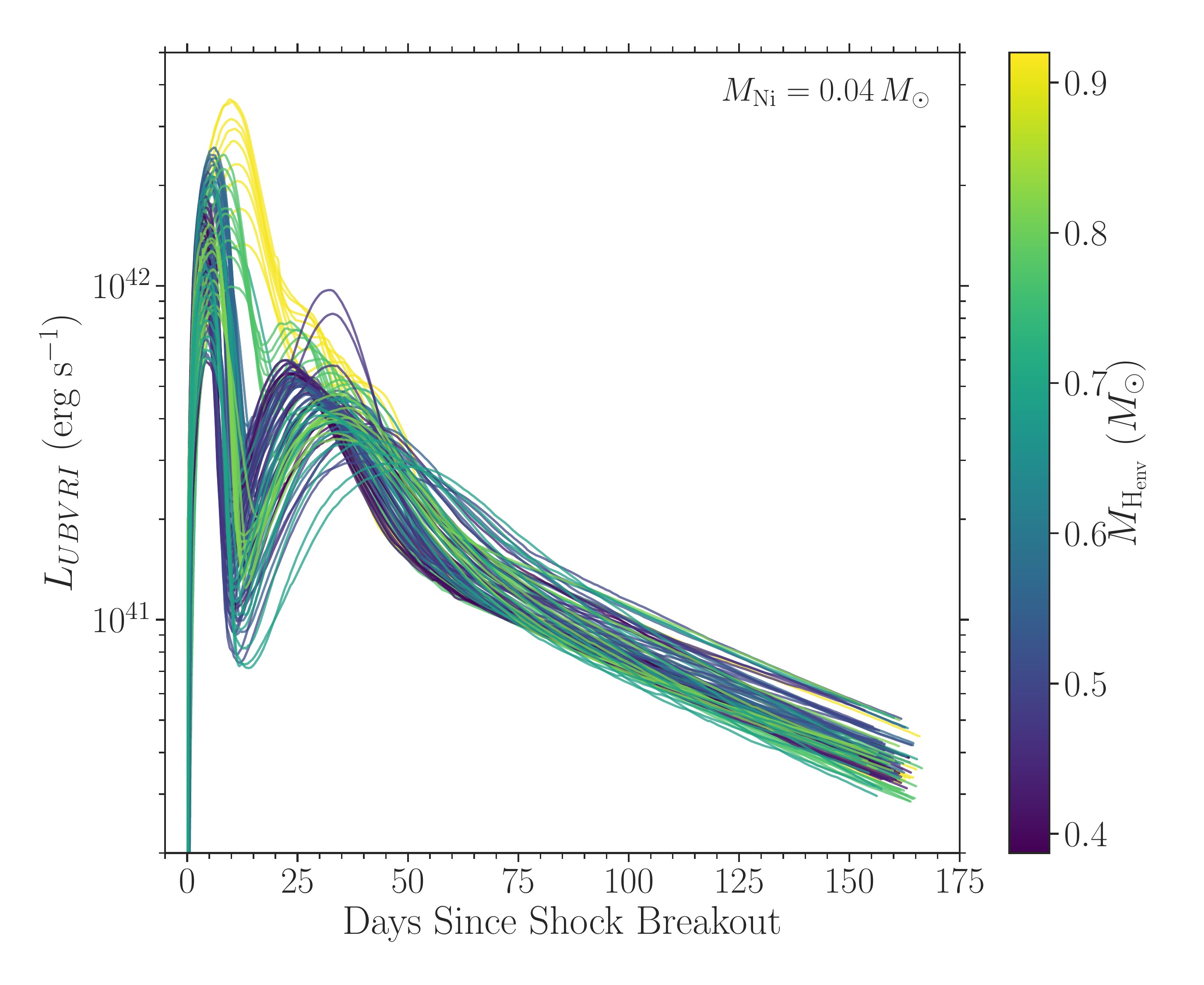}{0.49\textwidth}{\textbf{(d)} IIb ($0.39\,M_{\odot} \leq M_{\rm H_{\rm env}} \leq 0.91\,M_{\odot}$)}}
\caption{Same as Figure~\ref{fig:Model_Lopt}, but with $M_{\rm Ni}=0.04\,M_\odot$.
  } 
  \label{fig:Model_Lopt_004}
\end{figure*}

\begin{figure*}
 \centering
   \gridline{\fig{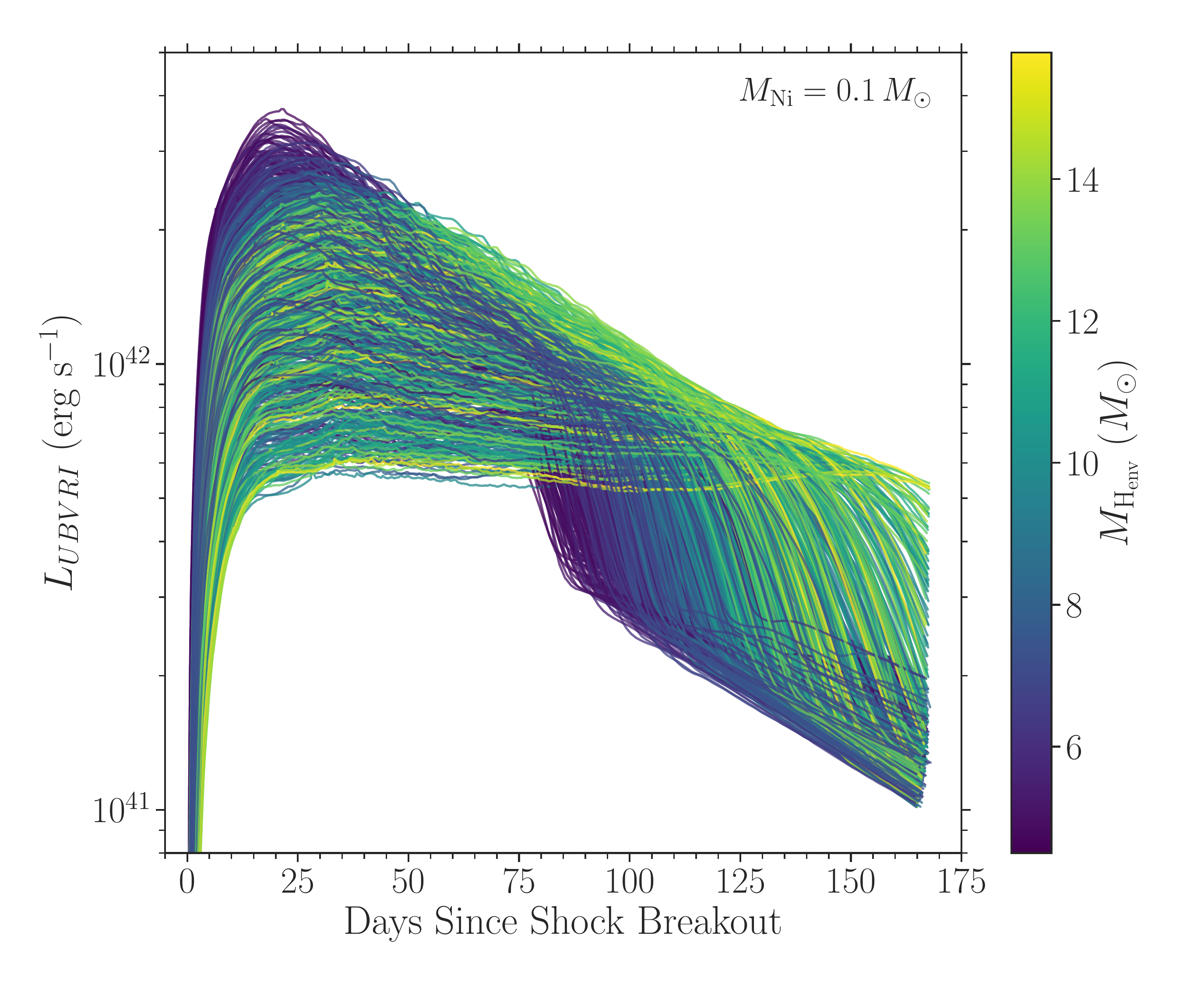}{0.49\textwidth}{\textbf{(a)} IIP ($4.5\,M_{\odot} \leq M_{\rm H_{\rm env}} \leq 15.8\,M_{\odot}$)}
          \fig{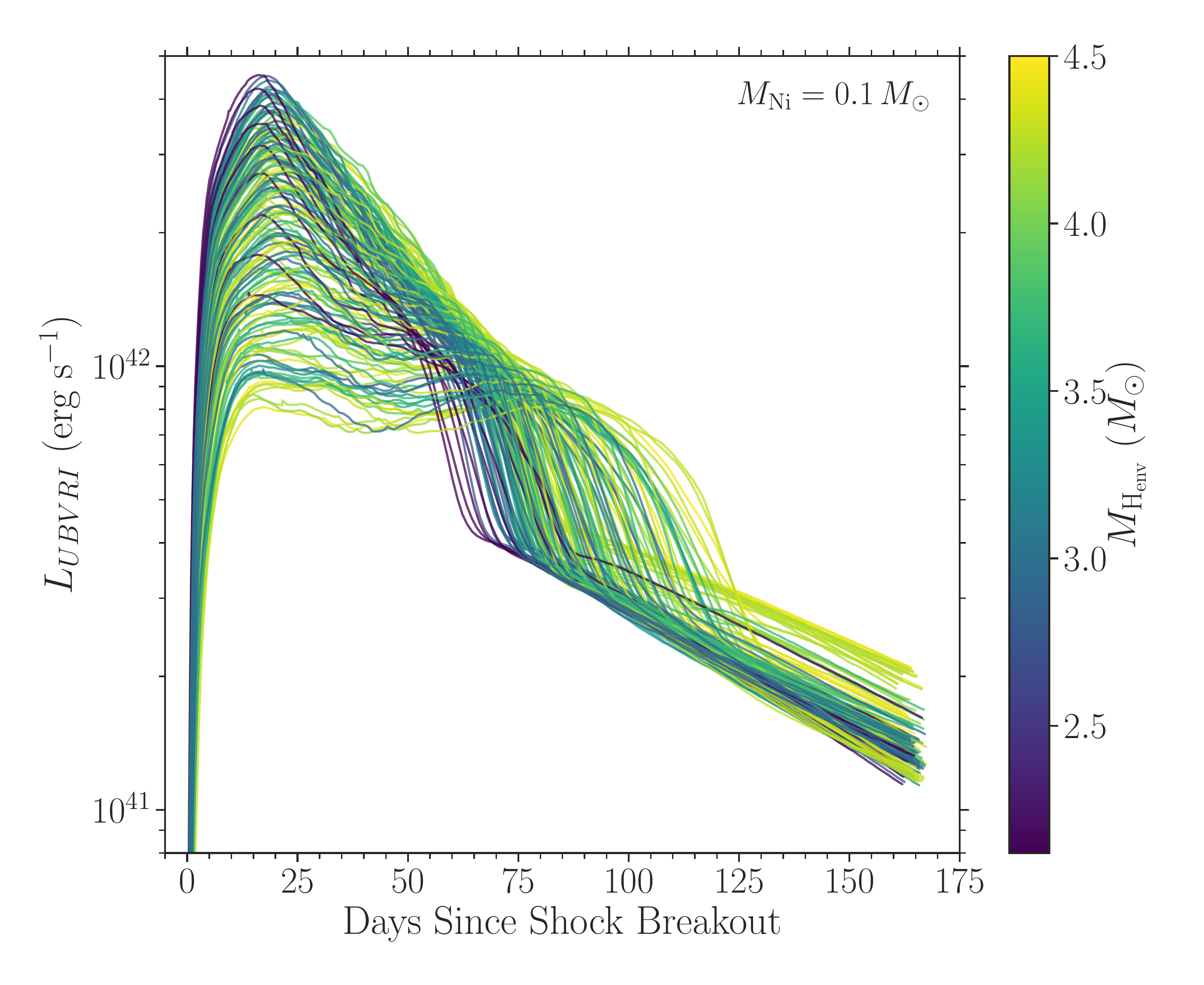}{0.49\textwidth}{\textbf{(b)} IIL ($2.12\,M_{\odot} \leq M_{\rm H_{\rm env}} \leq 4.5\,M_{\odot}$)}}
   \gridline{\fig{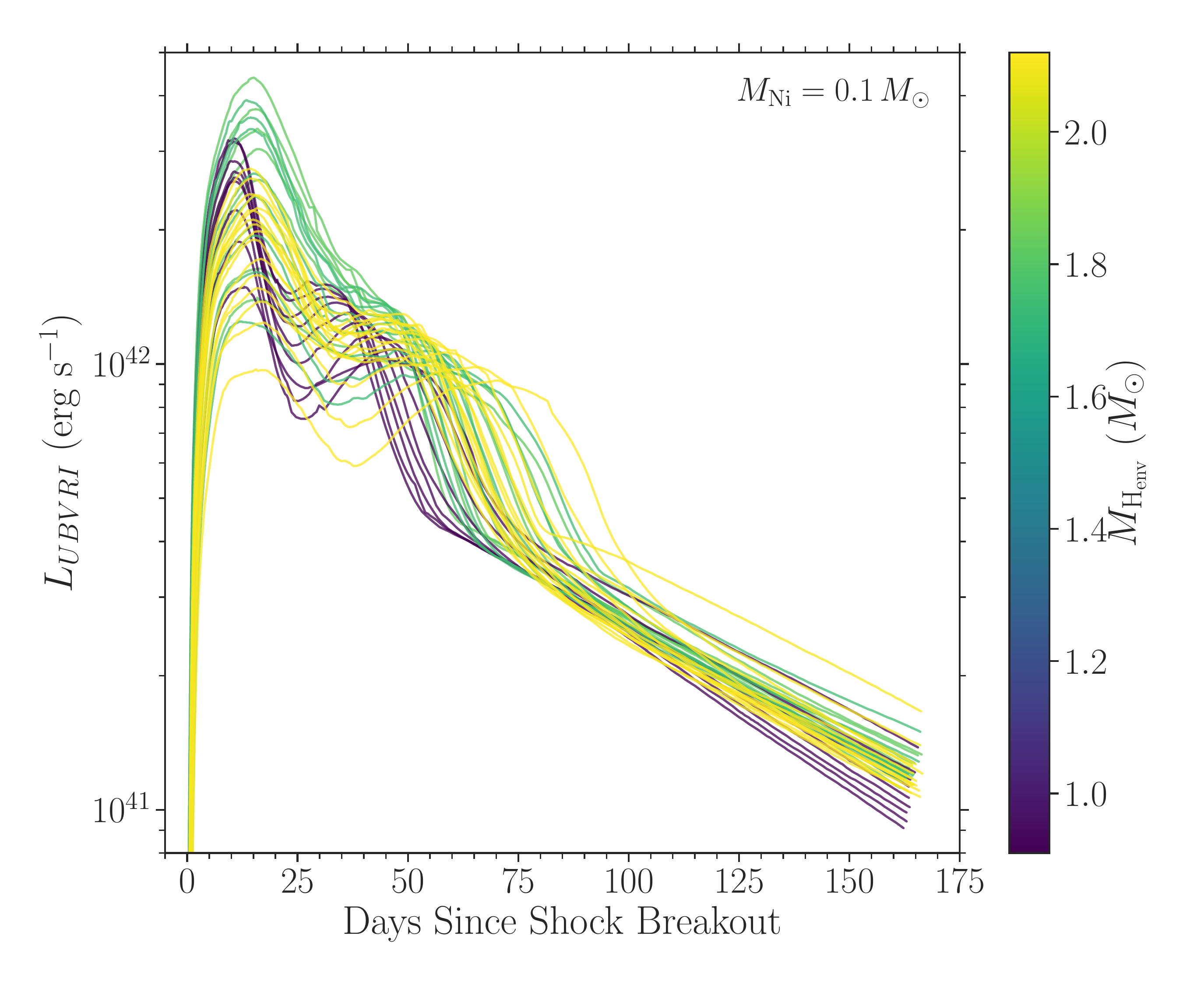}{0.49\textwidth}{\textbf{(c)} Short-Plateau ($0.91\,M_{\odot} \leq M_{\rm H_{\rm env}} \leq 2.12\,M_{\odot}$)}
          \fig{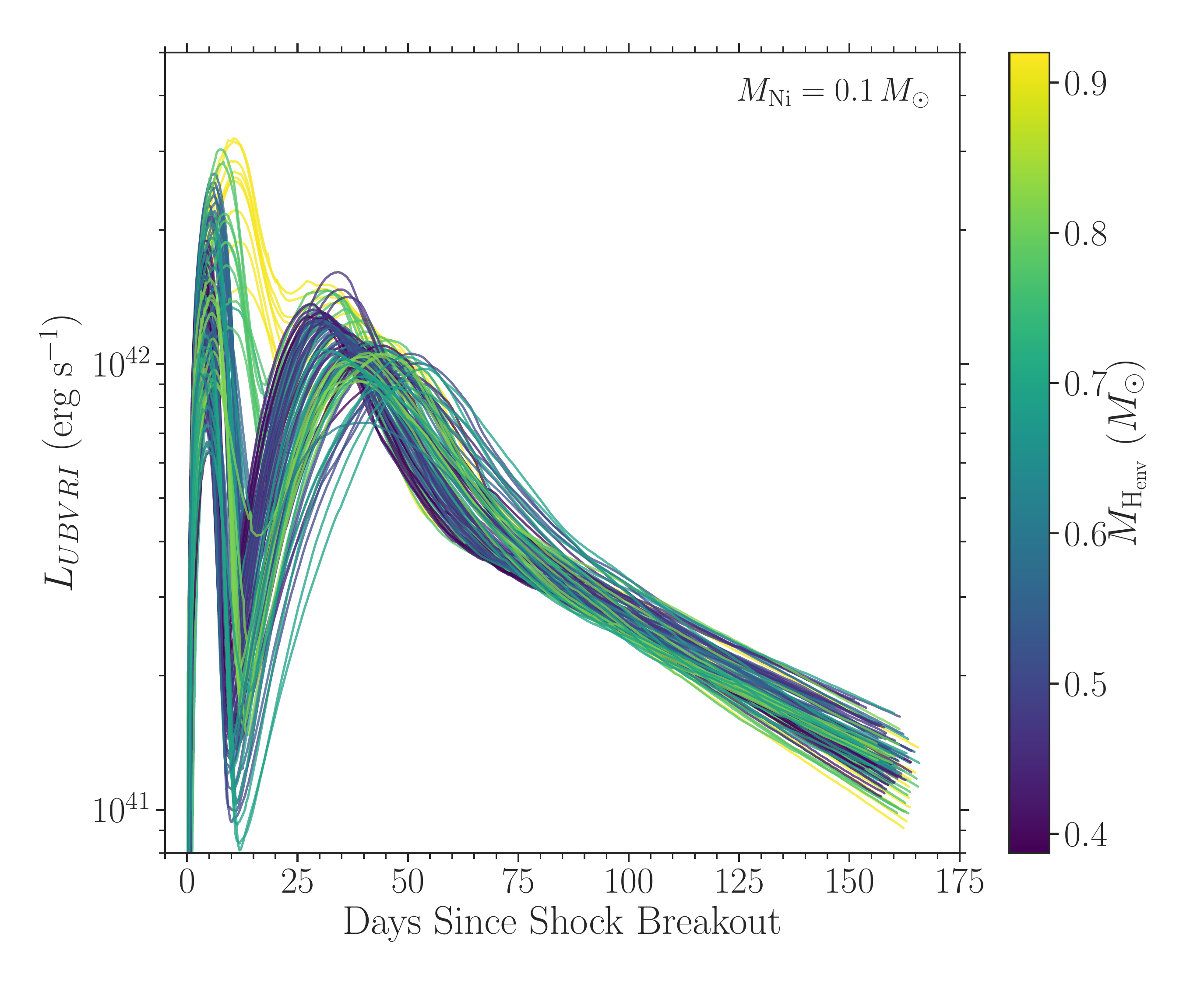}{0.49\textwidth}{\textbf{(d)} IIb ($0.39\,M_{\odot} \leq M_{\rm H_{\rm env}} \leq 0.91\,M_{\odot}$)}}
\caption{Same as Figure~\ref{fig:Model_Lopt}, but with $M_{\rm Ni}=0.1\,M_\odot$.
  } 
  \label{fig:Model_Lopt_010}
\end{figure*}

\begin{figure}
    \centering
    \gridline{\fig{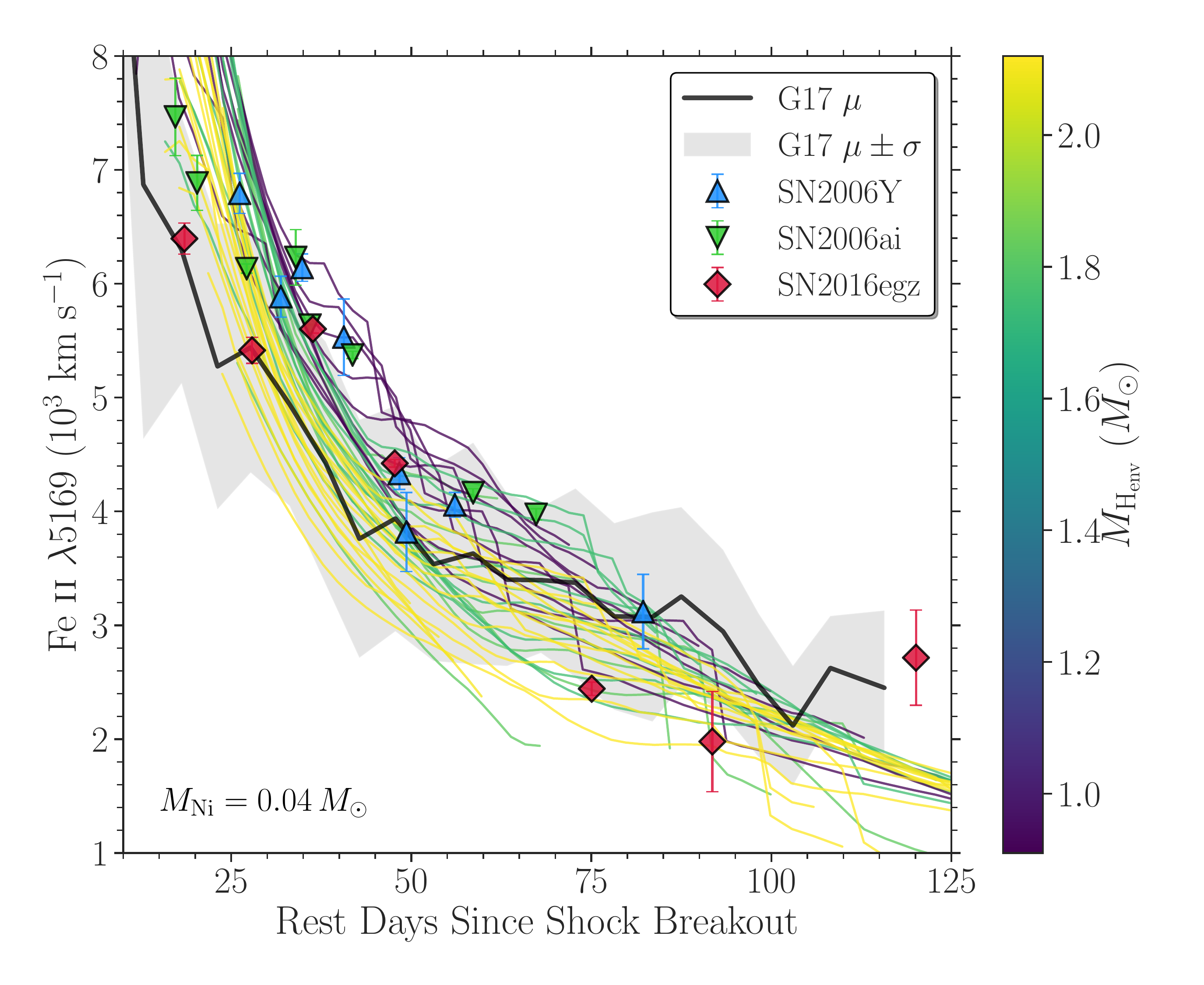}{0.45\textwidth}{\textbf{(b)} $M_{\rm Ni}=0.04\,M_\odot$}
          \fig{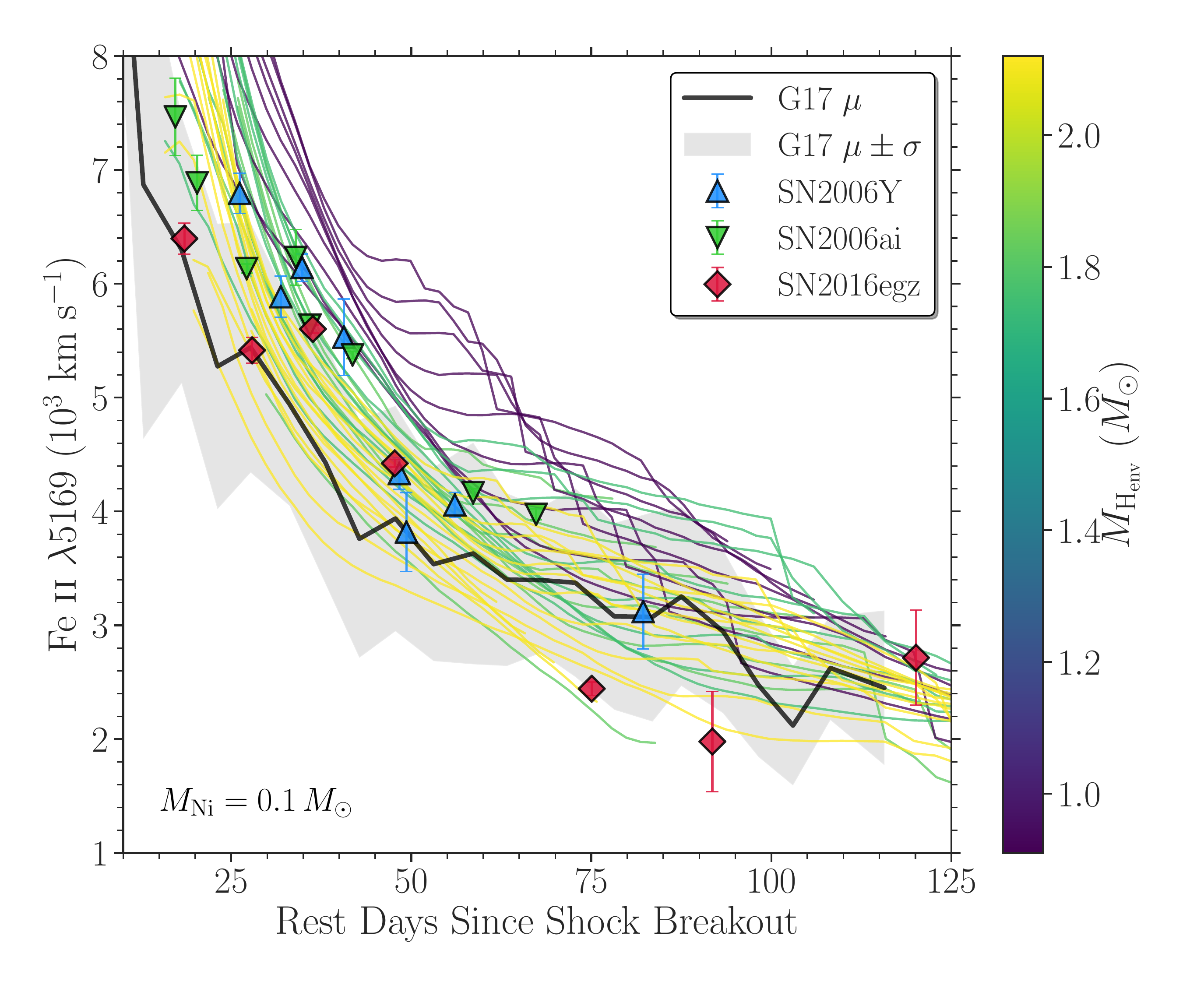}{0.45\textwidth}{\textbf{(a)} $M_{\rm Ni}=0.1\,M_\odot$}}
    \caption{Same as Figure~\ref{fig:VFe}, but with $M_{\rm Ni}=0.04$ and $0.1\,M_\odot$.
    }
    \label{fig:VFe_ext}
\end{figure}

As for the model grids with $M_{\rm Ni}=0.04$ and $0.1\,M_\odot$, the light-curve models are shown in Figures~\ref{fig:Model_Lopt_004} and \ref{fig:Model_Lopt_010}, and the short-plateau SN velocity models are shown in Figure~\ref{fig:VFe_ext}.

\bibliography{main}

\end{document}